%% file: TrafficProfiler_arxiv.tex
\documentclass[conference]{IEEEtran}

\makeatletter
\def\ps@headings{%
	\def\@oddhead{\mbox{}\scriptsize\rightmark \hfil \thepage}%
	\def\@evenhead{\scriptsize\thepage \hfil \leftmark\mbox{}}%
	\def\@oddfoot{}%
	\def\@evenfoot{}}
\makeatother
\IEEEoverridecommandlockouts
\usepackage{amsmath}
\usepackage{amsthm}
\usepackage{amsfonts}
\usepackage{times}
\usepackage{graphicx}
\usepackage{url}
\usepackage[noend]{algpseudocode}
\usepackage{xspace}
\usepackage{multirow}
\usepackage{subcaption}
\usepackage{thmtools, thm-restate}
\usepackage{xcolor}
\usepackage{bbm}
\usepackage{enumitem}
\usepackage{verbatim}

\newcommand{\myparagraph}[1]{\vspace{1pt} \noindent \textbf{#1}}

\newcommand{\etal}{{et al.}\xspace}

\newcommand{\ignore}[1]{}

\newcommand{\thesystem}{\textsc{PortFiler}\xspace}
\newcommand{\uvanetwork}{\textsc{Univ-1}\xspace}
\newcommand{\vtnetwork}{\textsc{Univ-2}\xspace}

\newcommand{\talha}[1]{{\textcolor{blue} {Talha says: #1}}}

\newcommand\pr{\ensuremath{\mathsf{PR}}}
\newcommand\prauc{\ensuremath{\mathsf{PR\mbox{-}AUC}}}

\begin{document}

\title{\thesystem: Port-Level Network Profiling for Self-Propagating Malware Detection}

%\author{Anonymous Submission                                                             }
\author{
	\IEEEauthorblockN{Talha Ongun, Oliver	Spohngellert, Benjamin	Miller, Simona	Boboila, Alina	Oprea, Tina	Eliassi-Rad}
	\IEEEauthorblockA{Northeastern University}
	\IEEEauthorblockN{Jason	Hiser, Alastair	Nottingham, Jack	Davidson, Malathi	Veeraraghavan}
	\IEEEauthorblockA{University of Virginia}
}

\maketitle

\begin{abstract}
	Recent self-propagating malware (SPM) campaigns compromised hundred of thousands of victim machines on the Internet. It is challenging to detect these attacks in their early stages, as adversaries utilize common network services, use novel techniques, and can evade existing detection mechanisms. We propose \thesystem (\underline{PORT}-Level Network Traffic Pro\underline {FILER}), a new machine learning system applied to network traffic for detecting SPM attacks. \thesystem\ extracts port-level features from the Zeek connection logs collected at a border of a monitored network, applies anomaly detection techniques to identify suspicious events, and ranks the alerts across ports for investigation by the Security Operations Center (SOC). We propose a novel ensemble methodology for aggregating individual models in \thesystem\ that increases resilience against several evasion strategies compared to standard ML baselines. We extensively evaluate \thesystem\ on traffic collected from two university networks, and show that it can detect SPM attacks with different patterns, such as WannaCry and Mirai, and performs well under evasion. Ranking across ports achieves precision over $0.94$ and false positive rates below $8 \times 10^{-4}$ in the top 100 highly ranked alerts. When deployed on the university networks, \thesystem\  detected anomalous SPM-like activity on one of the campus networks, confirmed by the university SOC as malicious. \thesystem\ also detected a Mirai attack recreated on the two university networks with higher precision and recall than deep-learning based autoencoder methods.

\end{abstract}

\begin{IEEEkeywords}
	malware detection, security analytics, self-propagating malware, traffic profiling
\end{IEEEkeywords}

\input{intro}

\input{background}

\input{overview}

\input{aggr_method}

\input{dataset}

\input{evaluation}

\input{deploy_aggr}

\input{discussion}

\input{related_work}

\input{conclusion}

\section*{Acknowledgements}
This research was sponsored by the contract
number W911NF-18-C0019 with the U.S. Army Contracting Command - Aberdeen Proving Ground (ACC-APG) and the Defense Advanced Research Projects Agency (DARPA), and 
by the U.S. Army Combat Capabilities Development Command
Army Research Laboratory under Cooperative Agreement Number
W911NF-13-2-0045 (ARL Cyber Security CRA). The views contained in this document are those of the authors and should not be interpreted as representing the official policies, either expressed or implied, of the ACC-APG, DARPA, Combat Capabilities Development Command Army Research Laboratory or the U.S. Government. The U.S. Government is authorized to reproduce and distribute reprints for Government purposes notwithstanding any copyright notation here on. This project
was also funded by NSF under grant CNS-1717634.

%\newpage
\bibliographystyle{abbrv}
\bibliography{portfiler}
%\newpage

\end{document}

%% file: intro.tex
\section{Introduction}

Self-propagating malware (SPM) is a prevalent class of malware that has recently gained in popularity among adversaries.  In 2016, Mirai~\cite{antonakakis2017understanding} infected more than 600K IoT devices and launched devastating denial-of-service attacks against high-profile websites.  In 2017, the WannaCry ransomware attack impacted more than 300K vulnerable devices in 150 countries in a few days~\cite{wannacry}. SPM attacks have been seen in the wild since 2001, with famous attacks such as Code Red being successful at widely propagating and infecting machines on the Internet~\cite{zou2002code}. Their recent surge in popularity can be attributed to an increased number of vulnerabilities being discovered in network services (such as the EternalBlue vulnerability in the SMB protocol exploited by WannaCry and NotPetya), the increasing number of connected devices on the Internet, and the new monetization mechanisms offered by ransomware.

Defending against SPM campaigns is challenging for multiple reasons. A widely-adopted response to stopping WannaCry and NotPetya was to block port 445 allocated to the SMB file-sharing service, a solution adopted today by many ISPs. However, many other SPM attacks leverage common network services that cannot be blocked entirely. For instance,  Mirai  performed scanning over multiple network protocols, including Telnet, HTTPS, FTP, SSH, and CWMP~\cite{antonakakis2017understanding}. As the scanning activities of SPM attacks blend in with large volumes of legitimate traffic, it becomes difficult to distinguish the malicious traffic and block it. Techniques based on machine learning have promise, but they suffer from limited ground truth for training machine learning models, and usually incur large false positive rates, which impact the investigation ability by human experts~\cite{SommerPaxson2010}. Furthermore, attackers behind SPM could employ evasive methods to hinder detection. While the WannaCry and Mirai attacks progressed  rapidly to have a significant global impact, future attacks can be more targeted or stealthier by propagating at slower rates. Thus, new techniques are needed to bridge the gap in the defense side against this new wave of automated, self-propagating attacks.

In this paper, we design \thesystem, a machine learning system that uses  network traffic captured at a border of an organizational network to detect and prioritize emerging SPM attacks. To the best of our knowledge, we are the first to design a robust ML system for SPM malware detection that uses port-based network-level information. The critical insight in this work is that SPM requires remote probing of a large number of machines to propagate effectively. This behavior can be viewed as an invariant characteristic of this class of malware attacks since the success of the operation significantly depends on the deployed propagation mechanism. To capture this behavior, \thesystem\ extracts a set of 35 features from the Zeek connection logs, mapping communication of internal hosts in the network with external destinations.  At training time, \thesystem\ learns the profile of the normal network communication on a set of monitored ports, without relying on traces of existing malware attacks. We design in \thesystem\ novel ensemble methods that train multiple unsupervised ML models and combine their anomaly scores into a single, unified score. The base models in the ensembles can be instantiated with density-based models such as Kernel Density Estimate (KDE), or unsupervised tree-based methods, such as Isolation Forest.  At testing time, \thesystem\  detects traffic anomalies that exhibit different behavior from training observations, and prioritizes the anomalies via score ranking for investigation by security analysts. Security Operations Centers (SOC) have limited manual investigation budgets, and having a low false positive rate in the top-ranked alerts is a strict requirement for an ML model to be deployed.

Another important requirement in the design of \thesystem\ is its resilience to  evasion attacks. We experiment with two evasion strategies that a motivated attacker might employ: reducing the rate of probing external destinations, and leveraging external destinations already visited by internal machines.  We show that standard ML methods degrade in performance as the malware becomes more evasive, but our proposed ensembles are much more robust against evasion. We compare the performance of \thesystem's ensembles with standard ML models and unsupervised deep learning models based on autoencoders, and show the advantages of our method.

Finally, we evaluate \thesystem\ using Zeek~\cite{zeek2020Jun} logs collected on two university networks. We use public malware traces for four SPM families~\cite{stratosphere} and generate our own WannaCry malware variants in a virtual environment. We show that both ensemble strategies result in increased resilience against evasion, and they have low false positive rates.  We evaluate the top ranked alerts at one of the university networks, and confirm with the SOC that a detection on port 445  is malicious. In addition, \thesystem\ detected a Mirai attack recreated on the two university networks with high precision and recall.

We summarize our contributions below:

\begin{itemize}[leftmargin=0.1in]
\item We design \thesystem, a new machine learning system for detecting SPM attacks. \thesystem\ learns the normal traffic profiles on selected ports in a network at training time without relying on traces of existing malware attacks, and can detect suspicious traffic anomalies generated by  SPM campaigns in testing.

\item We propose a novel ensemble methodology for increasing the resilience of \thesystem\ against several malware evasion strategies. For example, the ensemble method maintains an Area Under the Precision-Recall Curve of 0.93 for a WannaCry variant 64x slower than the original, an improvement of 31\% compared to standard KDE models.

\item We develop a methodology for ranking \thesystem\ alerts across ports and generating a unified list for investigation by SOC security experts. For the top 100 ranked alerts, the ensemble method achieves a precision above 0.94 with false positive rates below $8 \times 10^{-4}$ at detecting WannaCry. 

\item We evaluate \thesystem\ on more than 6 billion Zeek logs collected at two campus networks, and confirm our detections as malicious with the SOC. Additionally, we detected a recreated Mirai attack on the two university networks with high precision and recall.

\end{itemize}

The rest of the report is structured as follows. Section~\ref{sec:background}
we discuss the background of self-propagating malware attacks and threat model.
In Section~\ref{sec:system}, we present our methodology to aggregate 
network traffic as well as techniques we developed for anomaly detection. 
In Section~\ref{sec:experimental_setup}, we describe the experimental setup and the dataset used for evaluation. Section~\ref{sec:eval} presents detailed evaluation results for a variety of experiments. We discuss our results and limitations in Section~\ref{sec:discuss}, present related work in Section~\ref{sec:related} and conclude our paper in Section~\ref{sec:conc}.

%% file: background.tex
\section{Problem Definition and Threat Model}
\label{sec:background}

\myparagraph{Self-Propagating Malware (SPM).} In SPM attacks, an infected  machine attempts to propagate indiscriminately on the Internet, with the goal of spreading the infection widely. These attacks can take the form of ransomware, worm, trojan, and other attacks that cripple access to essential assets of users as well as organizations through exploiting vulnerabilities or configuration weaknesses in systems and networks. The adoption of powerful automated tools, along with a diverse set of vulnerable and publicly accessible services, has made self-propagating malware increasingly popular among adversaries. In this section, we set the stage for our work by providing background on specific properties of self-propagating malware attacks. 

SPM attacks involve two phases: (1) \emph{probing}, in which a large number of IPs are probed to identify potential victims; and (2)  \emph{propagation}, in which the malware attempts to infect the identified targets. After each successful infection, the malware continues the propagation to other IPs, and also monetizes the infection through different mechanisms.

We describe two famous SPM attacks, WannaCry and Mirai, that have been extremely successful at reaching hundreds of thousands of victims on the Internet. 
WannaCry leveraged the EternalBlue vulnerability in the Microsoft SMB service over port 445, resulting in more than 300,000 compromised devices in 150 countries in May 2017~\cite{wannacry}. 
During probing, WannaCry probes machines on the local network (\emph{internal probing}), and also probes pseudorandomly-generated IPs on the Internet (\emph{external probing}).
	 Once it identifies Windows machines vulnerable to the SMB exploit, WannaCry attempts to propagate to these machines.
	 
	 WannaCry is mostly known for its ransomware behavior: after propagation, it  encrypts the files on the victim machine and displays a ransom message to the user.
The Mirai campaign is mostly known for its large-scale DDoS attacks performed after getting a large victim base of IoT devices. To get access to these IoT devices, Mirai performed rapid probing on ports 23 and 2323 (later expanded to other ports) using pseudorandom IP addresses~\cite{antonakakis2017understanding}. After Mirai identifies a potential victim IP, it uses a brute-force login phase over Telnet, using common user names and passwords. If the login is successful, device-specific malware is downloaded to each victim. Once the campaign controls a large number of victim devices, they are instructed by the command-and-control center to launch DDoS attacks.

\begin{table*}[]
	\centering
	\vspace{2mm}
\begin{tabular}{lllll}
    \hline
	\textbf{Malware}        & \textbf{Ports}                                               & \begin{tabular}[l]{@{}l@{}}\textbf{Scan Rate}\\\textbf{(IPs per min)} \end{tabular} & \begin{tabular}[l]{@{}l@{}}\textbf{Entropy}\\\textbf{(IP octets)} \end{tabular} & \begin{tabular}[l]{@{}l@{}}\textbf{Infection }\\\textbf{Method}\end{tabular}  \\ 
	\hline
	\textbf{Mirai~\cite{Mirai}}          & {23, 2323}                                                    & 50-300k                                                                             & 7.9829                                                                           & Dictionary attack                                                             \\
	\textbf{Hajime~\cite{herwig2019measurement}}         & {23, 81}                                                     & 10k                                                                                 & 7.9822                                                                           & Vulnerabity Exploit: Dictionary attack, GoAhead-Webs credentials                                                  \\
	\textbf{Kenjiro~\cite{Spadafora2018Sep}}        & {80, 8080, 37215}  & 5k                                                                                  & 7.9832                                                                           & Vulnerabity Exploit: CVE-2016-6563, CVE-2017-17215                                                 \\
	\textbf{WannaCry~\cite{wannacry}}       & 445                                                          & 1750                                                                                & 7.9734                                                                           & Vulnerabity Exploit: CVE-2017-0143                                                                 \\
	\textbf{Hide and Seek~\cite{hideandseek}}  & {23, 9527} & 260                                                                                 & 7.9726                                                                           & Dictionary Attack, Vulnerabity Exploit: CVE-2017-11634       \\
	\hline                                     
\end{tabular}
	\caption{Comparison of SPM scanning patterns. We surveyed 5 malware families exhibiting various scanning rates (number of IPs contacted by a victim per minute) on different ports. A common characteristic of SPM malware is the probing of randomly distributed IP addresses, demonstrated by the entropy values. In our experiments, these parameters (i.e., scanning rate and ports)  are varied to capture a range of SPM behaviors. }
	\label{tab:spm_malware_comparison}
	%\vspace{-7mm}
\end{table*}

We highlight that while the end goal of these two campaigns is completely different -- WannaCry is a ransomware, while Mirai launches DDoS attacks --  both of them use self-propagating behavior to reach a large number of victims on the Internet. In fact, other SPM families we surveyed in Table~\ref{tab:spm_malware_comparison}  exhibit similar behavior as WannaCry and Mirai. We obtained public data for 5 different examples of SPM malware~\cite{stratosphere} and show the scanning rate (IPs probed per minute), as well as the entropy of the probed IP addresses (split into 4 octets) in the same table. We observe that all 5 malware families have entropy close to 8 and the entropy of the uniform distribution is $\log_2 256=8$, which implies that the IP addresses they probe are uniformly distributed. The main differences are in the scanning rate and ports, two properties that we vary in our experiments to capture a range of SPM behavior to evaluate our system against a variety of different potential SPM attacks.

\myparagraph{Problem definition and threat model.} 
We aim to detect SPM attacks in their very early stages of operation, during the initial probing and propagation phases, to prevent the spread of SPM attacks on the Internet. 
	SPM attacks could use a number of network ports to perform their malicious activity, including common ports such as 80 for HTTP and 443 for HTTPS. 
	Our goal is to detect SPM inside a network that is monitored at its border, a common setup used in industry. In particular, network logs between internal machines and external destinations are collected using standard monitoring software (e.g., Bro or Zeek~\cite{paxson1999bro}). 
	We argue that existing techniques for attack detection in networks are not immediately applicable to our problem of interest. 
	Rule-based methods are effective once an attack is already known. For instance, after the famous WannaCry campaign, most ISPs started blocking port 445~\cite{445block}. While this is an effective countermeasure against the specific WannaCry campaign, it fails to capture new SPM malware on other ports with different behavior. Other techniques to detect WannaCry rely on host-based information~\cite{kumar2018investigation, akbanov2019ransomware} and target mostly the ransomware aspect of the malware. 
	Botnet detection methods  based on machine learning (ML)~\cite{gu2008botminer,bilge2012disclosure,tegeler2012botfinder,alahmadi2020botection}  do not capture the patterns of SPM and need attack traces for training supervised learning models. Unsupervised threat detection techniques for enterprises, such as Beehive~\cite{yen2013beehive}, detect a large number of anomalies, not necessarily associated with SPM behavior. Our goal, in contrast, is to detect SPM while it propagates over the Internet, by using machine learning applied to network logs collected at the border of the monitored network.

We assume that the attack compromises one or several victims inside the monitored network and we aim to detect SPM behavior without knowledge of the port and service on which the malware propagates.  We rely on Zeek network logs collected at the border of the monitored network, which we assume are not compromised by the attacker. Thus, the information recorded in the logs is reliable and can be used for analysis. 
Our goal is to handle advanced SPM attackers, even those that employ evasion strategies against our detection methods. 
We consider several evasion strategies an attacker might perform. First, attackers could slow down the probing rate, without using much information about our detectors, at the cost of reducing the impact of the attack. Second, a more powerful attacker could observe the traffic on the targeted ports and leverage existing external IPs for scanning. This strategy evades some of the most relevant features used by our detectors, but it requires additional knowledge for the attack to be performed. 
We assume that attackers may have knowledge about the features and ML models used for detection. We design machine learning models for SPM detection that are robust against different evasion strategies.

\myparagraph{Challenges.} Some of the challenges in applying ML for SPM detection are similar to well-known challenges in network intrusion detection~\cite{apruzzese2018effectiveness,SommerPaxson2010}. In particular, there is limited availability of ground truth for SPM attacks, making the use of supervised learning difficult in this setting. We thus resort to unsupervised learning methods that can detect SPM behavior  on different ports. Furthermore, we aim to generate and prioritize alerts for investigation by SOC human experts, and reducing false positive rates for anomaly detection is a critical requirement, which is known to be challenging~\cite{apruzzese2018effectiveness,SommerPaxson2010}. Challenges specific to SPM malware detection include the ability of attackers to blend in with benign traffic to evade ML-based detectors using new vulnerabilities in common network protocols. There is a wide range of network ports and services on which attackers can mount SPM campaigns, and defenders might not have a priori knowledge of existing vulnerabilities. 

In particular, some of these ports (80, 443) exhibit large volume of traffic and high variation of traffic patterns, resulting in difficulty of predicting the legitimate behavior and distinguishing malicious activities on these ports. 
As usual, there is an asymmetry between attackers and defenders -- it is enough for attackers to exploit a single service  to be successful, while defenders need to protect all network services and ports.

%% file: overview.tex
\begin{figure*}[t]
	\centering
	\includegraphics[width=0.8\linewidth]{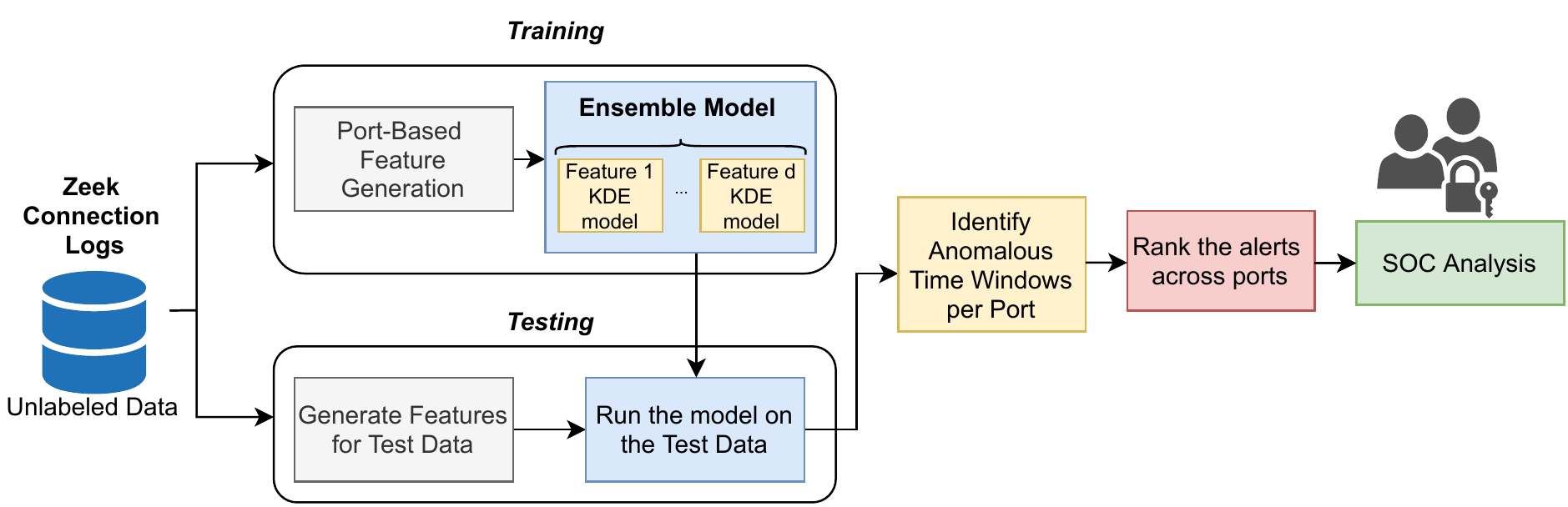}
	\caption{\thesystem\ System Overview. \thesystem\ extracts a set of 35 features for a fixed time window and port. During training, it learns  the legitimate distribution of network traffic on each port, using our ensemble methodology. During testing, anomaly scores are generated on new network logs, and the most anomalous connections are prioritized for SOC investigation. }
	\label{fig:system_diagram}
\end{figure*}

\section{\thesystem\ System Overview and Methodology}
\label{sec:system}
In this section, we provide an overview of the \thesystem\ system designed  to proactively detect SPM attacks. We then discuss our machine learning (ML) methodology for \thesystem.

\subsection{Overview}

Figure~\ref{fig:system_diagram} shows an overview of our system. \thesystem\ trains newly-introduced unsupervised ensemble models that learn the legitimate network traffic distribution, and applies them at testing time to identify and rank anomalies. A prioritized list of alerts is presented to SOC  for investigation.

\myparagraph{Network data monitoring.}
\begin{table}[]
	\begin{center}
		\begin{tabular}{|c|c|}
			\hline
			Field & Definition \\
			\hline
			$\mathtt{ts}$ & Timestamp \\
			$\mathtt{id.orig\_h}$ & Source IP address \\ 
			$\mathtt{id.orig\_p}$ & Source port \\ 
			$\mathtt{id.dest\_h}$ & Destination IP address \\
			$\mathtt{id.dest\_p}$ & Destination port \\
			$\mathtt{proto}$ & Transport protocol \\
			$\mathtt{duration}$ & Duration of connection \\
			$\mathtt{orig\_bytes}$ & Source payload bytes \\
			$\mathtt{resp\_bytes}$ & Destination payload bytes \\
			$\mathtt{orig\_pkts}$ & Source number of packets \\
			$\mathtt{resp\_pkts}$ & Destination number of packets \\
			$\mathtt{conn\_state}$ & Connection state \\
			
			\hline
		\end{tabular}
	\end{center}
	\caption{Fields from Zeek {\sf conn.log}.}
	\label{tab:connlog}
\end{table}
Zeek~\cite{zeek2020Jun} is a well-known network monitor that processes raw packet captures to generate network logs. For \thesystem, we use the Zeek connections logs ({\sf conn.log} files) that record connection metadata for both TCP and UDP connections crossing the border of two university networks we monitor. 
These logs produce an entry for each connection between a source (called originator IP) and destination (called responder IP).
The fields available in the {\sf conn.log} data include timestamp of connection start, source and destination IP and port, transport protocol, duration of connection, source and destination payload bytes and number of packets, as well as connection state as given in Table~\ref{tab:connlog}. The $\mathtt{conn\_state}$ field indicates one of the valid states according to the TCP state machine (e.g., S0, S1, SF, etc.).

\myparagraph{Model training.} 
For the ports of interest, \thesystem\ profiles the traffic and trains ensemble models for learning the regular communication patterns. We define a set of 35  features for each port and time window of fixed length. To learn the distribution of the traffic features per port, we introduce a new methodology for creating ensembles of anomaly detectors that combine anomaly scores produced by multiple ML models, increasing resilience against evasion. The individual models in the ensemble can be either density estimation models, such as Kernel Density Estimation (KDE)~\cite{silverman1986density} or tree-based models, such as Isolation Forest (IF)~\cite{liu2008isolation}. 

Models learn the distribution of network connections \emph{per port}, as different applications leverage different network ports and traffic profiles of various ports are different.

\myparagraph{Ranking alerts.} At testing time, we apply the trained ensemble models on new Zeek {\sf conn.log} data. We also develop methodology to rank the alerts across ports and prioritize them  for investigation by SOC. 
We evaluate the performance of ranking  and perform an analysis of the top detections of \thesystem\ in the two campus networks we monitor.

\myparagraph{Ethical considerations.} We obtained access to Zeek logs collected on two university networks. The IP addresses of the internal machines are anonymized in a consistent manner to protect personal information about the machines or users on the network. We performed all our analysis on servers within the university network, without downloading the data locally. The IRB  office at one of the universities reviewed our data collection and anonymization process and determined that our research does not qualify as Human Subject Research.

%% file: aggr_method.tex
%\section{Methodology}

%\thesystem generates aggregated features per port, by combining all connections generated by all internal IP in a fixed time window on that port. We describe the set of 37 features extracted for \trprofiler, the two unsupervised ML models that are trained, and our methodology for ranking the alerts. Ranking has two components: ranking alerts for each port, and ranking alerts globally across ports to prioritize the alerts a human analyst can investigate.

\input{features_traffic.tex}

\begin{figure}[t]
	\centering
	\includegraphics[width=\linewidth]{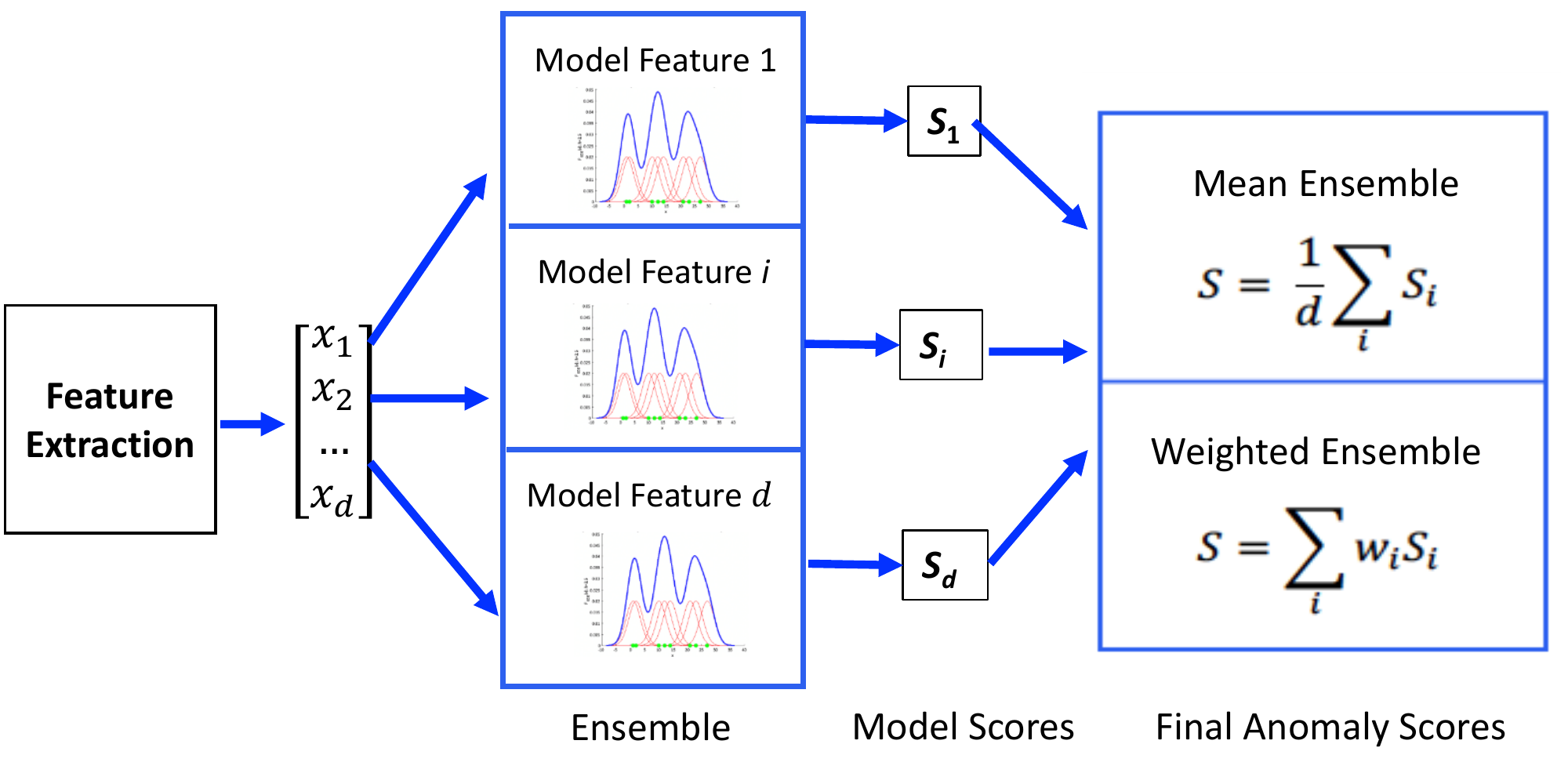}
	\caption{\thesystem\ overview of the novel ensemble methods we introduce. At training, features are extracted per port and time window. A separate ML model is trained on each of the $d$ features. At testing, new data is evaluated against all the $d$ models, to produce anomaly scores $S_1,\dots,S_d$. The scores are combined into a final anomaly score $S$ either by a mean ensemble (using equal weights) or weighted ensemble (using pre-defined weights).}
	\label{fig:ensembles}
	  \vspace{-4mm}
\end{figure}

\subsection{Novel Unsupervised Ensemble Models}
\label{sec:ens}

Network traffic profiling can be performed using a range of unsupervised ML models. One of our main design criteria, as mentioned, is the resilience of our models to advanced malware evasion strategies. As we show in our evaluation, standard  ML models fail in the face of  evasive SPM. This motivates us to introduce our new ensemble methodology for learning network traffic distribution, while achieving resilience against evasion. We implement and evaluate several existing ML models for  comparison to \thesystem's ensembles. These include a simple threshold-based baseline method, standard Kernel Density Estimate (KDE), Isolation Forest (IF), and an unsupervised deep learning method based on autoencoders.

KDE is a well-known method used to estimate probability densities of random variables. Given training data points $x_1,\dots,  x_n \in \mathbb{R}$ drawn from an unknown distribution with probability density $f$, KDE estimates the distribution's density by averaging the outputs of a kernel function $K$, applied to data centered at each point $x_i$ in the sample, and having the same bandwidth $h$. The estimated density $\hat{f}$ is:
$$\hat{f}(x) =\frac{1}{nh} \sum_{i=1}^n K \Large(\frac{x-x_i}{h}\Large) $$
A common choice for kernel functions is the Gaussian kernel, which uses the standard normal density as the kernel function. For KDE with Gaussian kernels, the bandwidth $h$ thus represents the variance of the normal density centered around each training data point. The method can be extended to multi-dimensional density estimation. In our setting each training example $x_i$, for $i \in \{1,2,\dots,n\}$ is $d$-dimensional, representing the set of features extracted per time window for a given port.

Once a KDE model is trained, the probability density function learned by the model is $\hat{f}$. At testing, we estimate the probability density for all new standardized samples $x$ (consisting of a set of 35 traffic features) by $\hat{f}(x)$.  If the probability density falls below a threshold $T$ ($\hat{f}(x)<T$), we deem $x$ as an anomaly. The threshold $T$ is determined at training time based on the legitimate data to fix the false positive rate.

Isolation Forest is a tree-based learning algorithm that attempts to isolate anomalies in data. The algorithm randomly chooses a feature and a split value within the value range of that feature, and applies this partitioning  recursively in order to ``isolate'' observations. The main intuition is that anomalies are easier to isolate than the in-distribution points. Each partition generates an isolation tree, in which anomalies can be partitioned with shorter paths than legitimate points. An Isolation Forest model constructs an ensemble of Isolation Trees, each generated independently. An anomaly score for a data point is defined by averaging the path lengths required to partition that point in each tree.
 We use Isolation Forest to model our training data, and then apply the model in testing to obtain the anomaly scores of the test data.

Ensembles of multiple ML models have been used extensively in supervised learning tasks for improved generalization, but not as much in unsupervised learning tasks like ours. The distribution learning task is particularly challenging due to feature correlation, which can degrade performance of multi-feature models. When analyzing the network logs, we observed that many of \thesystem\ features are naturally correlated. An ensemble of models trained on individual features addresses this issue, since each individual ML model learns a single-dimensional distribution, a much more tractable task. Similar to standard bagging methods for supervised learning, our proposed ensembles generalize better at testing time~\cite{Bagging}, which also implies more resilience under adversarial evasion, as shown in our experiments.

Figure~\ref{fig:ensembles} gives an overview of our proposed ensemble method for network traffic profiling.  During training, individual ML models are trained on each traffic feature and an ensemble of all these models is generated.  If training data samples $x_i=(x_{i1},\dots,x_{id})$ are $d$-dimensional for $i \in \{1,\dots,N\}$, we train $d$ unsupervised ML models: $f_1,\dots,f_d$, each $f_j$ taking feature $j$ of a sample as input. In our best-performing ensemble, we instantiate the individual base models $f_j$ with KDE models (but we experimented with ensembles of Isolation Forest  models as well).

We combine the anomaly scores generated by individual models into a single \emph{ensemble anomaly score} using weights $w_i$:
$$\mathsf{AnomalyScore}(x) = \sum_{i=1}^d w_i S_i$$

We consider two ensemble methods, which differ in model weight assignments:

 \myparagraph{\it 1) Mean Ensemble:} 
Each ML model contributes equally to the final anomaly score. This is useful when no a priori information is known about the attack.
 
 \myparagraph{\it 2) Weighted Ensemble:} 
 This uses a weighted combination of models for computing anomaly scores. This method has significant advantages in cases in which some information about the attack becomes available, and some features have higher relevance for a particular attack.

\myparagraph{Model weight computation.} An interesting property of  weighted ensembles is that they can be adapted to different attacks, by assigning higher weights to more relevant features. The model weights can be computed with a variety of methods. For instance, domain experts could assign weights manually based on attack forensic analysis. Instead, we wanted to find an automated procedure to assign model weights.

Initially, we experimented with unsupervised feature importance extraction for Isolation Forest, using the DIFFI method proposed by Carletti et al.~\cite{carletti2019explainable}. We evaluated this method through an exhaustive hyperparameter search and compared the results against prior knowledge of feature importance coefficients obtained with a supervised Random Forest approach. DIFFI was not able to correctly identify important features in our data\footnote{Based on communication with the authors, we determined that the unsupervised feature selection method based on subsampling may not be able to capture the characteristics of our data.}.
That led us to a supervised approach for weight computation, in which we assume the availability of an attack trace, which is a variant of the attack we are interested in (we generate multiple variants of WannaCry in our experiments by varying propagation rate and interval between probes). We use the labeled attack variant merged with one day of legitimate data and train a Random Forest classifier. We compute feature importance for the classifier and assign weights proportional to these values. This procedure identifies the features that have high importance for the attack, and assigns higher weights to the corresponding models in the ensemble.

\ignore{
We explore two methods of combining scores: 
\begin{itemize}
\item By mean: In this approach, the final score per (port, window) tuple is the \emph{mean} of all scores given by all single dimensional models on the same port and window.
\item By feature importance: In this approach, the final score per (port, window) tuple is a \emph{weighted sum} of scores given by all single dimensional models on the same port and window. The weighted sum coefficients represent the feature importance and are determined in a semi-supervised fashion by training a RandomForest classifier on a separate labeled dataset. This dataset is specifically used for this task only and is not used in the evaluation of the models.
\end{itemize}
}

\begin{figure}[t]
	\centering
	\includegraphics[width=\columnwidth]{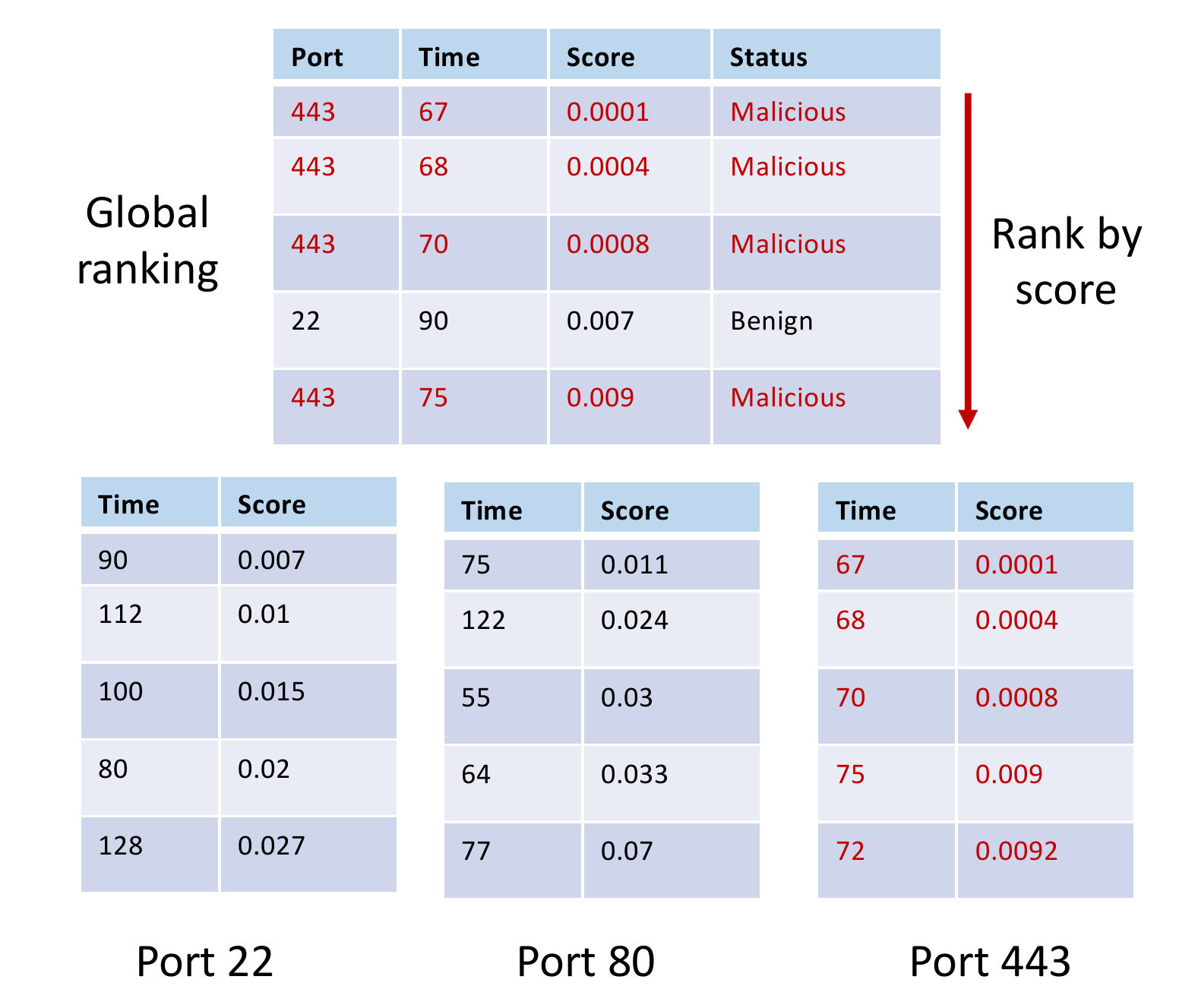}
	\caption{Illustration of global ranking across ports. Ranked alerts for port-based ML models are shown on the bottom. Globally ranked alerts across ports are shown on the top. Malicious activity is highlighted in red. A false positive occurs on port 22 in the top 5 globally ranked alerts.}
	\label{fig:ranking}
\end{figure}

\input{ranking.tex}

%% file: features_traffic.tex
\subsection{Port-Based Traffic Features}

We would like to capture the traffic patterns generated on the monitored network, and create features that distinguish SPM behavior from regular communication patterns.  An important consideration when defining our set of features is to profile the traffic characteristics of different applications running in a network. As applications are assigned specific ports, we choose to define the \thesystem\ features at the port level. This also enables our system to be lightweight, instead of computing features for each individual host in the network, which would be computationally expensive. We consider five ports of interest: 445 (the SMB port, used originally by the WannaCry malware), 80 (HTTP), 443 (HTTPS), 22 (SSH), and 23 (Telnet, used by Mirai). We selected these ports after analyzing several SPM malware (see Table~\ref{tab:spm_malware_comparison}), but our models can be easily extended to monitor a larger set of ports for SPM-like behavior.

\begin{table*}[]
	\begin{center}
		\begin{tabular}{|c|c|c|c|}
			\hline
			Category & Field & Operator & Definition \\
			\hline
			Traffic statistics & $\mathtt{id.dest\_h}$ & Distinct &  Number of distinct external $\mathtt{IP}$s \\
			& or $\mathtt{id.orig\_h}$ & & communicated with \\
			& & Distinct &  Number of distinct internal $\mathtt{IP}$s \\
			& & Count & Number of connections\\
			%			& & & communicated with \\
			& & Distinct &  Number of new distinct external $\mathtt{IP}$s\\
			%			& & & communicated with \\
			\hline
			Duration & $\mathtt{duration}$ & Var & Duration variance of connections \\
			& & Max & Max duration of connection \\
			& & Mean & Mean duration of connection \\
			& & Min & Min duration of connection \\
			\hline
			Bytes & $\mathtt{orig\_bytes}$   & Var & Bytes sent variance of connections \\
			& & Max &  Max bytes sent by an $\mathtt{IP}$ in a connection\\
			&  & Mean & Mean bytes sent by an $\mathtt{IP}$ \\
			& $\mathtt{resp\_bytes}$  & Var & Bytes received variance of connections \\
			& & Max &  Max bytes received by an $\mathtt{IP}$ in a connection\\
			& & Mean & Mean bytes received by an $\mathtt{IP}$\\
			& & Count & Number of connections with no bytes received \\
			\hline
			Packets & $\mathtt{orig\_pkts}$   & Var & Packets sent variance of connections \\
			& & Max &  Max packets sent by an $\mathtt{IP}$ in a connection\\
			&  & Mean & Mean packets sent by an $\mathtt{IP}$ \\
			& $\mathtt{resp\_pkts}$ & Var & Packets received variance of connections \\
			& & Max &  Max packets received by an $\mathtt{IP}$ in a connection\\
			& & Mean & Mean packets received by an $\mathtt{IP}$\\
			\hline
			Connection state &  $\mathtt{conn\_state}$ & Count & Number of connections for each state \\
			& & Count & Number of unsuccessful connections \\
			
			\hline
		\end{tabular}
	\end{center}
	\captionsetup{justification=centering,margin=0.5cm}
	\caption{Traffic features used in \thesystem\ per time window. These features are extracted for each of the monitored ports.}
	\label{tab:features_traffic}
\end{table*}
      
\ignore{In our first attempt, we consider the set of all destination IP addresses that $\mathtt{ip}$ communicates with: $S_{IP} = \{IP_1,\dots,IP_n\}$. From these we can define the set of /24 destination subnets that $\mathtt{ip}$ communicates with: $S_{subnet} = \{Sub_1,\dots,Sub_m\}$, with $m \le n$. If we define aggregated features per destination or subnet, we will encounter an issue when a host visits new IPs or new destinations. In that case, we need to add new features to our representation, which is not desirable in practice.}

 \ignore{To alleviate this problem, we instead define our aggregated features by destination port (corresponding to applications or network services). We define a set of popular application ports (i.e., HTTP - 80, HTTPS - 443, SSH - 22, DNS - 53, FTP, Other). }
      
We aggregate all logs for a fixed time window and extract a set of 35
features, given in Table~\ref{tab:features_traffic}. 
 These capture the following statistics on the set of connections on a particular port:

 \myparagraph{\bf Traffic statistics features:} 
We extract several traffic statistics features: number of distinct internal and external IPs communicating on that port, number of connections, and number of new distinct external IPs (that have not been contacted before) per port. We expect to observe an increase in these features  during periods with high SPM activities.

 \myparagraph{\bf Duration features:} 
We extract max, min, variance and mean of connection duration values during each time window. SPM connections have a lower duration than most legitimate traffic.

 \myparagraph{\bf Bytes and packets features:} 
 We extract max, variance, and mean of sent and received byte and packet values. The distribution of bytes sent and received during SPM infection might be different compared to normal periods, and similarly for the packet distribution. We also define ``number of connections with no bytes received'' as most of SPM connections are for non-responding IPs and result in zero bytes received.

 \myparagraph{\bf Connection state:} 
We define the number of connections for each state (e.g., S0, S1, OTH, etc.). In particular, a large number of failed connections might be indicative of a large number of SPM probing and propagation attempts. We also define the number of failed connections, a feature that aggregates multiple connection failure states.

We highlight an engineered feature that consistently shows highest feature importance in our evaluation. \ignore{This feature was designed to capture the SPM propagating behavior that probes many IP destinations in a time window. } The \emph{new external IPs} feature represents the number of new external IPs, (i.e., IPs that have not been seen before) contacted per port within each time interval. To compute this feature, we create a history  of visited external IPs and update it over time. This feature captures the randomness in SPM propagation behavior, which results in many new, previously unvisited IP destinations. 
For each external IP, we verify whether it has been visited before by checking the history. If it has not been seen, we count it as a ``new IP'', and add it to the history as well. The new IPs feature has proven very useful in \thesystem, as shown in Section \ref{sec:eval}, since these attacks propagate by probing random (therefore new, previously unvisited) IP destinations.

%% file: ranking.tex
\subsection{Ranking}
\label{subsec:ranking_method}
Motivated by constraints on SOC investigation time, our goal is to prioritize the alerts generated by our models and provide the highly ranked alerts to human experts. When each port is analyzed individually, ranking alerts is based directly on anomaly scores: the lower the score, the more anomalous a sample is.
In reality, SOC analysts might not know in advance which ports are exploited by SPM,  and analyzing alerts on each port is time consuming. To address this issue, we rank the alerts \emph{across all ports} to generate a unified list of the most suspicious alerts an analyst should investigate. The standard KDE model returns probability densities, which are aggregated by the ensemble, but are not directly comparable across ports. To address this issue, we normalize the probability densities by computing the Complementary Cumulative Distribution Function (CCDF) for each sample $x$. 
We develop methodology for ranking alerts for each port, as well as combining them across  ports to create a single, unified list of prioritized alerts.

\myparagraph{Per-port ranking.} 
During testing, we run the ML models on every port and obtain an estimation of how anomalous the feature vector collected at each time window is. The standard KDE model estimates the probability density function of a random variable (in our case, the feature vectors). The lower the probability density, the more anomalous a sample is. Isolation Forest returns the anomaly score of each sample directly. This score represents the path length from the root node to the terminating node. The shorter the path, the more anomalous a sample is.  

With ensemble models, the anomaly scores generated by individual models are combined into a single ensemble anomaly score as described in the previous section. We thus rank the alerts  of the test day for a port based on the lowest anomaly scores given by the model. 

\myparagraph{Global ranking across ports.} 
For ensembles using KDE as the base ML model, we  compute weighted combinations of CCDF scores for ranking. If $f_1,\dots,f_d$ are the $d$ base models in the ensemble, the score of a data point $x=(x_1,\dots,x_d)$ is: 
\begin{equation}
 \resizebox{1.0\hsize}{!}{%
$C(x) = \sum_{i=1}^d w_i \mathsf{CCDF}(x_i) = \sum_{i=1}^d w_i (1- \int_{-\infty}^{x_i} f_i(x)) dx$
}
\nonumber
\end{equation}
Thus, we aggregate the CCDF values using the ensemble weights. 
We then rank the alerts by CCDF values, with the most anomalous ones having the lowest CCDF. 
We use this normalized anomaly score to rank across ports and prioritize the alerts with the lowest scores across all ports. 
For ensembles using Isolation Forest as the base ML model, normalization is not necessary, since the score represents the path length from the root node to the terminating node (rather than a probability density). The shorter the path, the more anomalous a sample is. Thus, we can compare and rank the weighted ensemble scores directly across ports.

The advantage of performing global ranking is to reduce the human effort required for post-analysis, since an ordered, unified list is presented for further investigation.

See an example of a prioritized list of alerts ranked across ports in Figure~\ref{fig:ranking}.

\ignore{Intuitively, points with the lowest CCDF have a very low probability according to the trained model, and are ranked at the top.

\begin{equation}
CCDF(x) = 1 - P(X \leq x) = 1- \int_{-\infty}^x \hat{f}(x) dx
\nonumber
\end{equation}
}

%% file: dataset.tex
\section{Experimental Setup}
\label{sec:experimental_setup}

\begin{table}[t]
	\begin{center}
		\scalebox{0.8}{
\begin{tabular}{c|c|c|}
	\cline{2-3}
	& \multicolumn{2}{c|}{\textbf{Number of Events}} \\ \hline
	\multicolumn{1}{|c|}{\textbf{Day}}     & \textbf{\uvanetwork}      & \textbf{\vtnetwork}     \\ \hline
	\multicolumn{1}{|c|}{\textbf{Sept 3}}  & 638 million                & 292 million              \\ \hline
	\multicolumn{1}{|c|}{\textbf{Sept 4}}  & 656 million                & 289 million               \\ \hline
	\multicolumn{1}{|c|}{\textbf{Sept 5}}  & 637 million                & 278 million               \\ \hline
	\multicolumn{1}{|c|}{\textbf{Sept 6}}  & 536 million                & 238 million               \\ \hline
	\multicolumn{1}{|c|}{\textbf{Sept 7}}  & 370 million               & 165 million               \\ \hline
	\multicolumn{1}{|c|}{\textbf{Sept 8}}  & 416 million               & 195 million               \\ \hline
	\multicolumn{1}{|c|}{\textbf{Sept 9}}  & 660 million               & 284 million               \\ \hline
	\multicolumn{1}{|c|}{\textbf{Sept 10}} & 675 million             & 286 million               \\ \hline
\end{tabular}}
	\end{center}
	\caption{Dataset statistics for the Zeek {\bf conn.log} collected from the two university networks in September 2019.}
	\label{tab:dataset_stat}

\end{table}

\subsection{Datasets for Evaluation}
\label{sec:dataset}

\myparagraph{Network logs at two university networks.} The network traffic used for evaluation is the anonymized Zeek {\sf conn.log} data provided by two universities: University of Virginia (\uvanetwork) and Virginia Tech (\vtnetwork).

Our experimental setup consists of one week of training and one day of testing, resulting in a total of 1.5 TB of data and 9.64 billion events for \uvanetwork, and 1.2 TB of data and 9.69 billion events for \vtnetwork. 
Table \ref{tab:dataset_stat} provides the number of events on the five ports (22, 23, 80, 443, 445) that our system profiled. 

For each day of training and testing, we extract the 35 features 
described in Table \ref{tab:features_traffic} 
on the five ports of interest. 
Processing one day to extract features takes on average 1.7 hours for \uvanetwork traffic and 0.4 hours for \vtnetwork traffic.

\begin{figure}[]
	\centering
	\begin{subfigure}[b]{0.49\linewidth}
		\includegraphics[width=\linewidth]{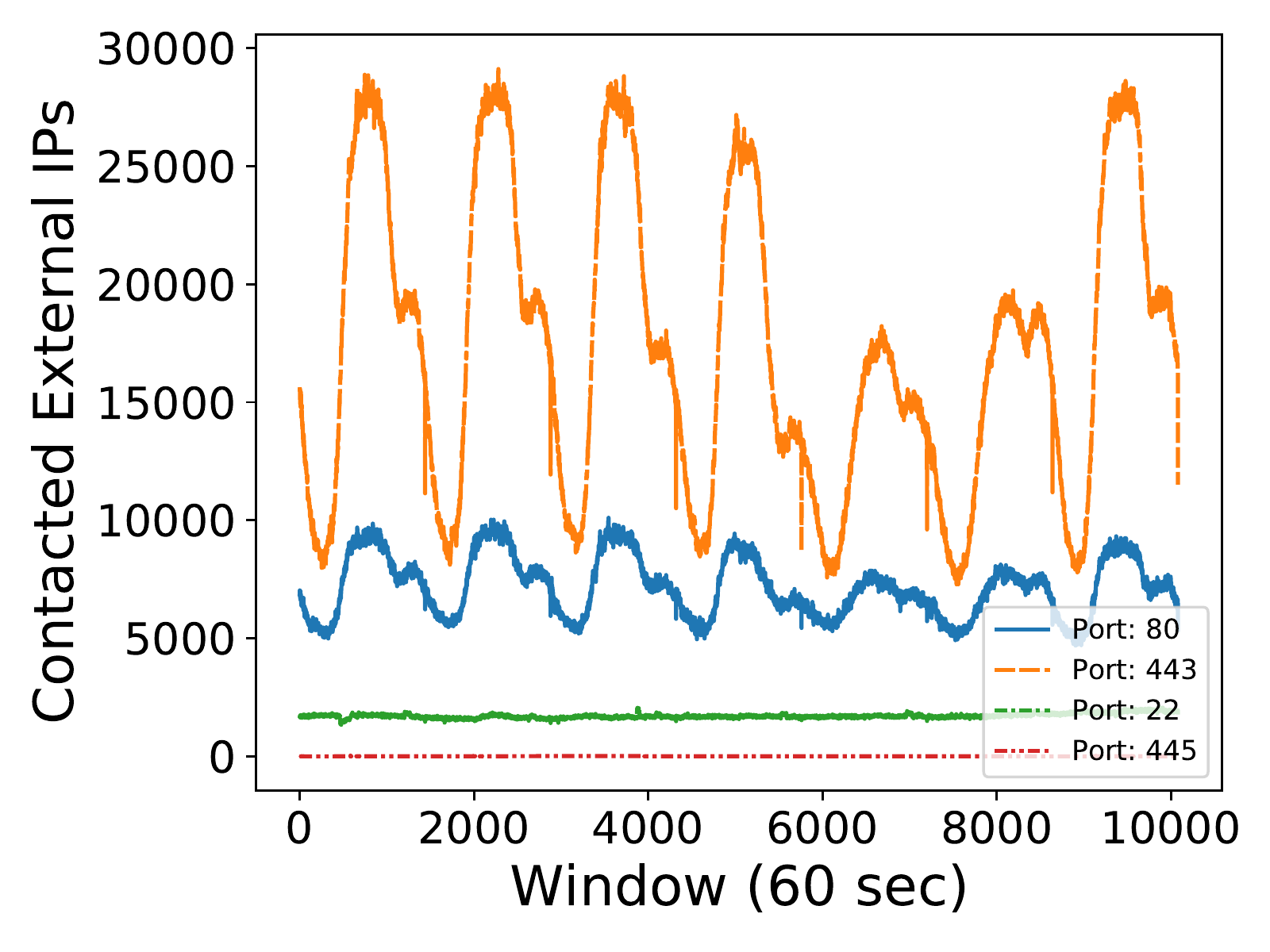}
	\end{subfigure}
	\begin{subfigure}[b]{0.49\linewidth}
		\includegraphics[width=\linewidth]{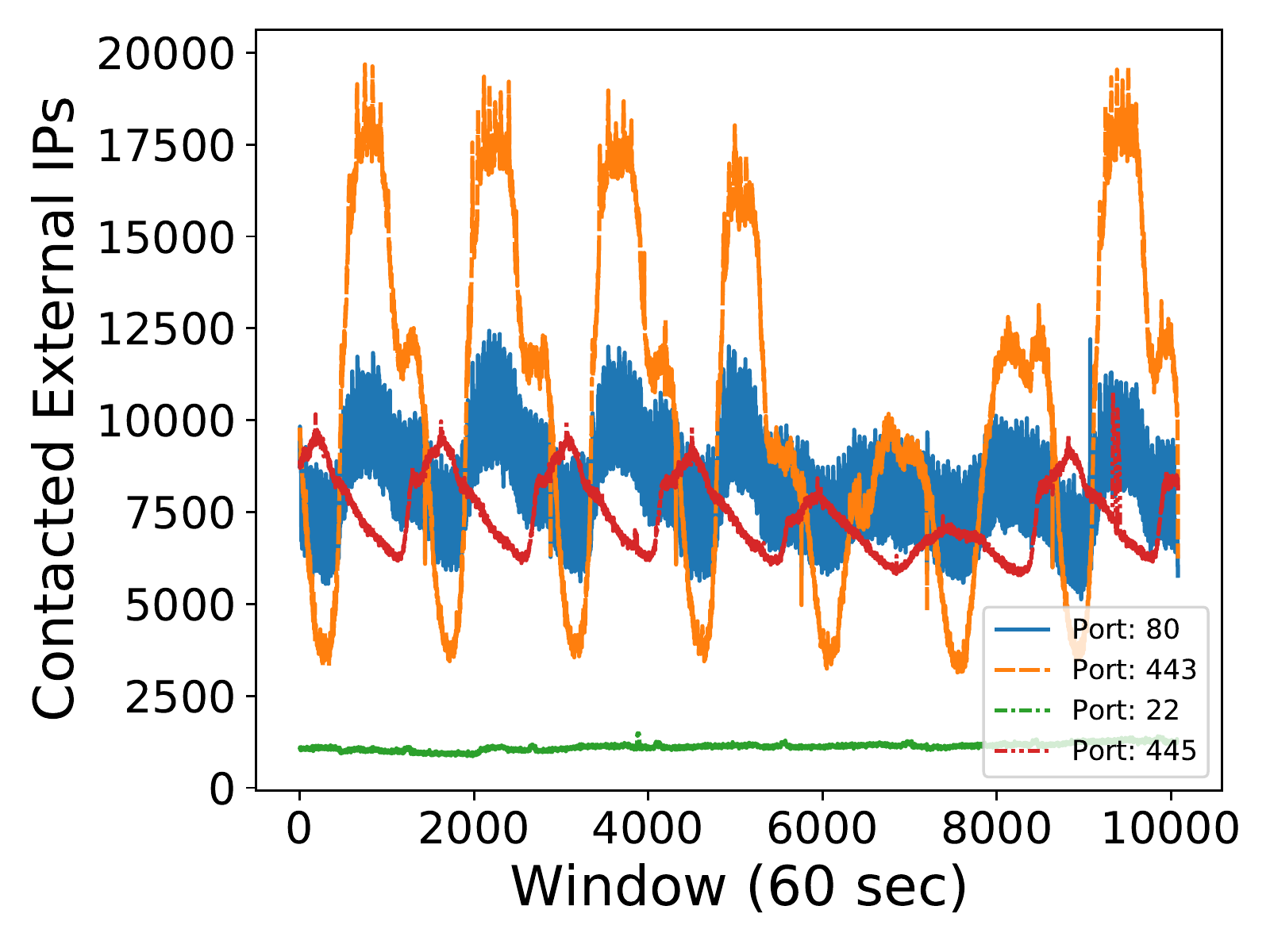}
	\end{subfigure}
		        \vspace{-1\baselineskip}
	\caption{Unique external IPs contacted in one-minute intervals in \uvanetwork (left) and \vtnetwork traffic (right), on ports 22, 80, 443, and 445.}
	\label{fig:ext_traffic}
	
\end{figure}

Figure \ref{fig:ext_traffic} illustrates the traffic patterns at \uvanetwork and \vtnetwork along a full week from Sept 3rd to Sept 9th 2019. The $x$-axis plots the index  of the one-minute window, while the $y$-axis shows the number of distinct external IPs contacted within each time window. The traffic patterns are similar for the two sites, with a higher volume of traffic occurring during weekdays, and less traffic over the weekend. Port 443 withstands the largest volume of traffic compared to the other ports, in line with the data shown in Table~\ref{tab:dataset_stat}. The main difference is on port 445, which is still open for traffic within some of the campuses at \vtnetwork, while \uvanetwork blocks most of it.

\begin{table}[]
	\centering
	\scalebox{0.8}{
	\begin{tabular}{c|c|c|c|c|}
		\cline{2-5}
		& \textbf{\begin{tabular}[c]{@{}c@{}}Scanning \\ Port\end{tabular}} & \textbf{\begin{tabular}[c]{@{}c@{}}Speed \\ (IPs/min)\end{tabular}} & \textbf{\begin{tabular}[c]{@{}c@{}}Infected\\ Hosts\end{tabular}} & \textbf{Duration} \\ \hline
		\multicolumn{1}{|c|}{\textbf{WannaCry}} & 445                                                               & $\sim$14K                                                          & 48                                                                & 116 mins          \\ \hline
		\multicolumn{1}{|c|}{\textbf{Mirai}}    & 23                                                                & $\sim$330K                                                         & 1                                                                 & 62 mins           \\ \hline
	\end{tabular} }
	\caption{Statistics of the malicious datasets.}
	\label{tab:malicious_dataset}
	%\vspace{-7mm}
\end{table}

\myparagraph{Malicious Datasets.} We obtained access to a public dataset including the SPM families from Table~\ref{tab:spm_malware_comparison}~\cite{stratosphere}. We determined that all malware families have entropy of the external IPs close to that of the uniform distribution, confirming that the IP addresses they connect to are chosen pseudorandomly. The scanning rate and ports vary across families. In these traces, the attack is limited to a single infected IP and usually has small duration (less than an hour).  We determined that WannaCry is a representative SPM family, with relatively slow scanning rate compared to others, and decided to use it to generate more realistic SPM samples. 
We reverse engineered the WannCry malware and identified two parameters that can be controlled: the number of threads used for scanning and the time interval between scans. 
We set up a virtual network with 50 virtual machines running on three physical machines. Each VM ran a vulnerable version of Windows, which can be exploited by EternalBlue. We isolate the virtual environment by blocking the external network traffic, but we allow internal connections between the virtual machines.  
We infect one of the virtual machines with WannaCry and then let the malware perform internal and external scanning. Upon getting responses from internal IPs, the malware attempts to propagate to these virtual machines. 
We run the original WannaCry attack for two hours 
and infect 48 machines on a subnet. 

 We also generate  other variants with different scanning rates  to test the resilience of our methods to evasion. We capture full pcaps of the network traffic, from which we generate Zeek {\sf conn.log}. 

In our data collection setup, all connections to external IPs done by WannaCry virtual machines are blocked by the virtual environment. We choose to use the successful SMB traffic between internal machines, and emulate it as  external traffic. We achieve this by generating a random external IP address and mapping the destination IP to it, while keeping the same source IP.  This design choice was motivated in order to include examples of successful SMB infections in our malicious data. We reiterate that obtaining real malicious network traffic from machines infected with SPM malware is challenging due to ethical considerations.

The second malicious dataset we use it for Mirai. Mirai is a self-propagating botnet that propagates using a number of different methods targeting vulnerable IoT devices. It performs scanning on different ports, and tries to gain unauthorized access using password spraying on open ports. We use part of a public dataset from Stratosphere Laboratory~\cite{stratosphere}, where they provide a sample of the Mirai which contains a scanning behavior on multiple ports. We extract the scanning behavior on port 23 (Telnet) to evaluate our models. Table \ref{tab:malicious_dataset} shows the overview of the two malicious datasets we use for evaluation.

\subsection{Evasion Strategies}
\label{sec:evasion_dataset}

Self-Propagating-Malware, like WannaCry and Mirai, performs aggressive scanning to spread and infect as many IPs as possible in a short time. As a result, our anomaly detection models can detect these attacks with high accuracy, as demonstrated in Section~\ref{sec:eval}.

In reality, advanced attackers can employ evasion strategies to make the detection of SPM traffic more difficult. We experiment with two  evasion strategies:

\myparagraph{Method 1: Slowing down the probing rate.} An adversary with no knowledge about the ML models could attempt a straightforward strategy to lower the scanning rate and become more stealthy.  The original WannaCry attack contacts on average 14K external IPs per minute and runs for two hours. We generate new variants, where each infected machine decreases its probing rate by a factor of 2, 4, 8, 16, 32, 64, or 128 by dropping a fraction of connections uniformly at random. 

This results in the same number of infected hosts, scanning the same amount of time, but with fewer probing connections.

\myparagraph{Method 2: Leveraging IP destinations from history.} One of the most relevant features in our models is the number of ``new IP'' destinations, which are IP addresses not visited before on that particular port. SPM is likely to generate a large number of new IPs if the probed IPs are generated at random, as done by both WannaCry and Mirai. An attacker with knowledge of our feature set could attempt to evade this feature by probing already visited  IP addresses. 

In a way,  IP addresses from history are probed ``for free'', without increasing the count on the new IP feature.  
This assumes that the adversary is powerful and has the ability of observing the network traffic of internal machines for some time period. 
Starting from the malware variant 4 times slower than the original, we select variants that use a factor 2, 4, 8, 16, 32, 64, or 128 new IPs, while the rest are are previously-visited IPs from the history.
The variants have the same number of infected hosts and similar scanning rate, the only difference being that some destination IPs are replaced with IPs from history.

%% file: evaluation.tex
\section{Evaluation}
\label{sec:eval}
We perform an in-depth evaluation of our proposed ensemble methods against several baselines for detecting SPM and show the higher resilience of ensembles against evasion strategies. Given that our data is imbalanced, we use Area Under the Precision Recall Curve (\prauc) for evaluation. We are also interested in minimizing the false positive rate to account for the limited budget of SOC investigations. We leverage Python3 and sklearn  machine learning packages for training and testing ML models. Codebase for the system is available online\footnote{ https://github.com/tongun/PORTFILER}. We use a High-Performance-Computing (HPC) platform provided at \uvanetwork to collect and analyze data. 

\input{eval_baselines}

\input{eval_ensembles}

%% file: eval_baselines.tex
\subsection{Baselines and Standard ML Detectors}
\label{sec:baseline}

\myparagraph{Setup.} For these experiments, we merge the malicious traces at testing time to generate controlled experiments with attack ground truth. We highlight that this process is done only during  testing,  but we do not train our models on  malicious traces. We overlay the malicious traces onto the network traffic from the testing day by selecting random internal IPs in the network, and merging the legitimate and malicious {\sf conn.log} data. When merging the malicious traffic, we preserve all the attributes of the connections.

\begin{figure}[th]
    %\vspace*{-4mm}
	\centering
	\begin{subfigure}[b]{0.49\linewidth}
		\includegraphics[width=\linewidth]{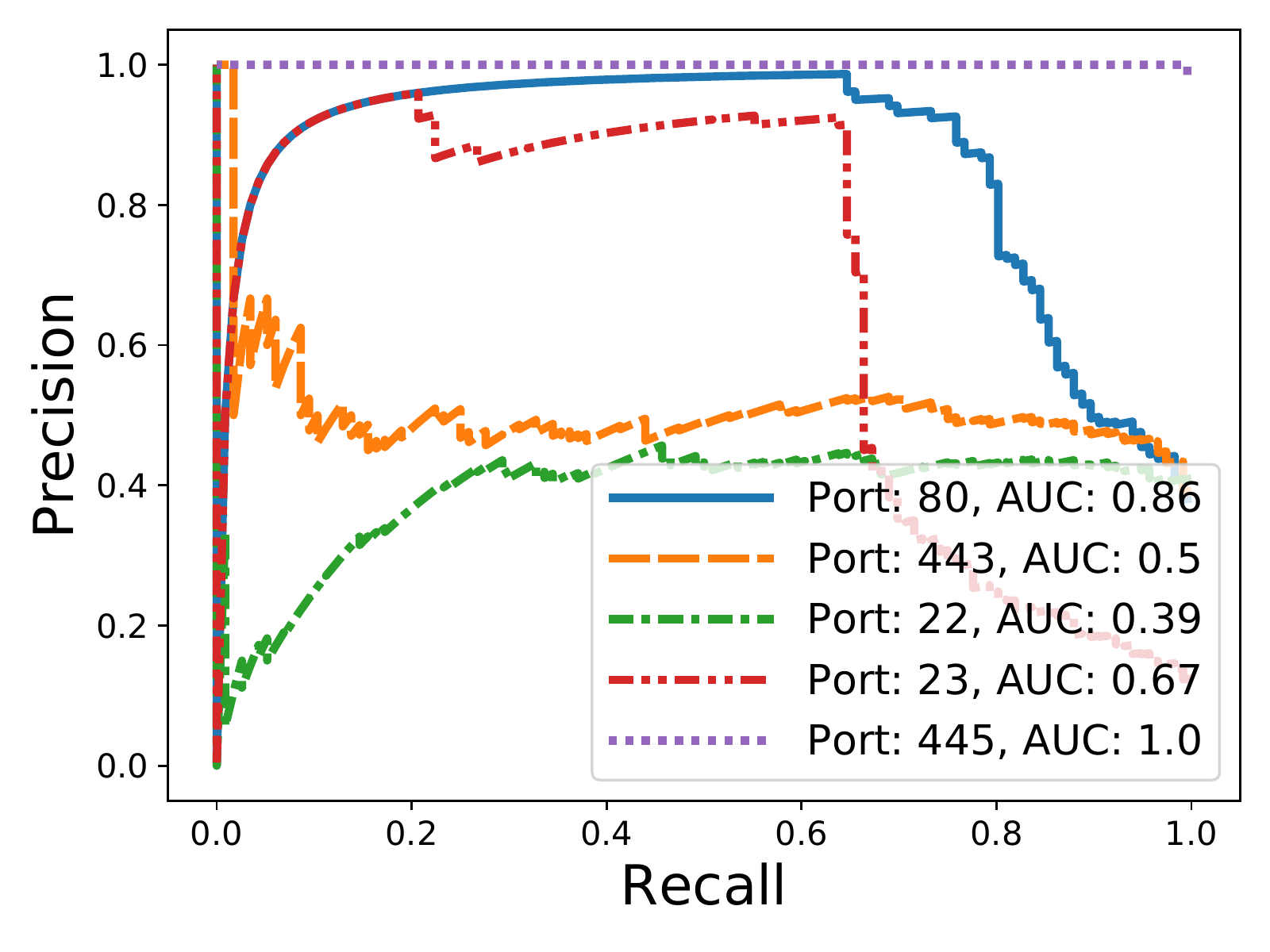}
		
	\end{subfigure}
	\begin{subfigure}[b]{0.49\linewidth}
		\includegraphics[width=\linewidth]{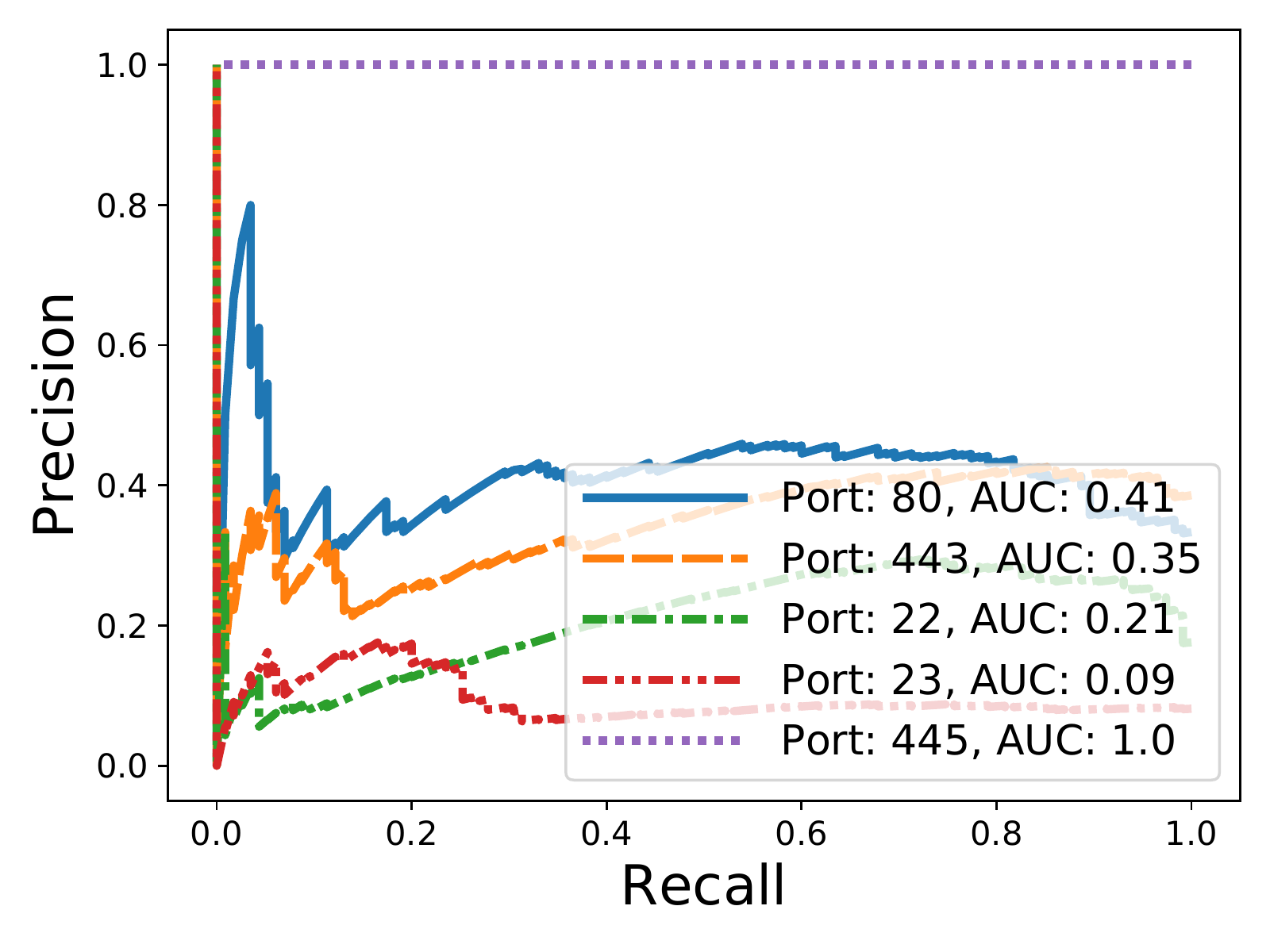}
	\end{subfigure}
	%\vspace{-1\baselineskip}
	\caption{Precision-Recall curve showing poor performance to detect original (left) and 8x slower WannaCry (right) using the threshold-based baseline detector at \uvanetwork.}
	
	\label{fig:baseline}
	\vspace{-4mm}
\end{figure}

\myparagraph{Baseline Threshold Detector.} We first experiment with a threshold-based baseline detector that marks a time window as anomalous if the number of connections on the monitored port exceeds a  threshold. The connection threshold is varied in order to obtain a full precision-recall (\pr) curve. Figure~\ref{fig:baseline} shows the \pr\ curves of the threshold-based approach on the original WannaCry variant and on an 8-times slower variant. 
The threshold-based detector performs poorly even on the original WannaCry variant. For instance, on port 22 the \prauc\ is as low as 0.39, while for port 443 the \prauc\ is 0.5. The only port where the threshold-based approach performs well is 445 because it is blocked at \uvanetwork\ and there is very little regular traffic. These numbers reduce considerably for the 8x slower variant, resulting in an \prauc\ of just 0.09 on port 23. This approach disregards the connection properties, and the volume-based signal is mostly lost at the port level.
This demonstrates that the problem of detecting SPM attacks cannot be solved trivially with a threshold-based detector, and it also motivates us to employ ML-based approaches for learning the distribution of network traffic on each port.  

\begin{figure}[]
    %\vspace*{-4mm}
	\centering	
	\begin{subfigure}[b]{0.49\linewidth}
		\includegraphics[width=\linewidth]{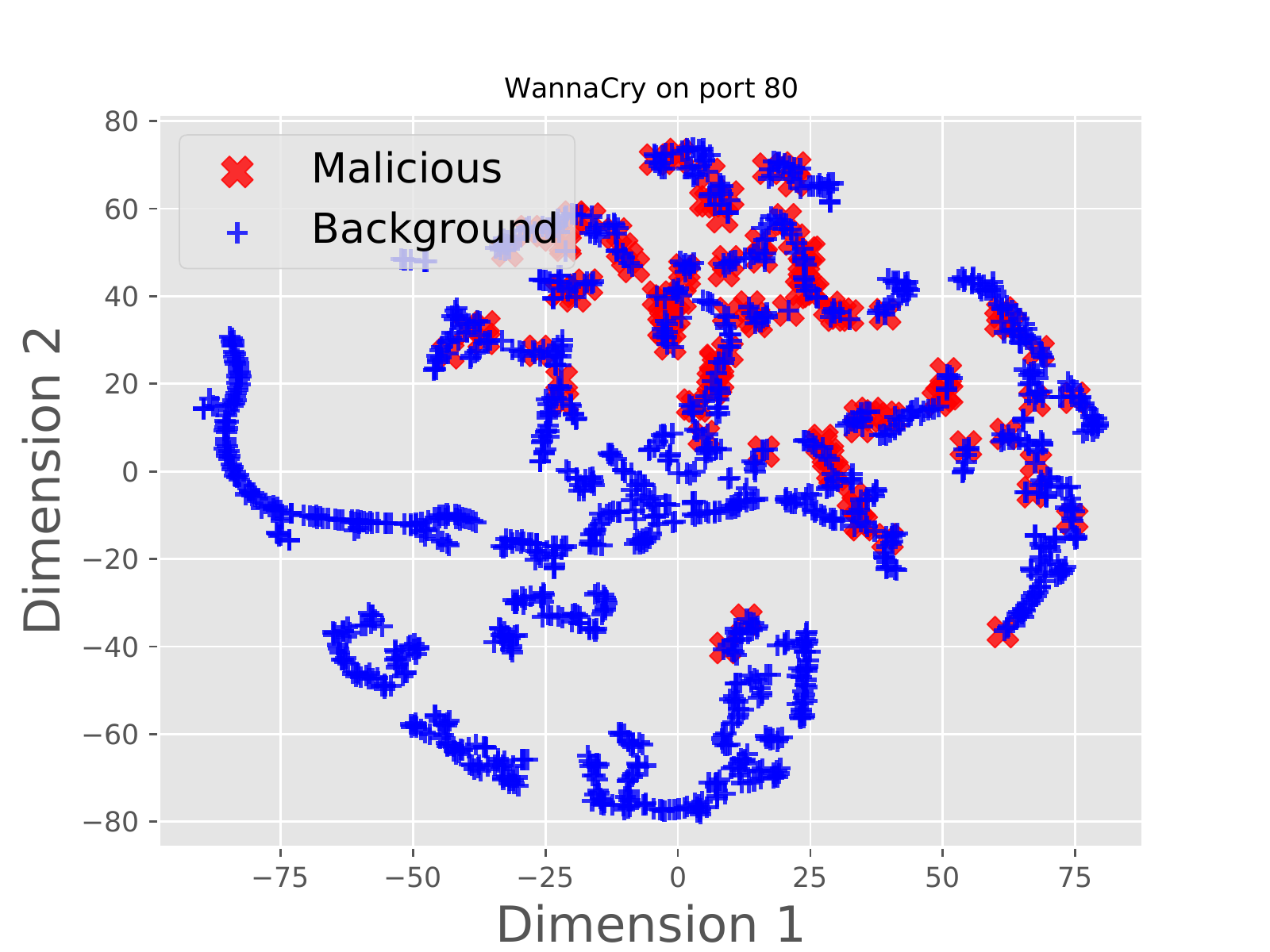}
	\end{subfigure}
	\begin{subfigure}[b]{0.49\linewidth}
		\includegraphics[width=\linewidth]{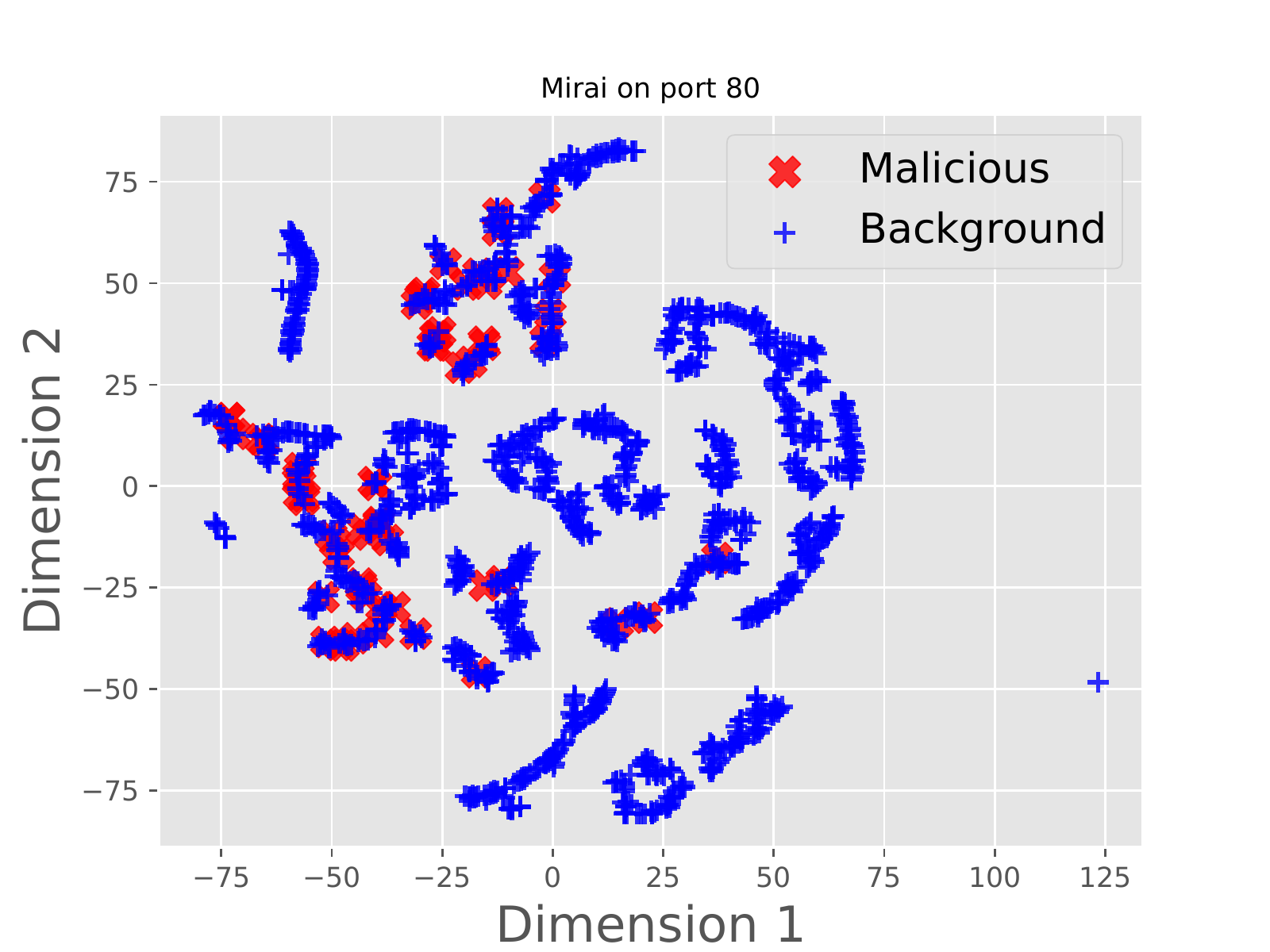}
	\end{subfigure}	
	%\vspace{-1\baselineskip}
		\begin{subfigure}[b]{0.49\linewidth}
		\includegraphics[width=\linewidth]{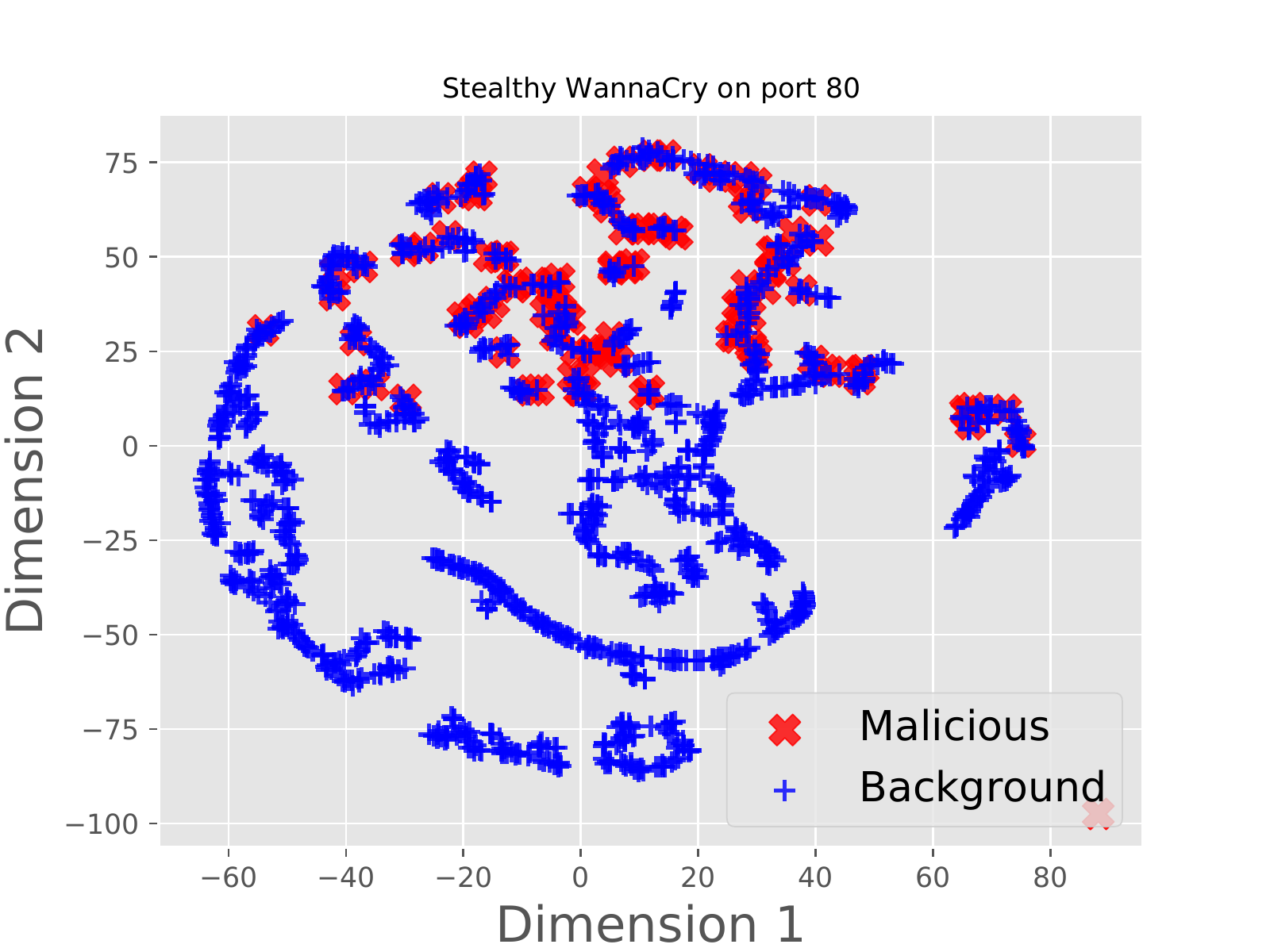}
	\end{subfigure}
	\begin{subfigure}[b]{0.49\linewidth}
		\includegraphics[width=\linewidth]{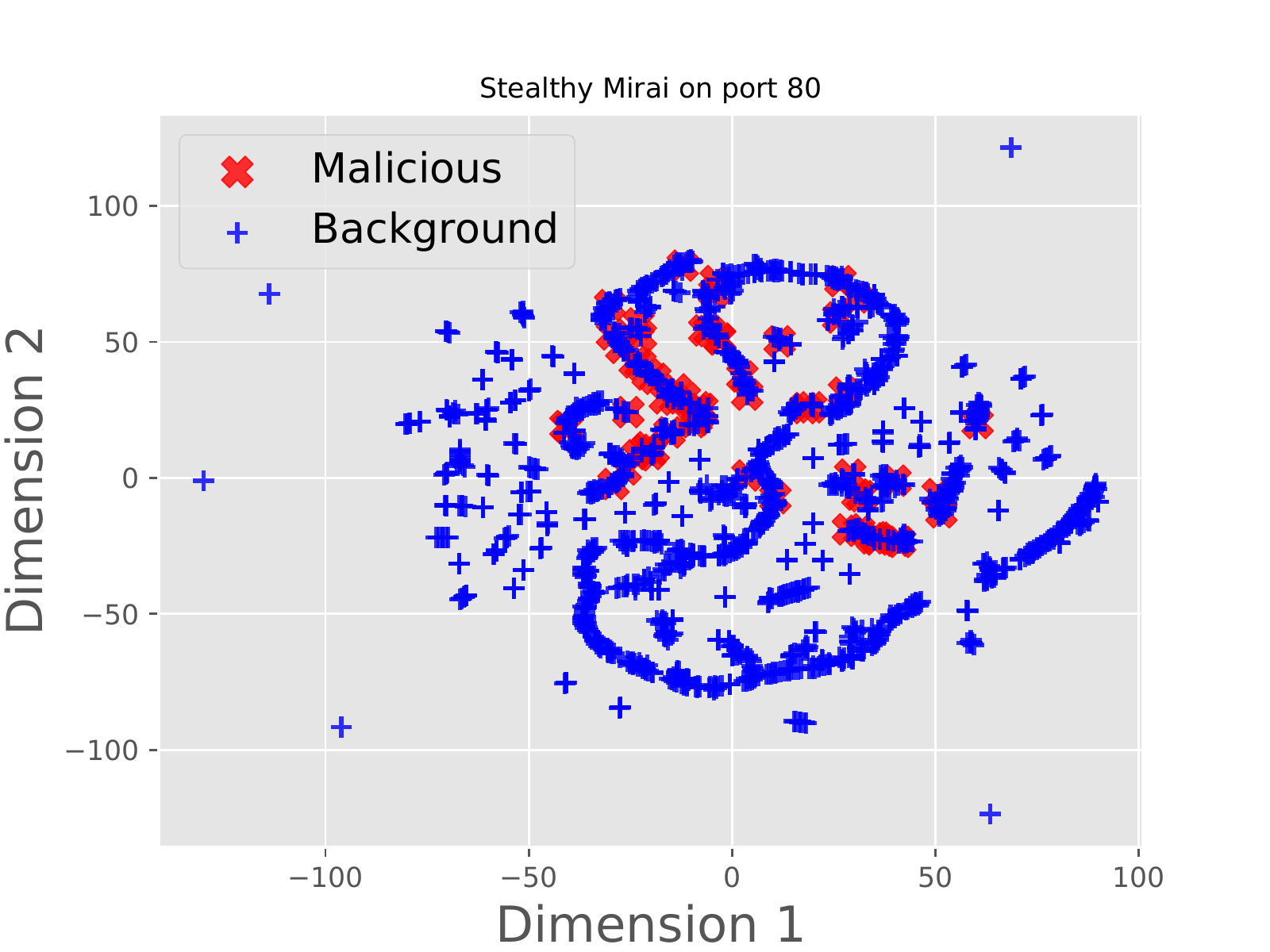}
	\end{subfigure}

	\caption{t-SNE visualization of the 35-feature representation of original (top) and slow (bottom) WannaCry (left) and Mirai (right) at \uvanetwork on different ports. Malicious samples overlap with background samples, indicating that off-the-shelf anomaly detection algorithms are less likely to find malicious samples with high accuracy.}
	
	\label{fig:tsne_wc}
	\vspace{-4mm}
\end{figure}

\myparagraph{Dimensionality Reduction.} Dimensionality reduction methods, such as PCA, UMAP, and t-SNE,
 are used extensively for anomaly detection. We investigate if our malicious and benign samples become separable after projection to a lower-dimensional space.
We performed experiments with  both t-SNE and UMAP, but only present results for t-SNE. Figure~\ref{fig:tsne_wc} shows a scatterplot of the two principal components of malicious and background samples after t-SNE for slow variants (1/128 rate) of both WannaCry and Mirai on port 80. We observe that most of the malicious samples overlap with background samples, and there is no clear separation. The results were similar for other ports and SPM variants. These results indicate that off-the-shelf anomaly detection algorithms are less likely to find malicious samples with high accuracy, and more guided approaches are required for our system.

\myparagraph{Baseline ML Models.} As simple methods are ineffective for our problem of detecting SPM attacks, we are motivated to employ ML-based approaches for learning the distribution of network traffic on each port. We evaluate the standard KDE and Isolation Forest (IF) ML models trained with our 35 features previously described, using the \uvanetwork\ dataset, after doing an extensive hyper-parameter search. In all our experiments, we selected a time window for feature extraction of one minute (after experimenting with several values: 30 sec, 1 min, 2 min, 5 min, and 10 min). Thus each day of data has 1440 time windows per port.  We  discuss here results on \uvanetwork, but results on \vtnetwork\ are similar. 

\begin{figure}[t]
	\centering
		\begin{subfigure}[b]{0.49\linewidth}
		\includegraphics[width=\linewidth]{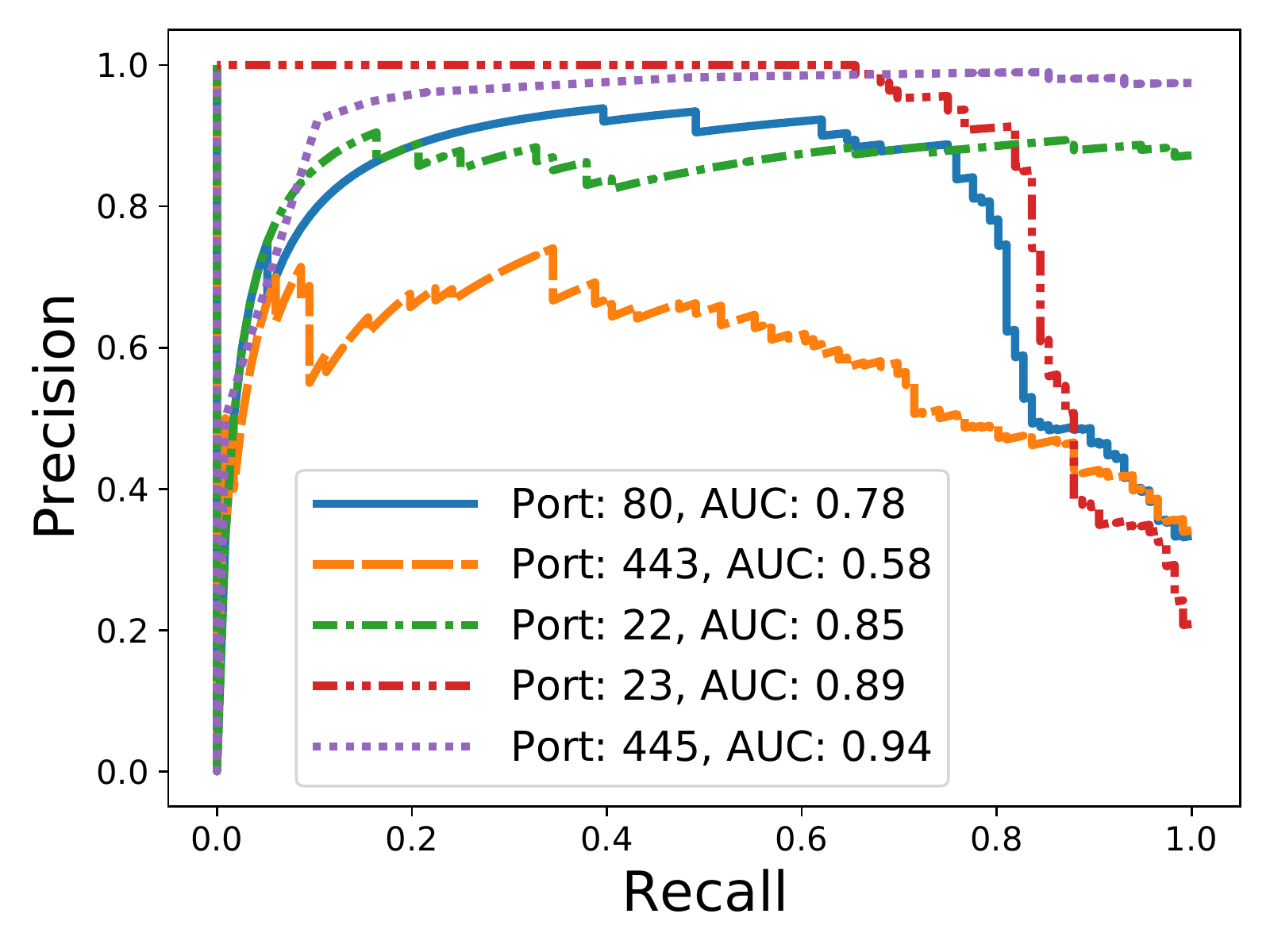}
		
	\end{subfigure}
	\begin{subfigure}[b]{0.49\linewidth}
		\includegraphics[width=\linewidth]{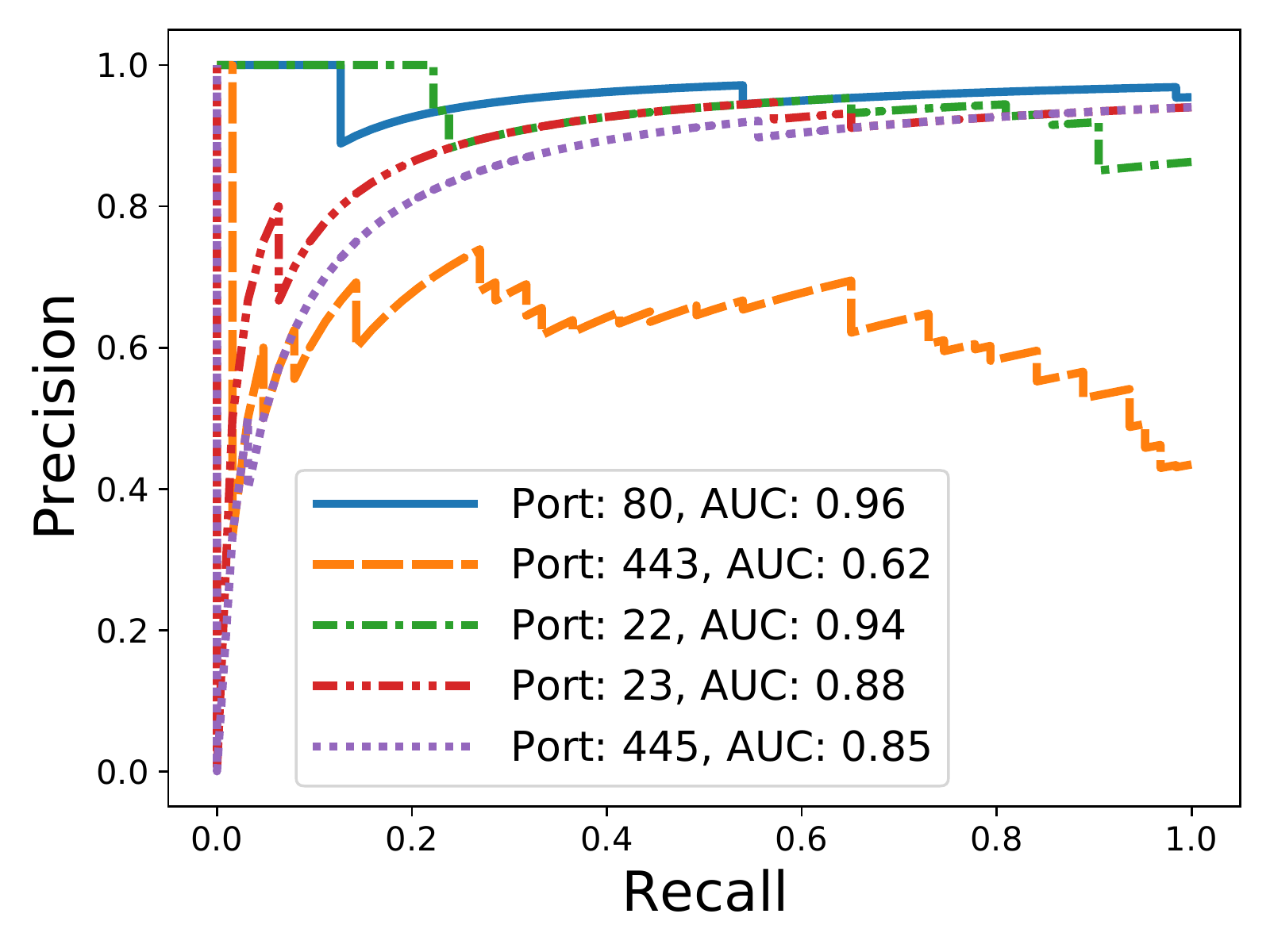}
		
	\end{subfigure}
			        \vspace{-1\baselineskip}
	\caption{Isolation Forest model: Precision-Recall curves for WannaCry (left) and Mirai (right). The model detects Mirai better due to faster SPM activity.}
	\label{fig:isolation_pr}
		\vspace{-4mm}
\end{figure}

\begin{figure}[t]
  \centering
  \begin{subfigure}[b]{0.49\linewidth}
	\includegraphics[width=\linewidth]{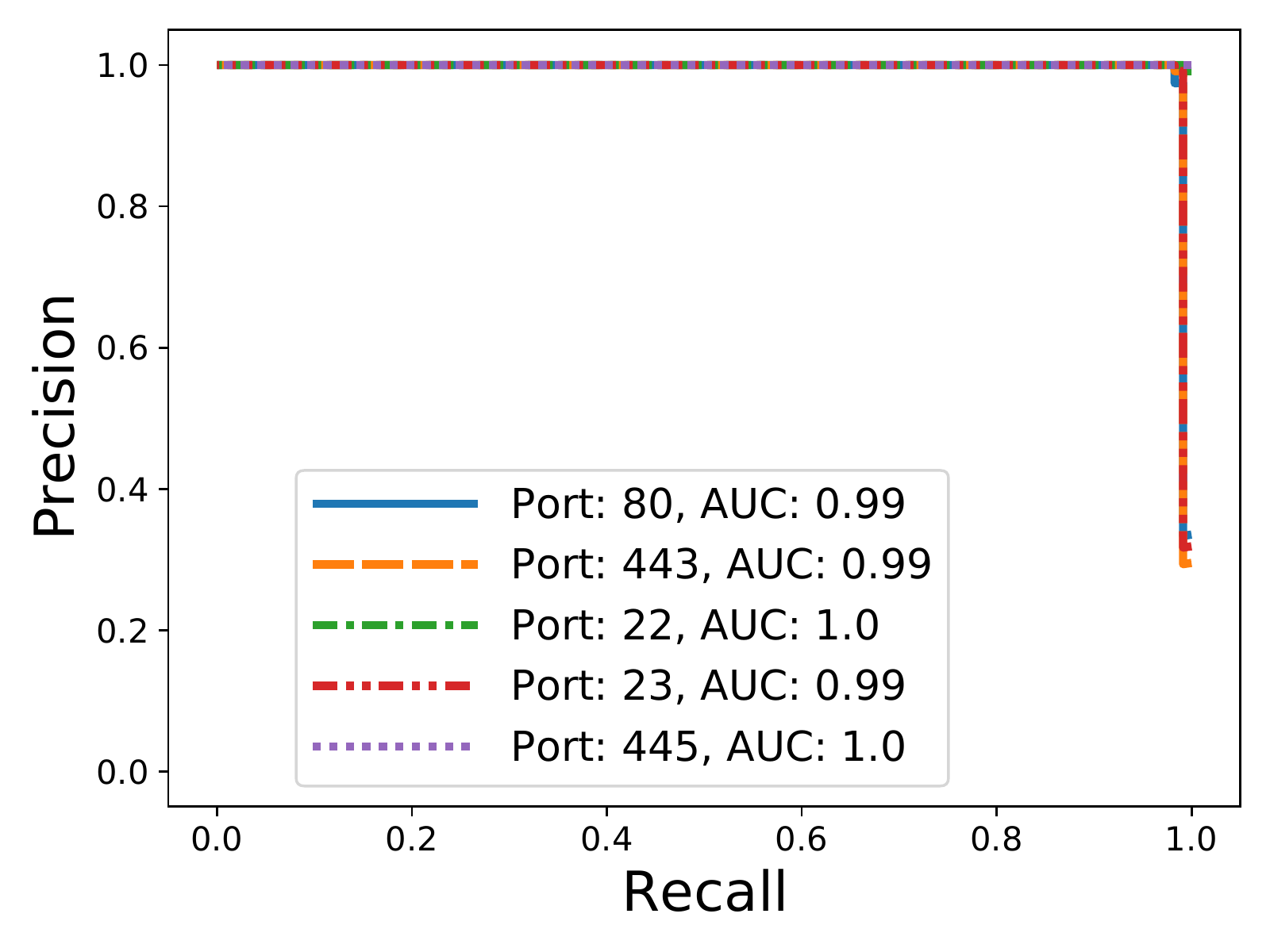}

	\end{subfigure}
	\begin{subfigure}[b]{0.49\linewidth}
		\includegraphics[width=\linewidth]{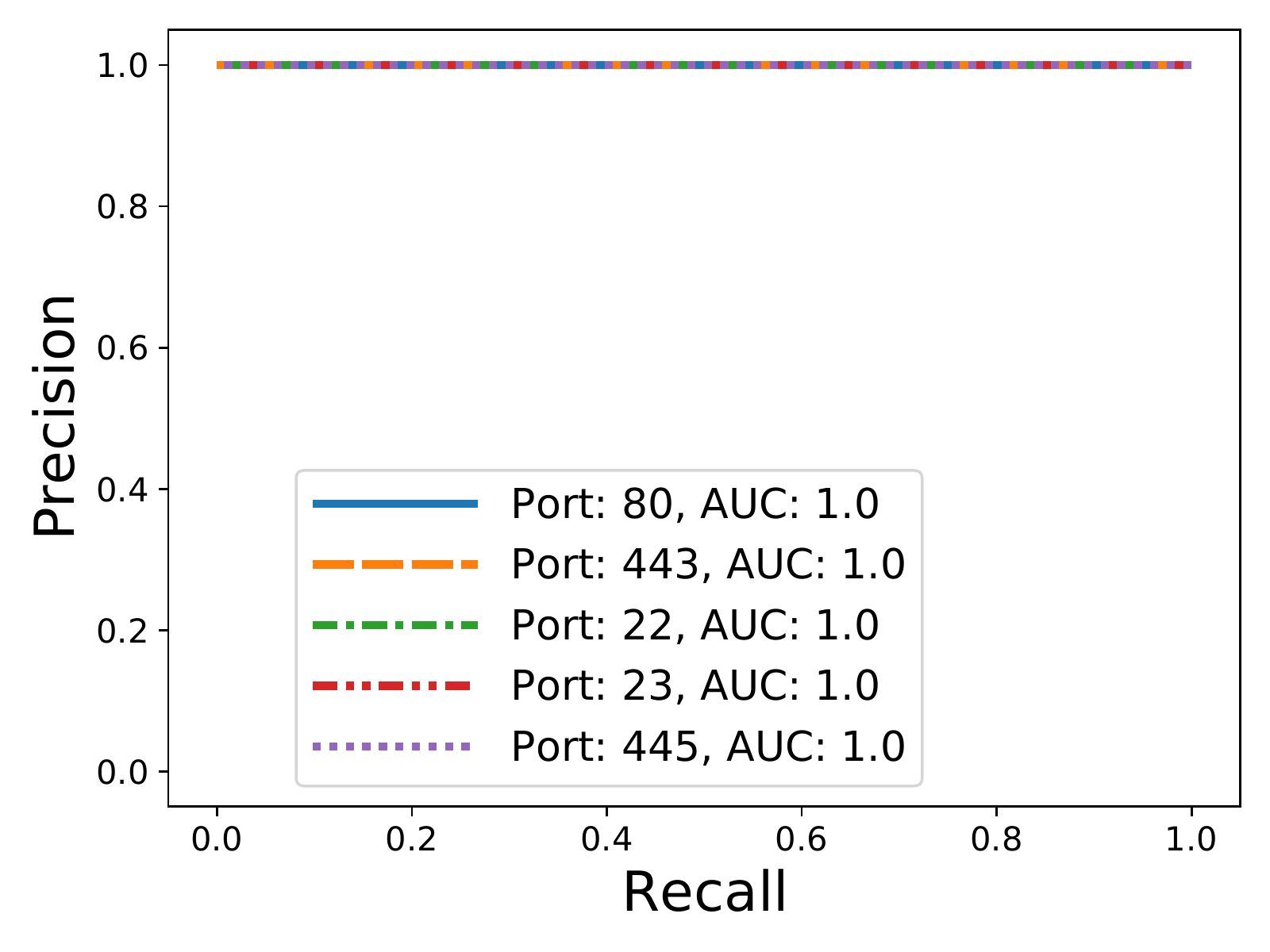}

	\end{subfigure}

  \caption{KDE model: Precision-Recall curves for WannaCry (left) and Mirai (right). The model detects both malware successfully.}
  \label{fig:kde_pr}
\end{figure}

We found that the KDE model performs better than IF, and both methods improve upon the threshold-based detector as demonstrated in Figure~\ref{fig:kde_pr} and Figure~\ref{fig:isolation_pr}. IF has the lowest \prauc\ on port 443, which is the highest volume port, with \prauc\ reaching 0.58 for WannaCry and 0.62 for Mirai. KDE has \prauc\ scores  higher than 0.99 on all ports for the original WannaCry and Mirai attacks. Similarly, we obtained \prauc\ score of 1.0 using the Kenjiro and Hajime malware samples described in Table~\ref{tab:spm_malware_comparison}. 

\begin{figure}[]
	\centering
	%\vspace*{-4mm}
	\begin{subfigure}[b]{0.49\linewidth}
		\includegraphics[width=\linewidth]{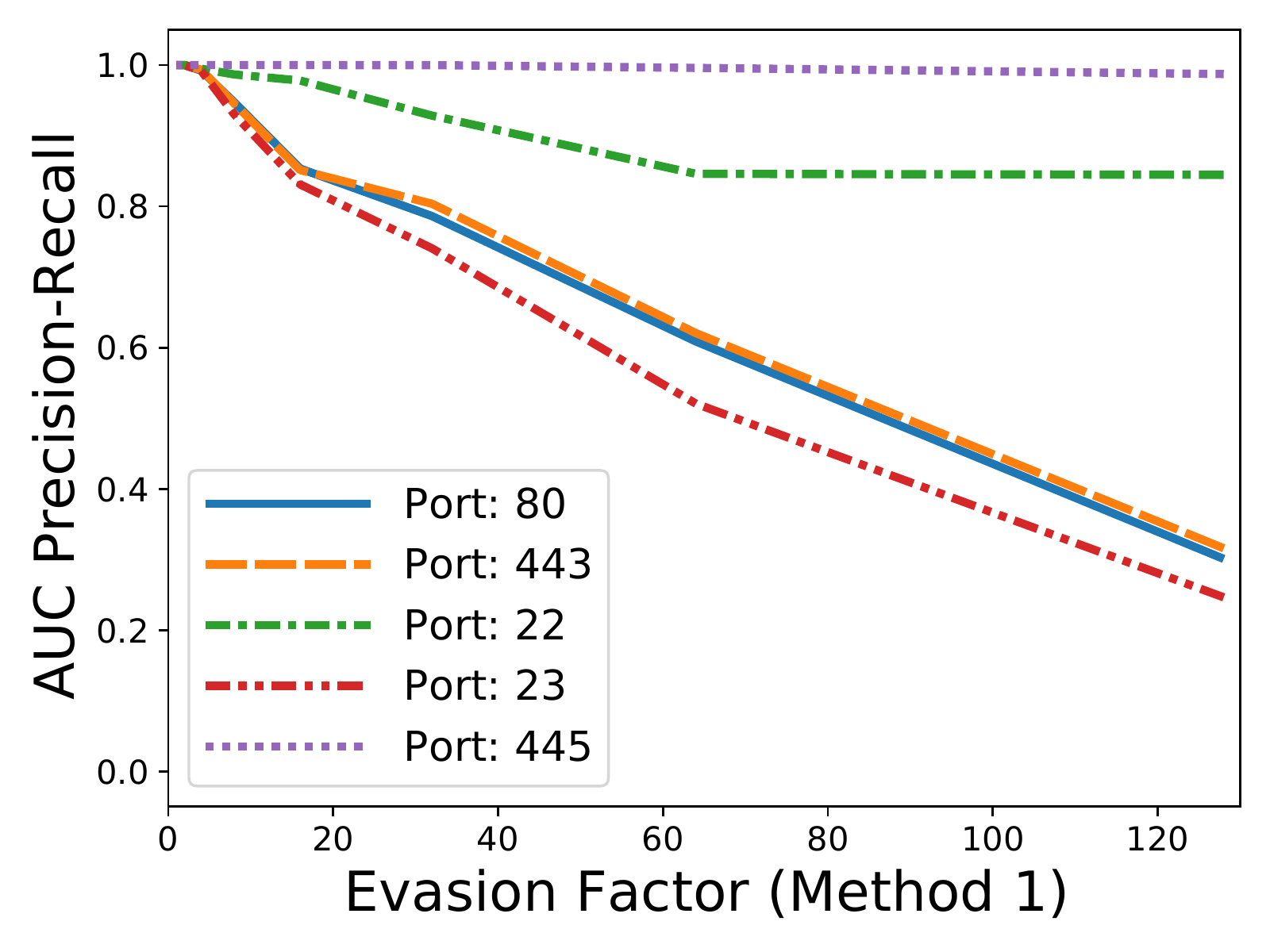}
		
	\end{subfigure}
	\begin{subfigure}[b]{0.49\linewidth}
		\includegraphics[width=\linewidth]{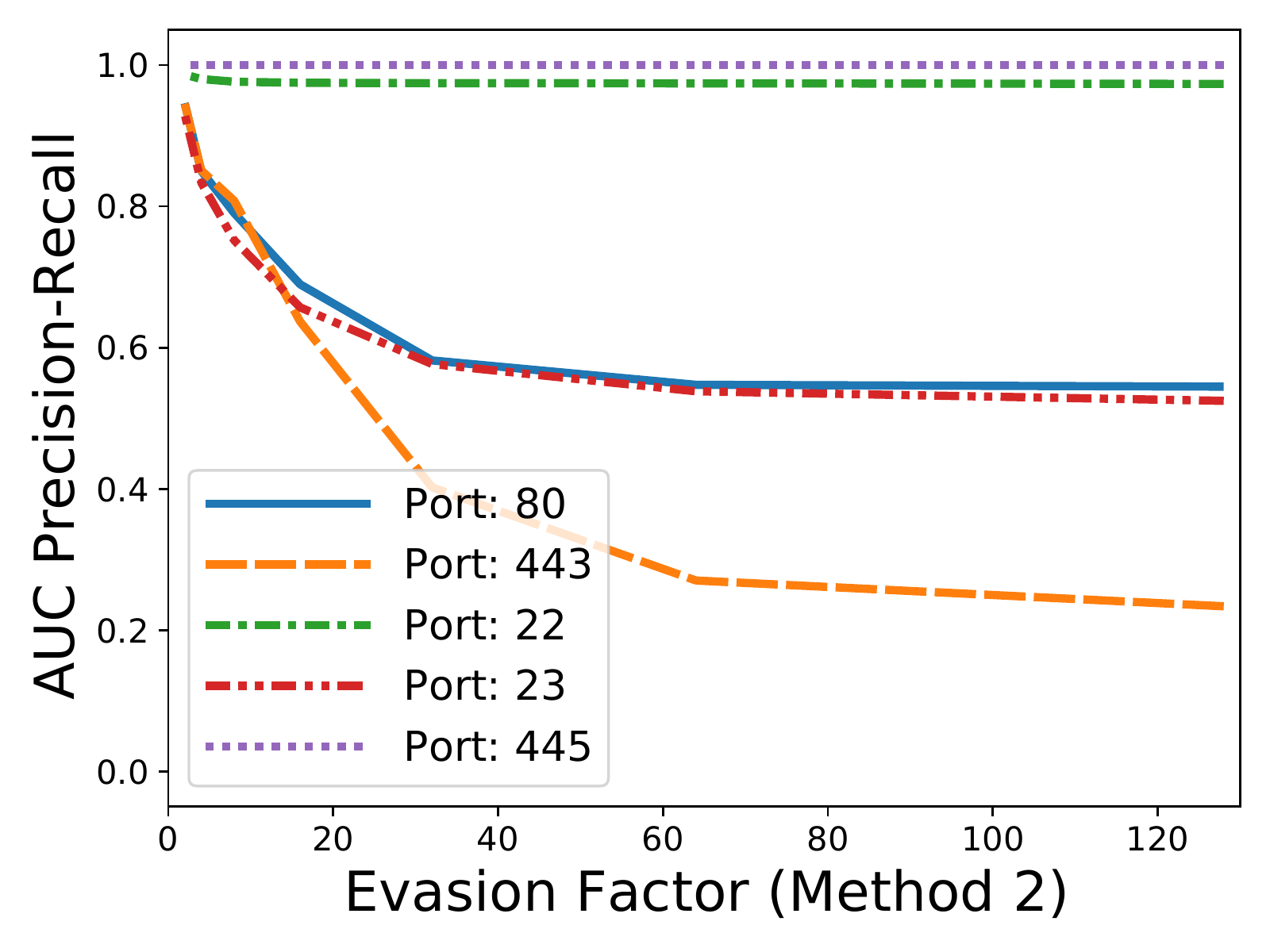}
	\end{subfigure}
	\begin{subfigure}[b]{0.49\linewidth}
	\includegraphics[width=\linewidth]{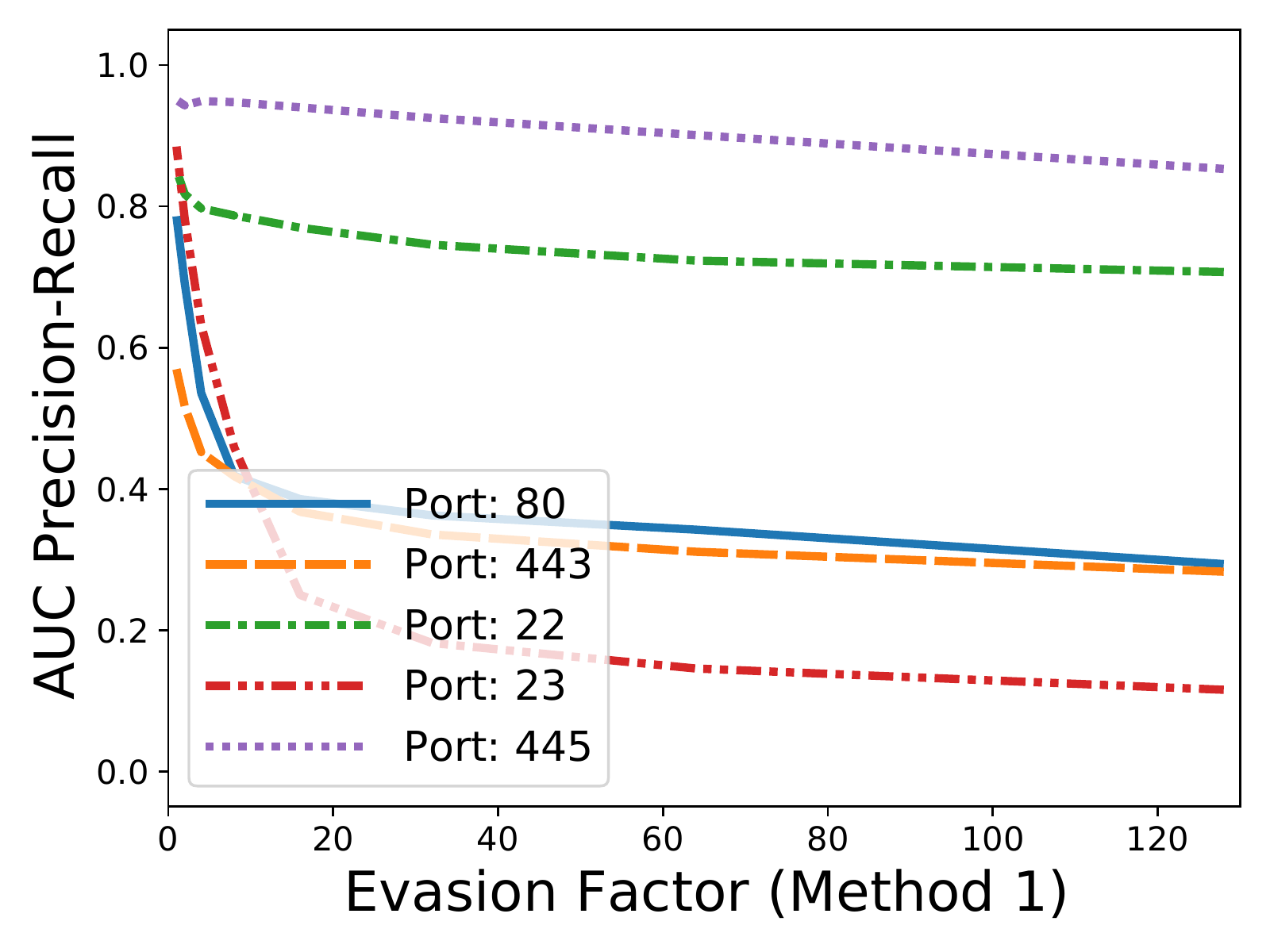}
	
\end{subfigure}
\begin{subfigure}[b]{0.49\linewidth}
	\includegraphics[width=\linewidth]{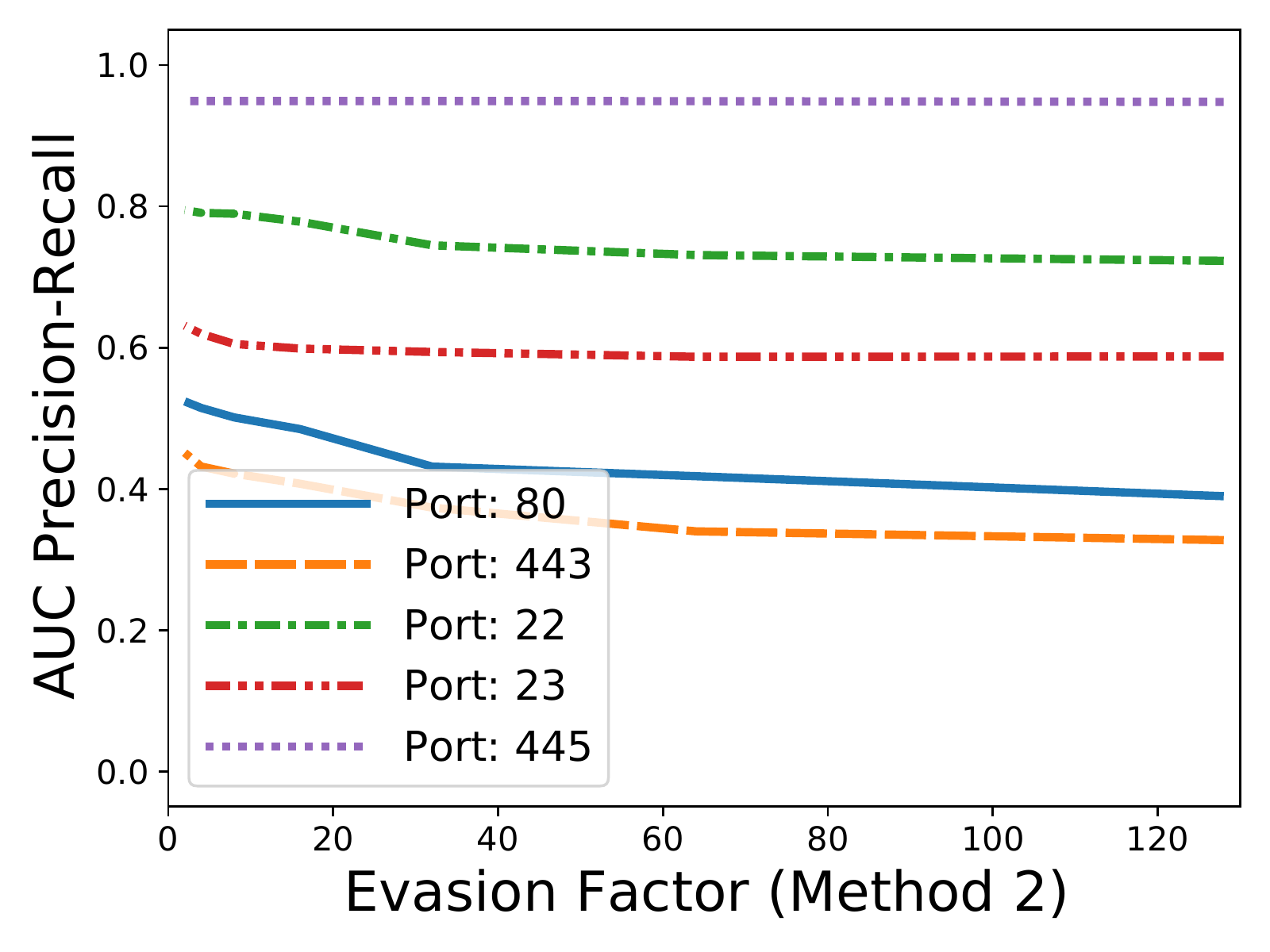}
\end{subfigure}

	%\vspace{-1\baselineskip}
	\caption{Baseline KDE (top) and Isolation Forest (bottom) models. Robustness against evasion is generally low for Evasion Method 1 (slowing down the probing rate) and 2 (leveraging IP destinations from history). Performance drops significantly as the evasion factor increases.}	
	\label{fig:evasion_kde}

\end{figure}

However, neither of these standard ML models are resilient against the evasion strategies described in Section~\ref{sec:evasion_dataset}: \emph{Method 1: Slowing down the scanning rate}, and \emph{Method 2: Leveraging IP destinations from history}. To evaluate evasion resilience, we use the representative WannaCry malware we generated and its evasive variants. Figure~\ref{fig:evasion_kde}  shows the change in the KDE \prauc\ score as we decrease the probing rate of WannaCry (\emph{Method 1}), and  the new IP rate (\emph{Method 2}). It is apparent that both Isolation Forest and the standard KDE model are not resilient to these evasion methods. For example, by slowing down by a factor of 64, the \prauc\ score of KDE reaches around 0.2 on ports 23, 80, and 443. While ports 445 and 22 show less impact, the scores decrease more significantly on ports with large amount of traffic. We obtained similar results for Mirai, and we believe that other SPMs will exhibit similar behavior under evasion.

\begin{figure}[t]
	\centering
	\begin{subfigure}[b]{0.49\linewidth}
		\includegraphics[width=\linewidth]{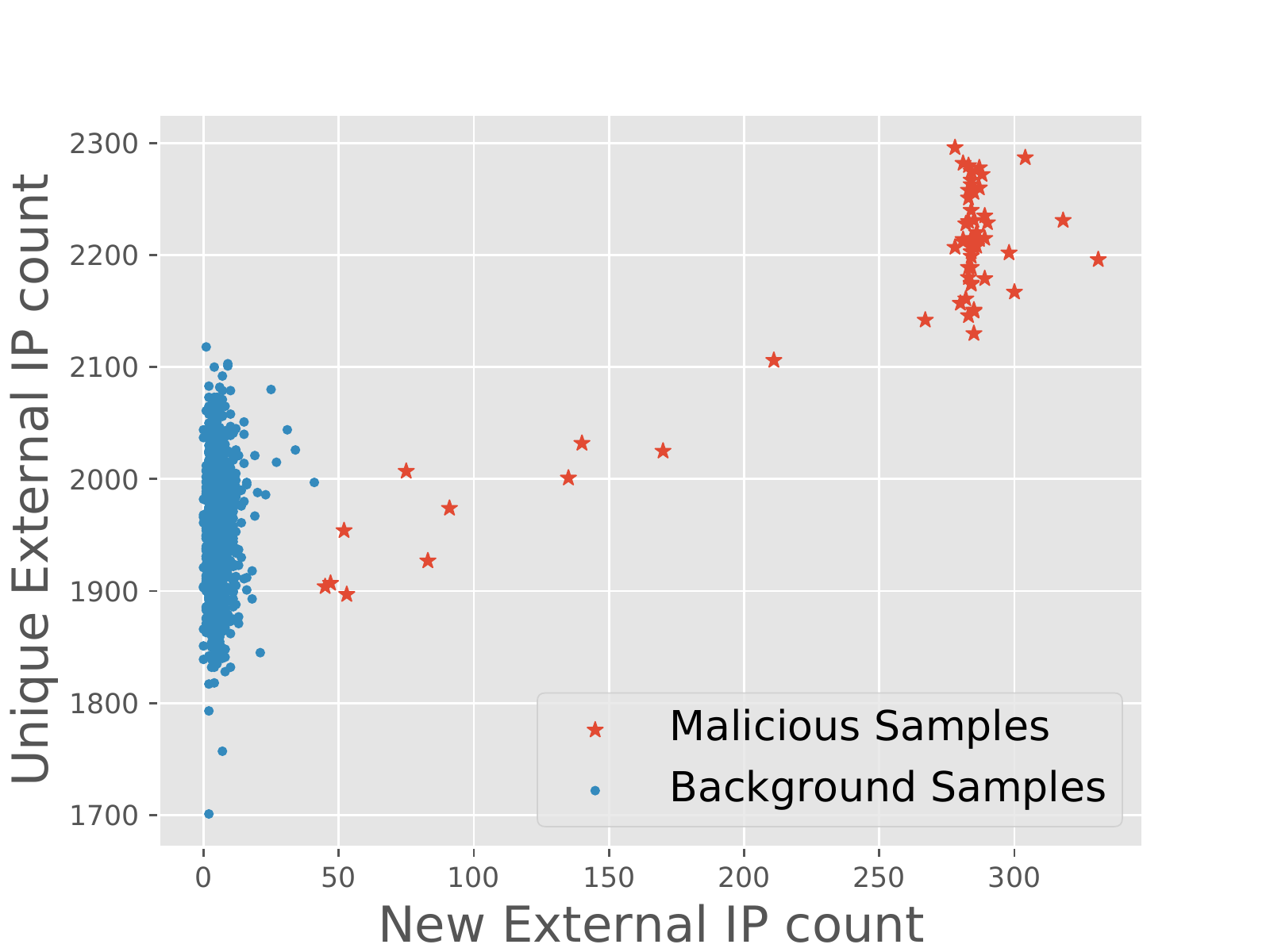}
		
	\end{subfigure}
	\begin{subfigure}[b]{0.49\linewidth}
		\includegraphics[width=\linewidth]{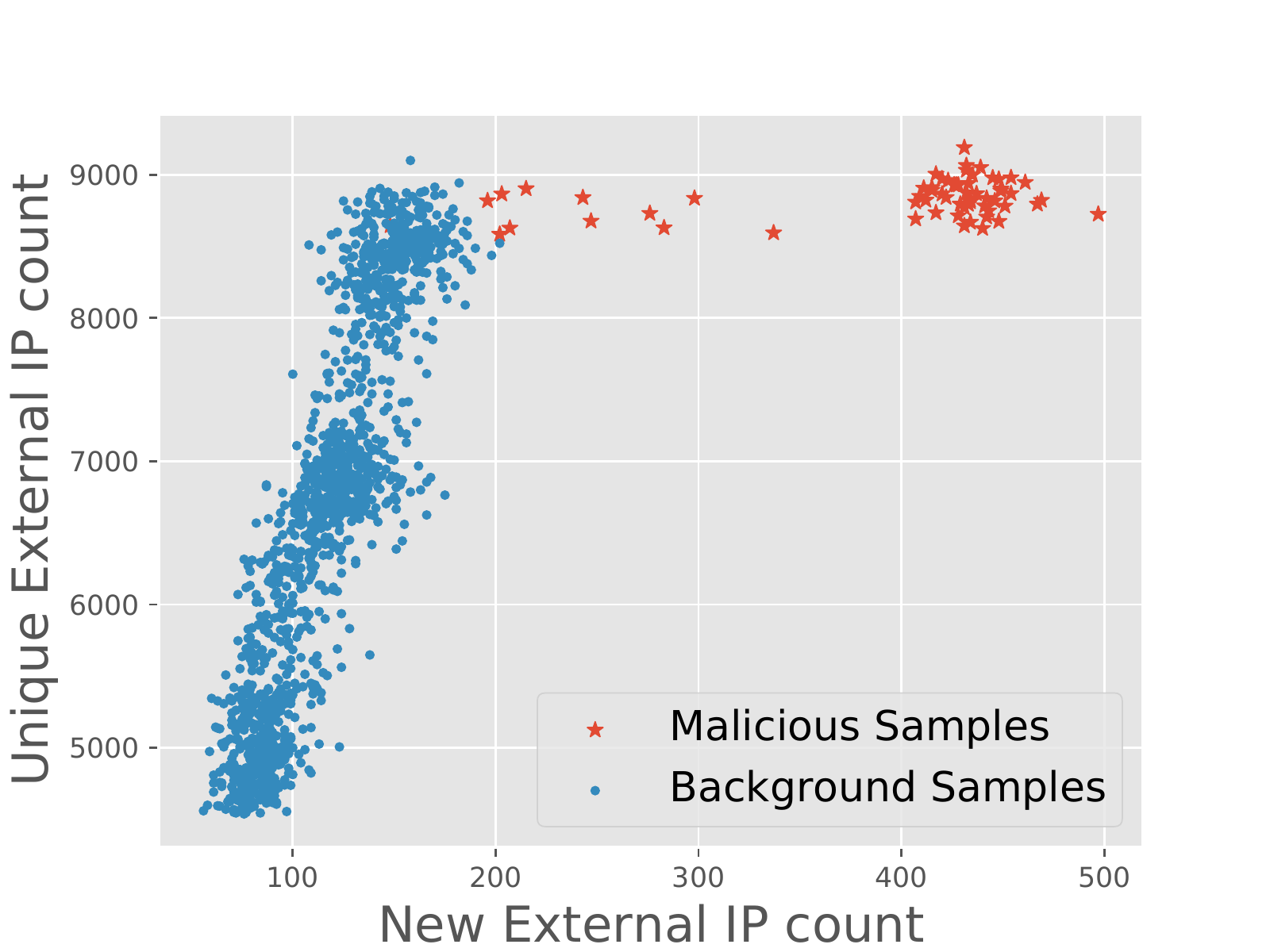}
		
	\end{subfigure}
	\vspace{-1\baselineskip}
	\caption{Comparison of new external IP count and distinct external IP count for benign and malicious WannaCry samples, for port 22 (left) and port 80 (right).}
	\label{fig:newIPs}

\end{figure}

\myparagraph{New external IP feature.}
Detecting an SPM attack depends on the ability of our feature set to capture the attack characteristics. One particularly useful feature for self-propagating malware is the ``new external IP''.
This feature measures the number of novel IPs within each time window, with respect to the history of IPs visited on that port. It captures the ``randomness'' aspect of malware that propagates by contacting randomly generated external IPs. Figure \ref{fig:newIPs} shows a comparison of the number of new external IPs versus unique external IPs in testing for two ports: 22 and 80, using WannaCry samples. On both ports we can see the separation of ``normal windows'' from ``malicious windows'', resulting in two clusters. This separation occurs along the $x$-axis (i.e. the axis showing the new IPs), which indicates that the new IPs feature is instrumental in identifying the attack windows in the traffic.

%% file: eval_ensembles.tex
\subsection{\thesystem: Ensemble Models}
\label{subsec:robeval}

\begin{table}[]
	\centering
	\begin{tabular}{|c|c|}
		\hline
		\textbf{Feature}         & \textbf{Importance Score} \\ \hline
		new\_external\_ips       & 0.3613                    \\
		external\_ips            & 0.3570                    \\
		zero\_resp\_bytes\_count & 0.0375                    \\
		mean\_duration           & 0.0263                    \\
		min\_duration            & 0.0206                    \\
		RSTR\_count              & 0.0164                    \\
		REJ\_count               & 0.0157                    \\
		S0\_count                & 0.0154                    \\
		RSTO\_count              & 0.0130                    \\
		failed\_conn\_count      & 0.0120                    \\ \hline
	\end{tabular}
	%\captionsetup{justification=centering,margin=0.cm}
	\caption{Feature importance values obtained by Random Forest model for WannaCry. These scores are used as weights in the ensemble models.}
	\label{tab:feature_imp}
\end{table}

\myparagraph{Weight computation for ensembles.} We explain here how we compute feature weights for the weighted ensembles. We assume that a WannaCry trace was shared with the monitored network. The original WannaCry variant uses 128 threads for probing, and  a time interval of 100ms between successive probing requests (we denote this \textbf{wc-128-100ms}). We generated the following variants: \textbf{wc-8-500ms}, \textbf{wc-8-1s}, \textbf{wc-8-5s}, \textbf{wc-8-10s}, and \textbf{wc-8-20s}, after reverse engineering a WannaCry binary. All of these variants have 8 threads, and use a larger timing between connections compared to the original variant, so overall the number of probed IPs is much less than in the original variant. As it is desirable to vary the inter-arrival time between successive connections, we use all of these variants, and  merge them at different times into the training day. We then take the labeled data on the training day (after merging the WannaCry traffic) and train a Random Forest classifier with 100 trees. Our goal is to determine feature importance and, in particular, which features have highest importance at distinguishing malicious and benign traffic. We list the feature importance we obtained from the Random Forest classifier in  Table~\ref{tab:feature_imp}. They confirm our intuition that the new IP feature is ranked first, and features such as zero bytes returned in the connection and failed connection are ranked at the top. We only include the features with the highest importance in the table.

\myparagraph{Resilience to Evasion.}  We designed  ensembles of multiple models with the goal of increasing the system's resilience to evasion.

\begin{figure}[]
	\centering
	\vspace*{-2mm}
	\begin{subfigure}[b]{0.49\linewidth}
		\includegraphics[width=\linewidth]{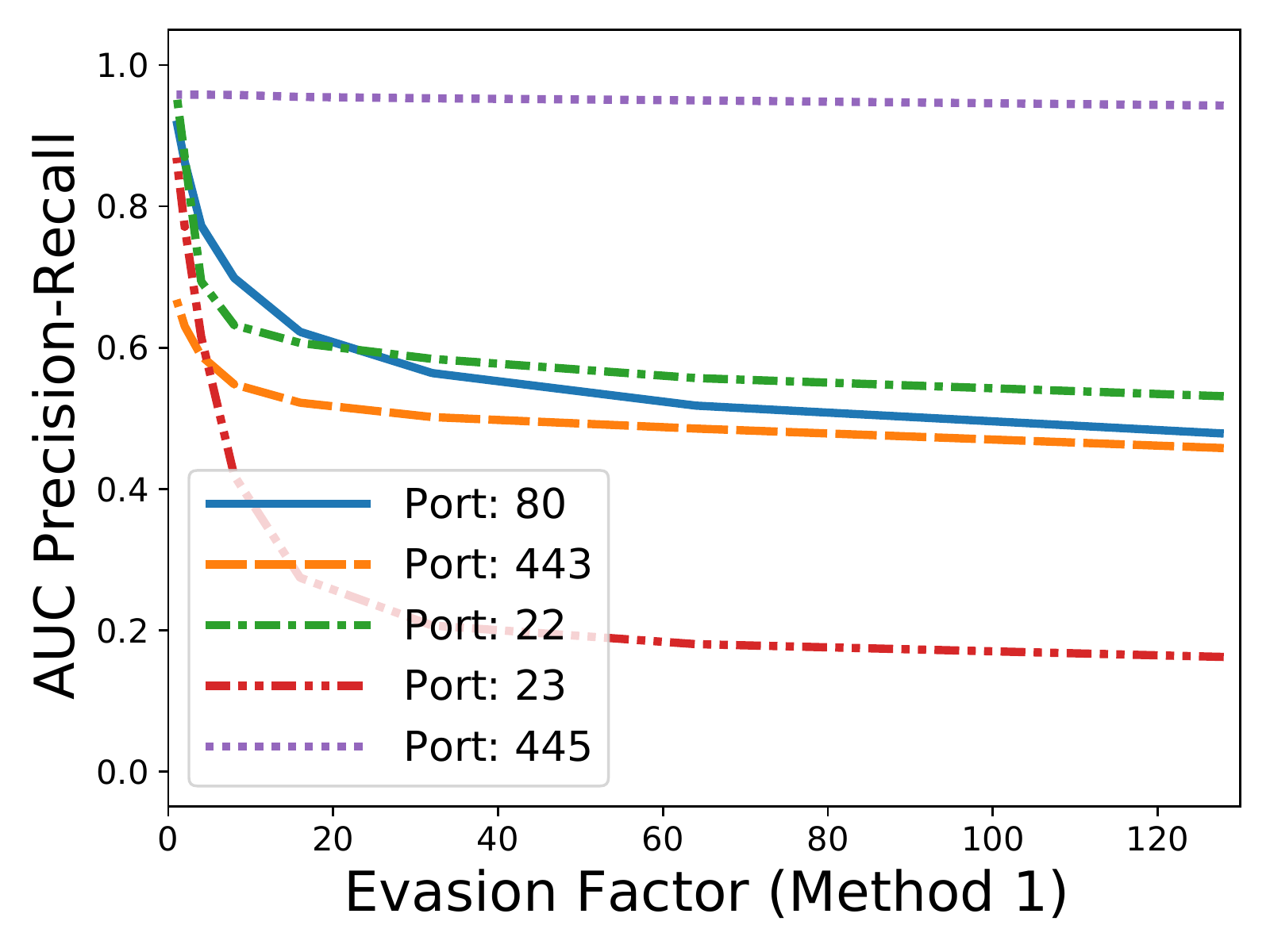}
		
	\end{subfigure}
	\begin{subfigure}[b]{0.49\linewidth}
		\includegraphics[width=\linewidth]{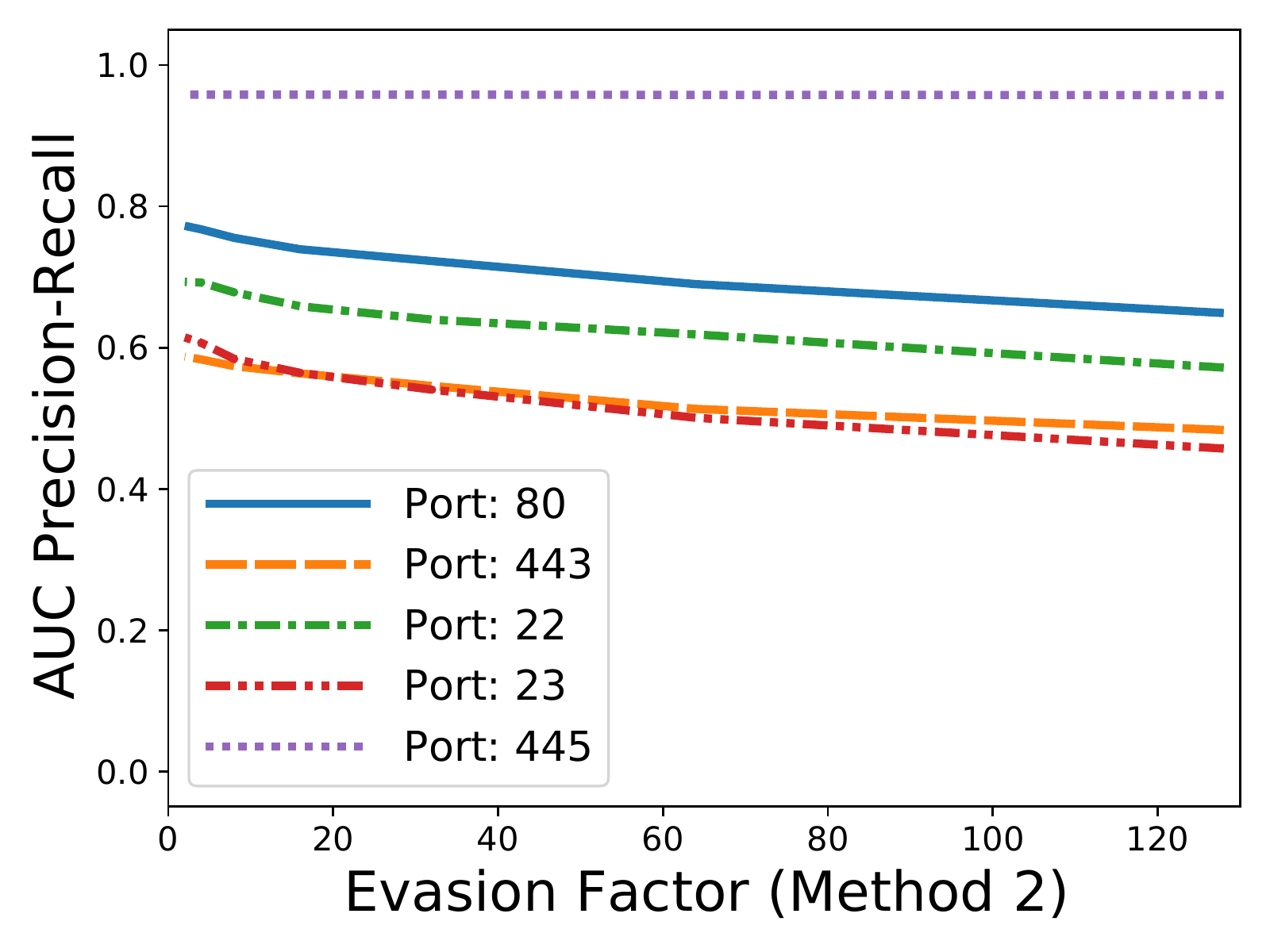}
	\end{subfigure}

	\begin{subfigure}[b]{0.49\linewidth}
	\includegraphics[width=\linewidth]{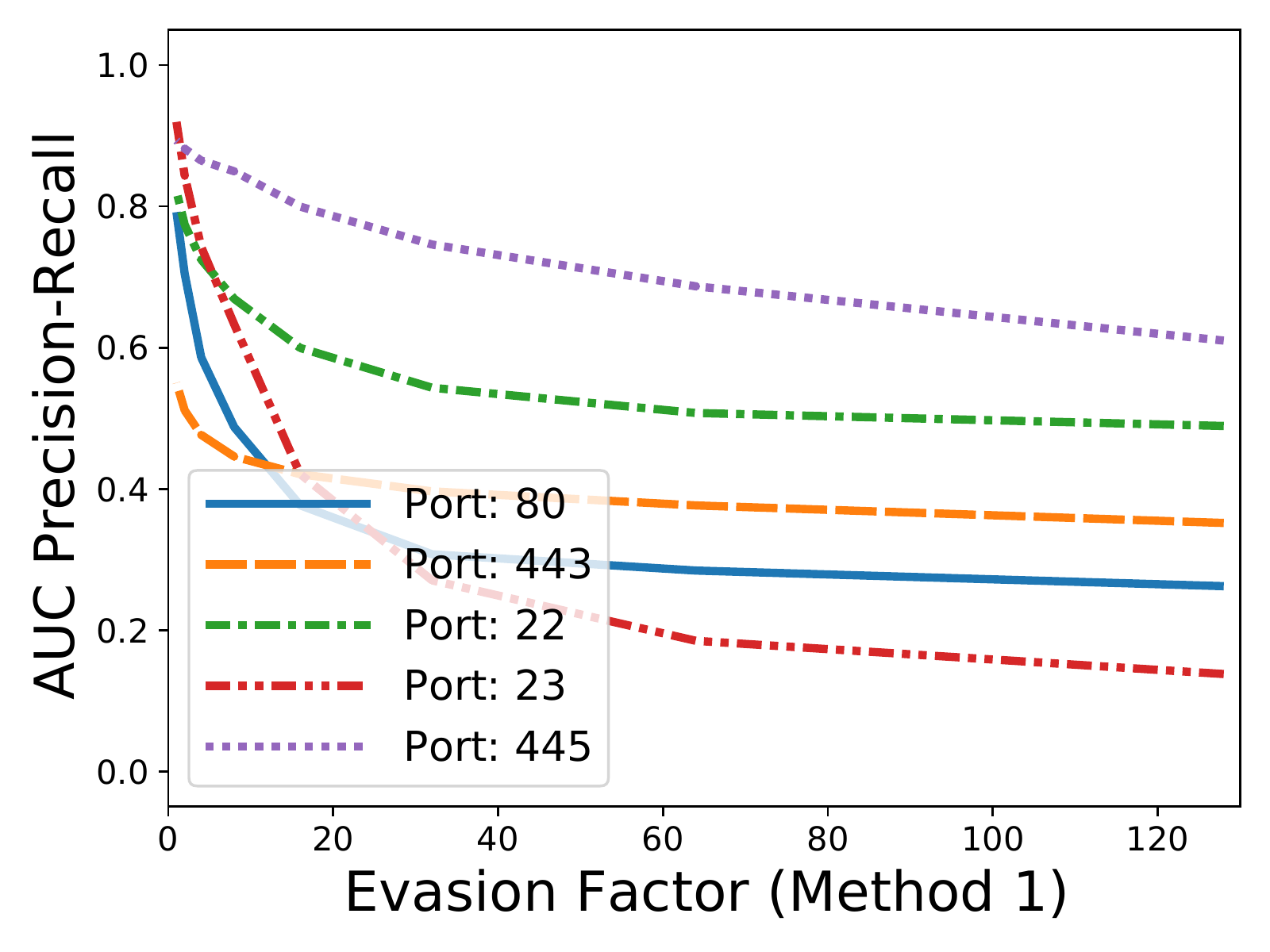}
	
\end{subfigure}
\begin{subfigure}[b]{0.49\linewidth}
	\includegraphics[width=\linewidth]{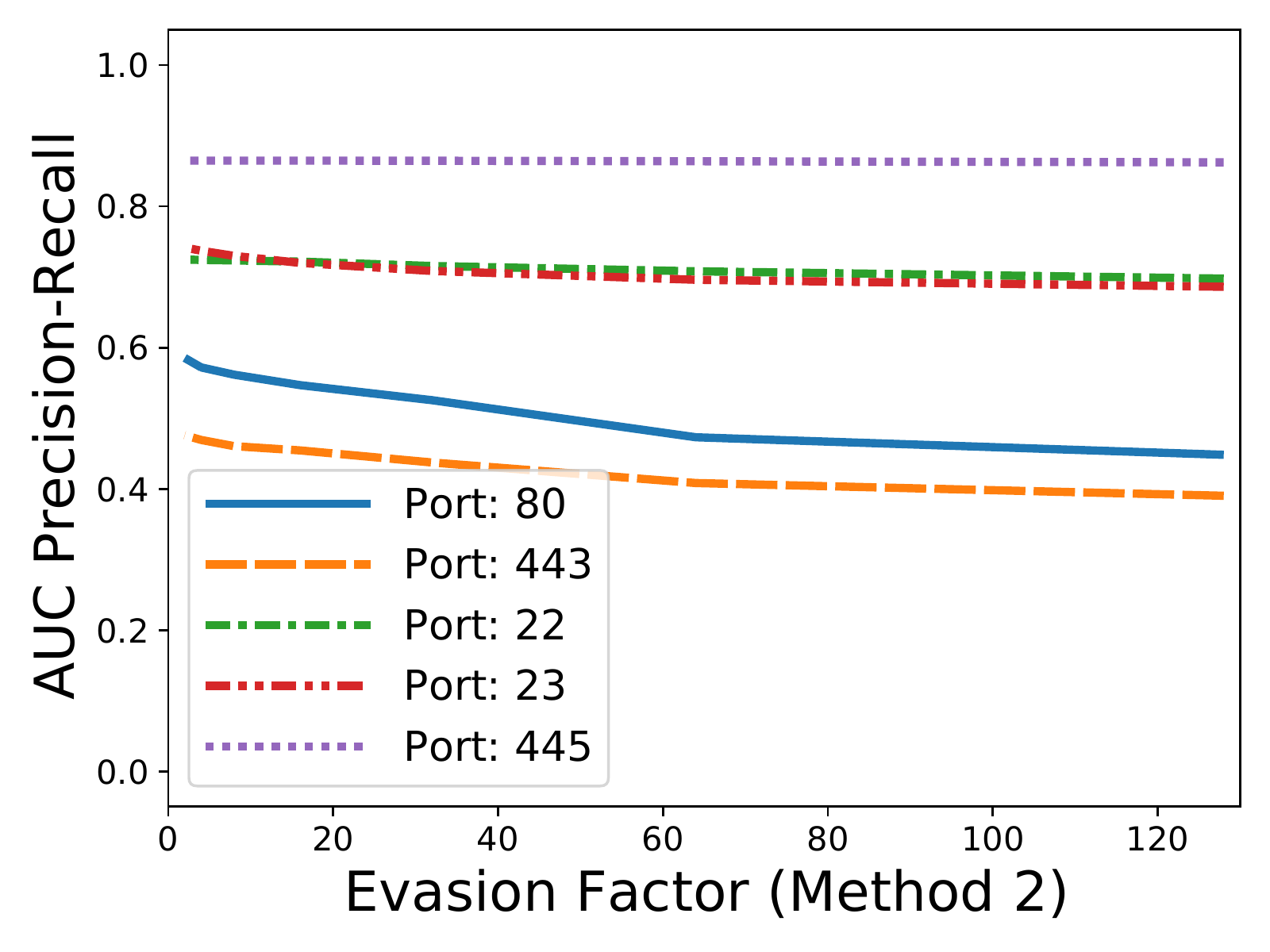}
\end{subfigure}
	%\vspace{-1\baselineskip}
	\caption{Mean Ensemble of KDEs (top) and Isolation Forest (bottom). Mean ensemble model is more resilient against evasion compared to baseline ML models.  Evasion Method 1 (slowing down the probing rate) and 2 (leveraging IP destinations from history) are shown.}
	\label{fig:evasion_mean_ensemble}
	%	\vspace{-4mm}
\end{figure}

\begin{figure}[]
	\centering
	\vspace*{-2mm}
	\begin{subfigure}[b]{0.49\linewidth}
		\includegraphics[width=\linewidth]{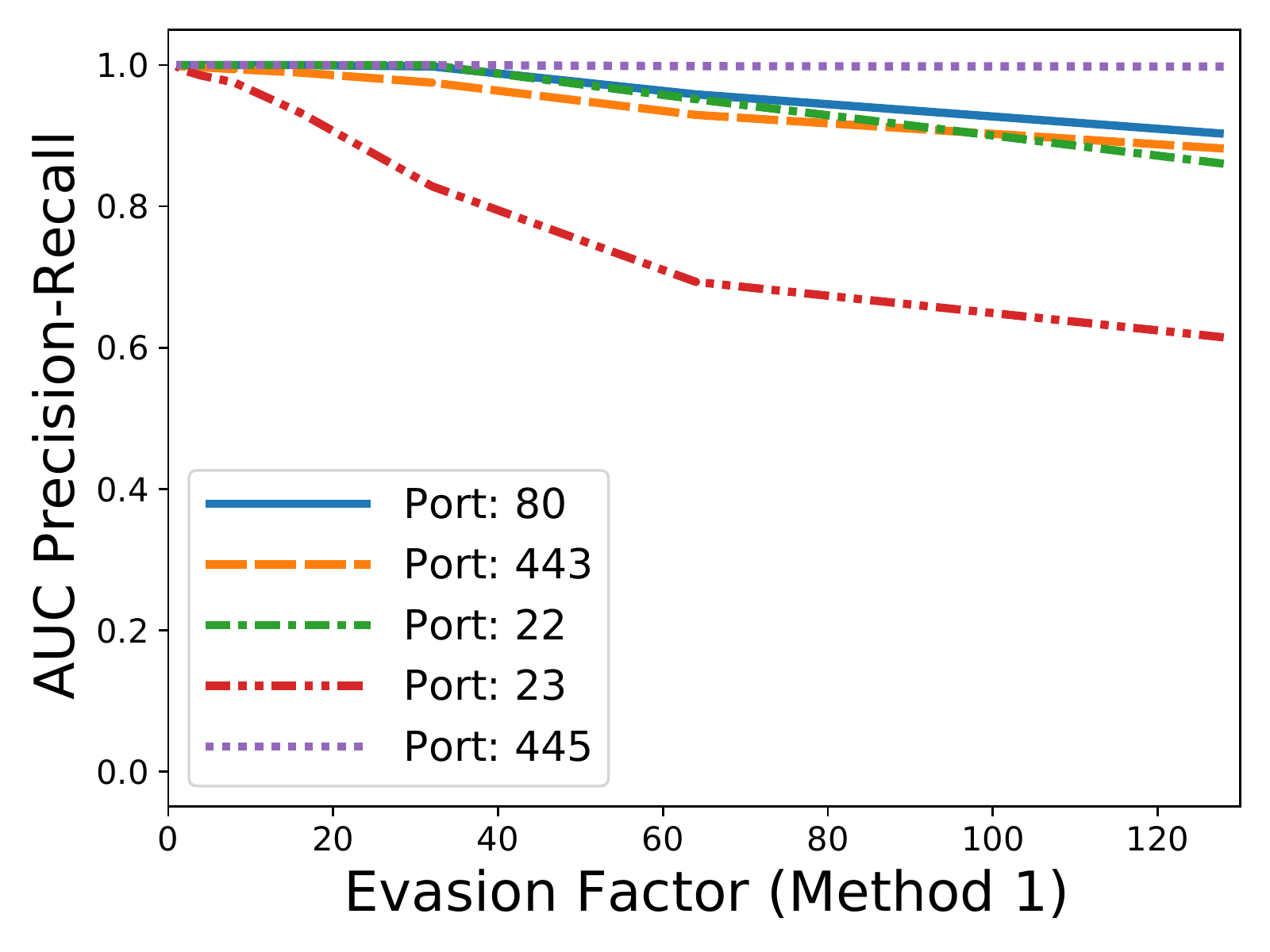}
		
	\end{subfigure}
	\begin{subfigure}[b]{0.49\linewidth}
		\includegraphics[width=\linewidth]{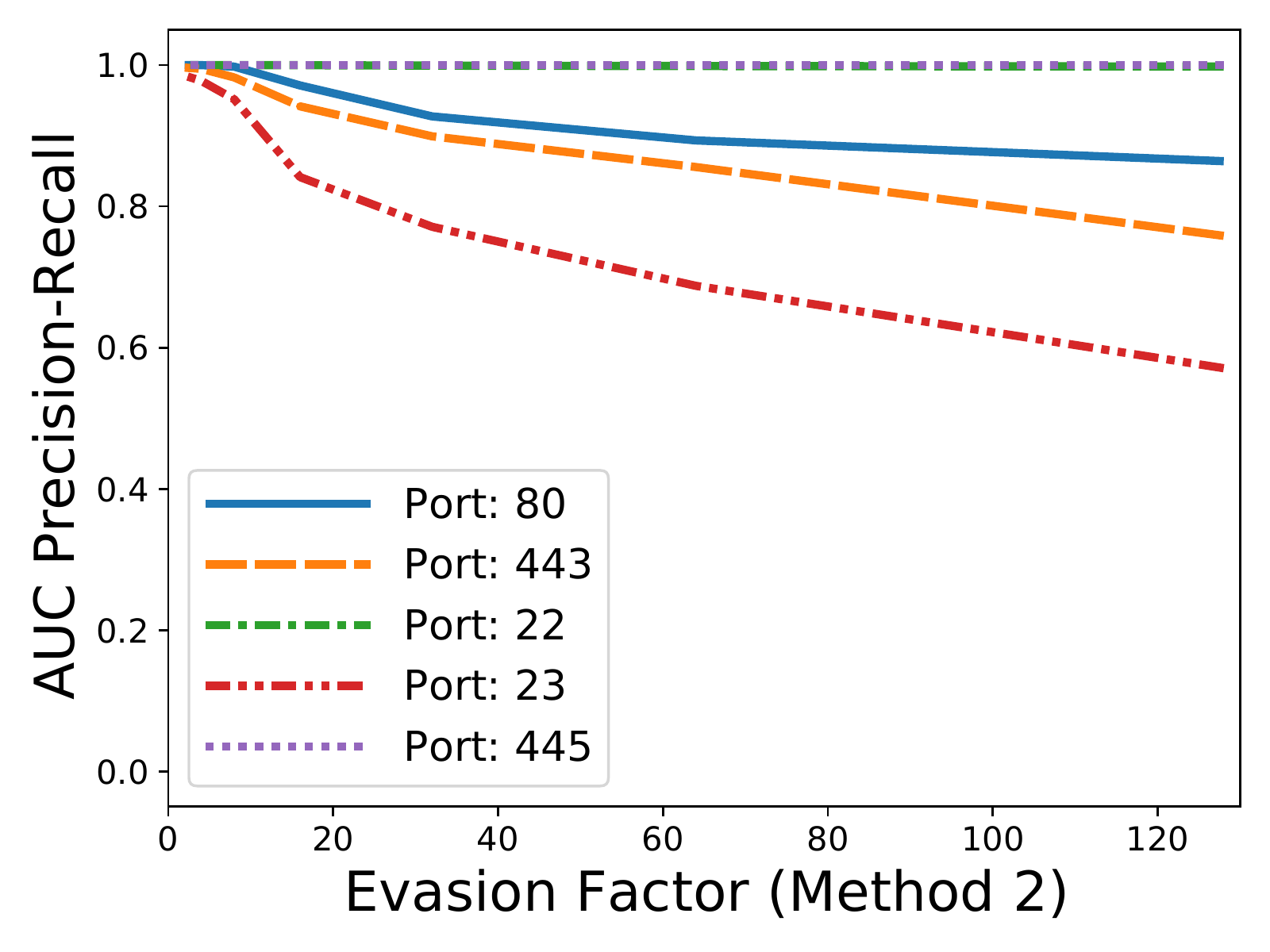}
	\end{subfigure}
	
	\begin{subfigure}[b]{0.49\linewidth}
		\includegraphics[width=\linewidth]{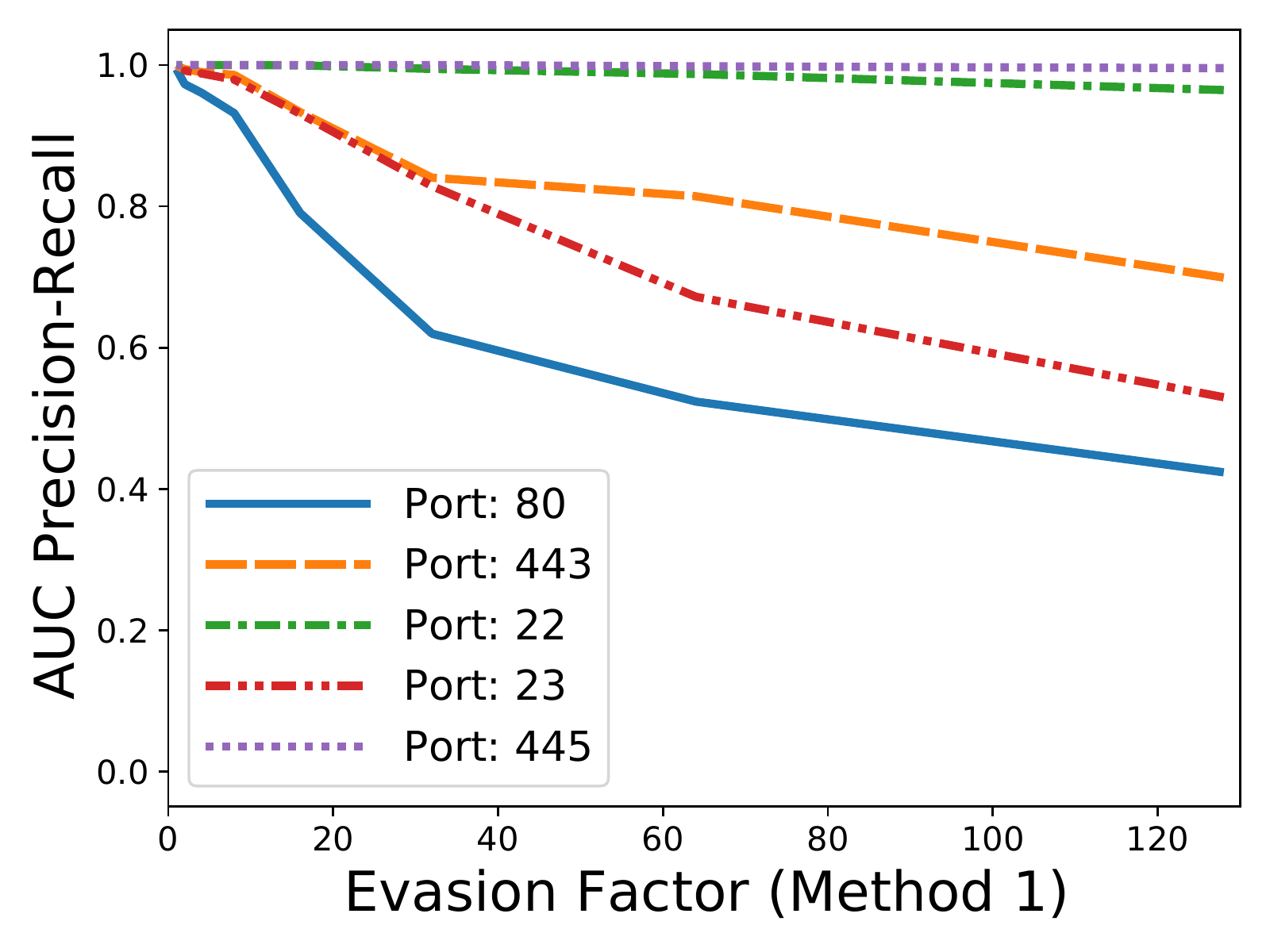}
		
	\end{subfigure}
	\begin{subfigure}[b]{0.49\linewidth}
		\includegraphics[width=\linewidth]{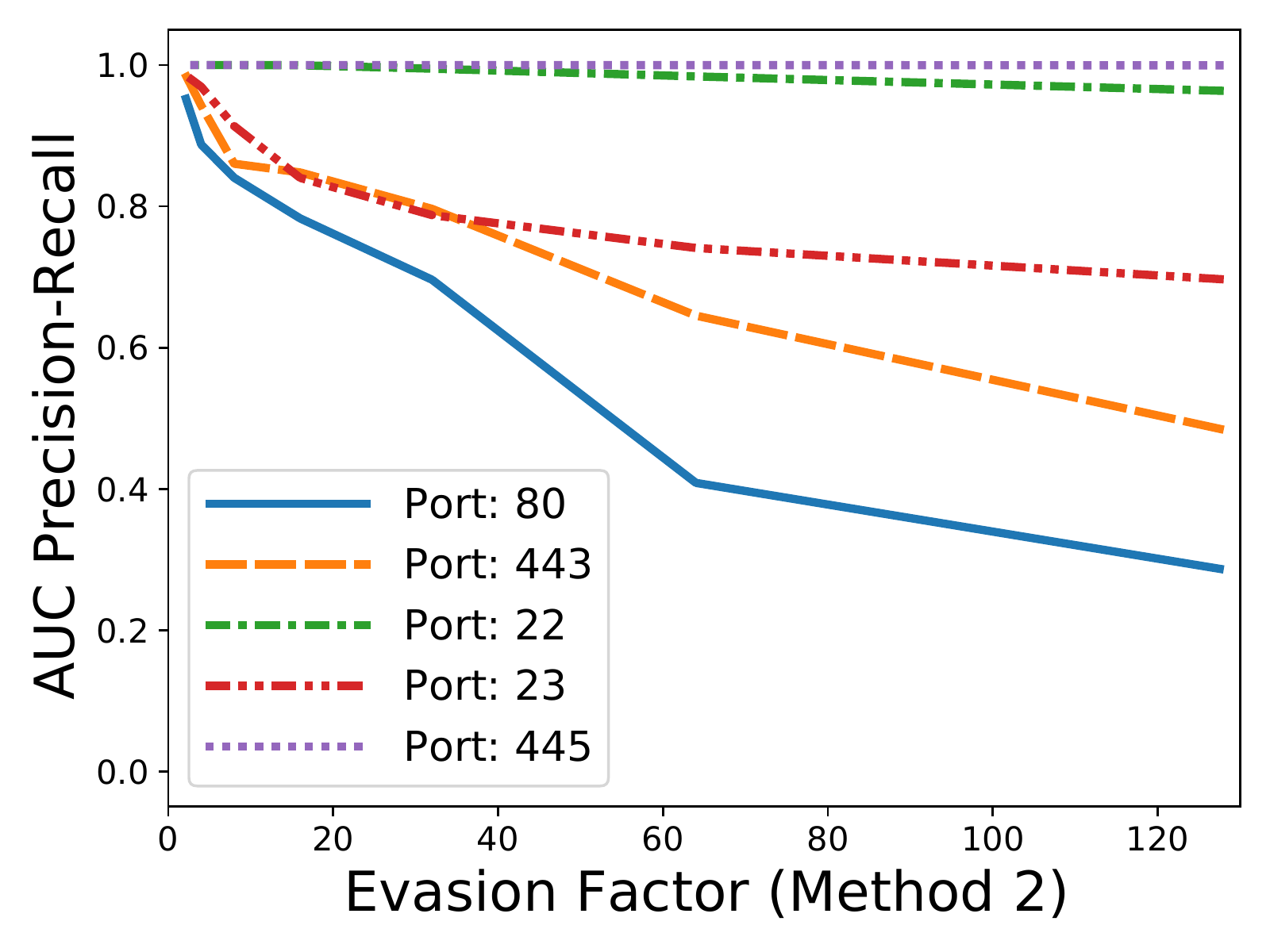}
	\end{subfigure}
	%\vspace{-1\baselineskip}
	\caption{Weighted Ensemble of KDEs (top) and Isolation Forest (bottom). Ensemble model is more resilient against evasion as we learn how to weight each model properly.  Evasion Method 1 (slowing down the probing rate) and 2 (leveraging IP destinations from history) are examined.}
	\label{fig:evasion_ensemble}
	%	\vspace{-4mm}
\end{figure}

 Mean ensembles using uniform weights are presented in Figure~\ref{fig:evasion_mean_ensemble} for both KDE (top) and Isolation Forest (bottom) models. This method is generally more resilient against both evasion strategies than the baseline ML models on port 80 and 443, with no knowledge about the attack, and it is more stable as the evasion factor increases. Experiments in Figure~\ref{fig:evasion_ensemble} use weighted ensembles, where the weights represent normalized feature importance coefficients. 

Figure~\ref{fig:evasion_ensemble} shows that the weighted ensemble is the most resilient technique against both evasion strategies on all ports. For instance, at a propagation rate 128 times slower than the original WannaCry, it still maintains \prauc\ above 0.8 for ports 80 and 443. 

\begin{figure}[]
	\centering
	\begin{subfigure}[b]{0.49\linewidth}
		\includegraphics[width=\linewidth]{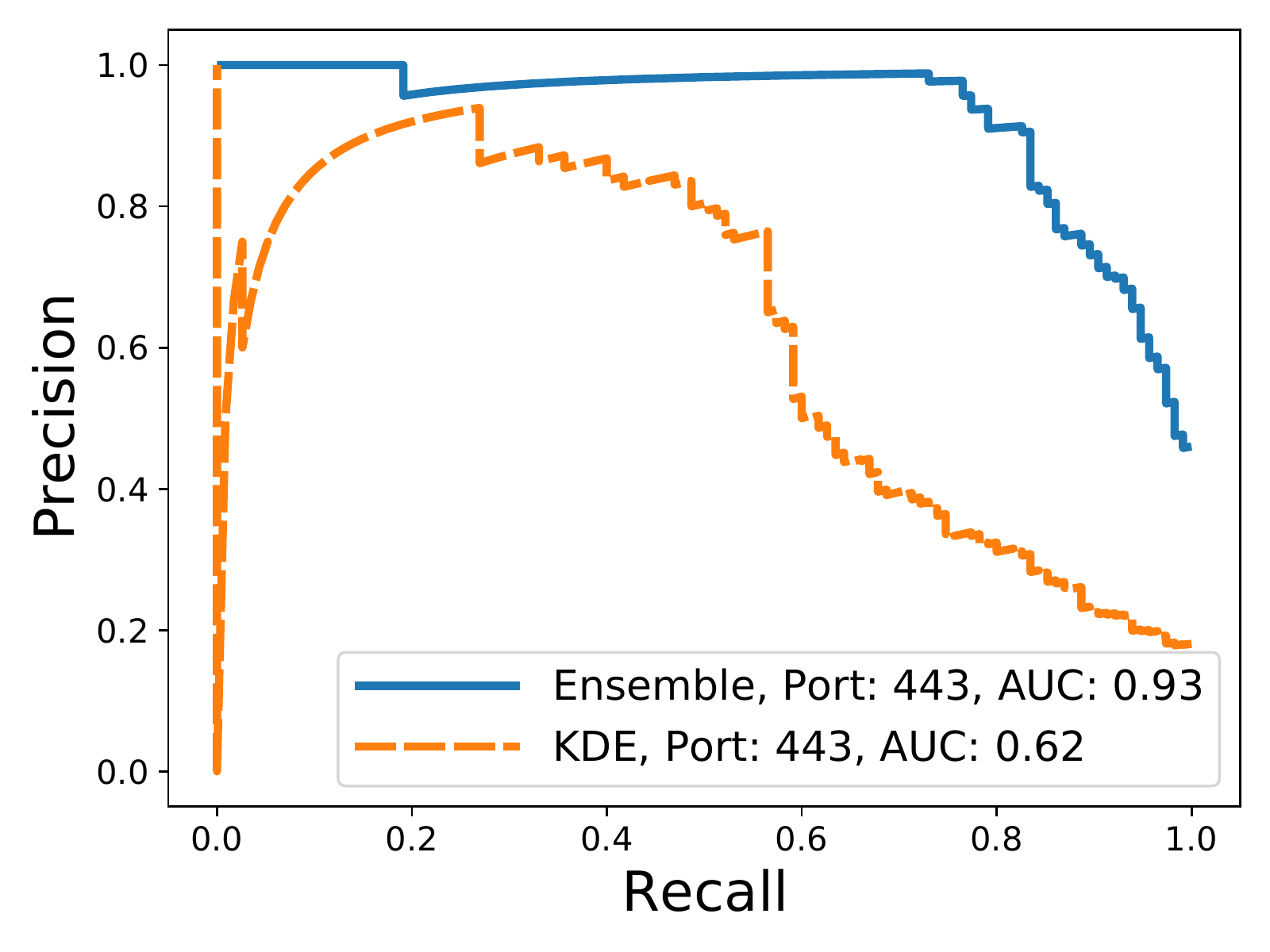}
		
	\end{subfigure}
	\begin{subfigure}[b]{0.49\linewidth}
		\includegraphics[width=\linewidth]{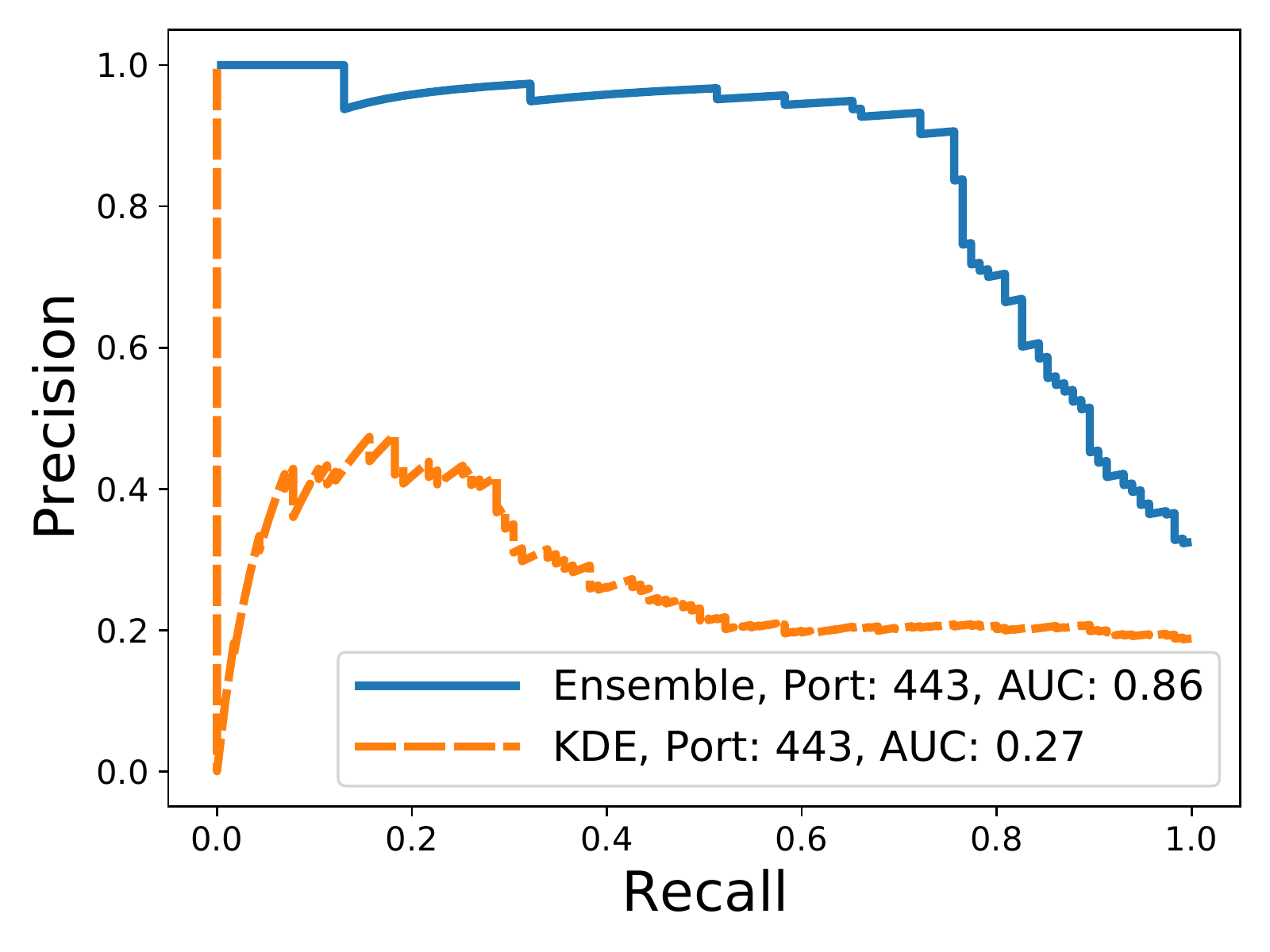}
	\end{subfigure}
	%\vspace{-1\baselineskip}
	\caption{Precision-Recall curves demonstrate the improved performance of weighted ensembles compared to the baseline KDE model on port 443 using the WannaCry variant with  evasion factor 64. Graphs for both evasion methods are included: Evasion Method 1 — slowing down the probing rate (left)  and Evasion Method 2 — leveraging IP destinations from history (right).}	
	\label{fig:evasion_comparison}
	\vspace{-4mm}
\end{figure}

\begin{figure*}[]
		\centering
\begin{subfigure}[b]{0.27\linewidth}
	\centering
	\includegraphics[width=\textwidth]{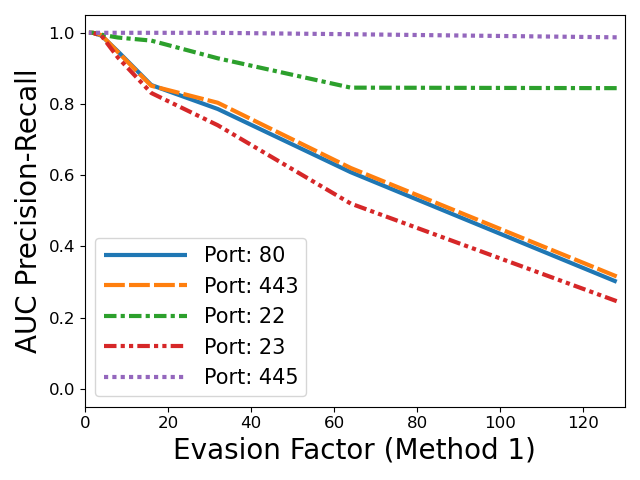}
	
\end{subfigure}
\hspace{0.5cm}
\begin{subfigure}[b]{0.27\linewidth}
	\centering
	\includegraphics[width=\textwidth]{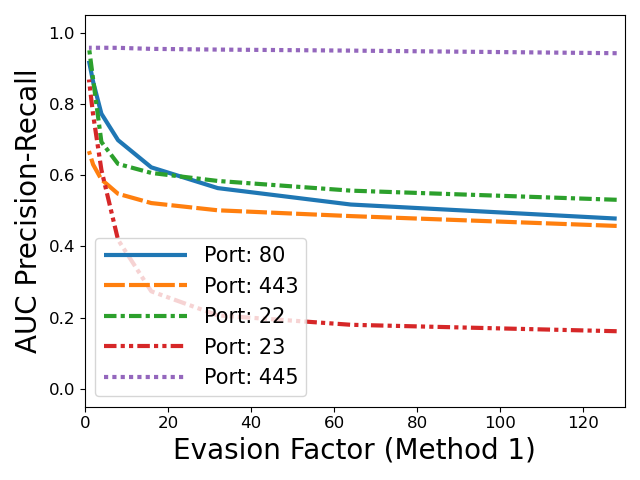}
	
\end{subfigure}
\hspace{0.5cm}
\begin{subfigure}[b]{0.27\linewidth}
	\centering
	\includegraphics[width=\textwidth]{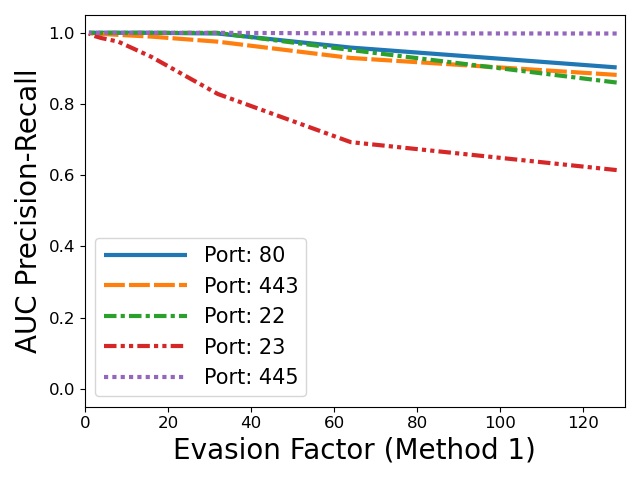}
	
\end{subfigure}
	%\vspace{-1\baselineskip}
	\centering
	\begin{subfigure}[b]{0.27\linewidth}
		\centering
		\includegraphics[width=\textwidth]{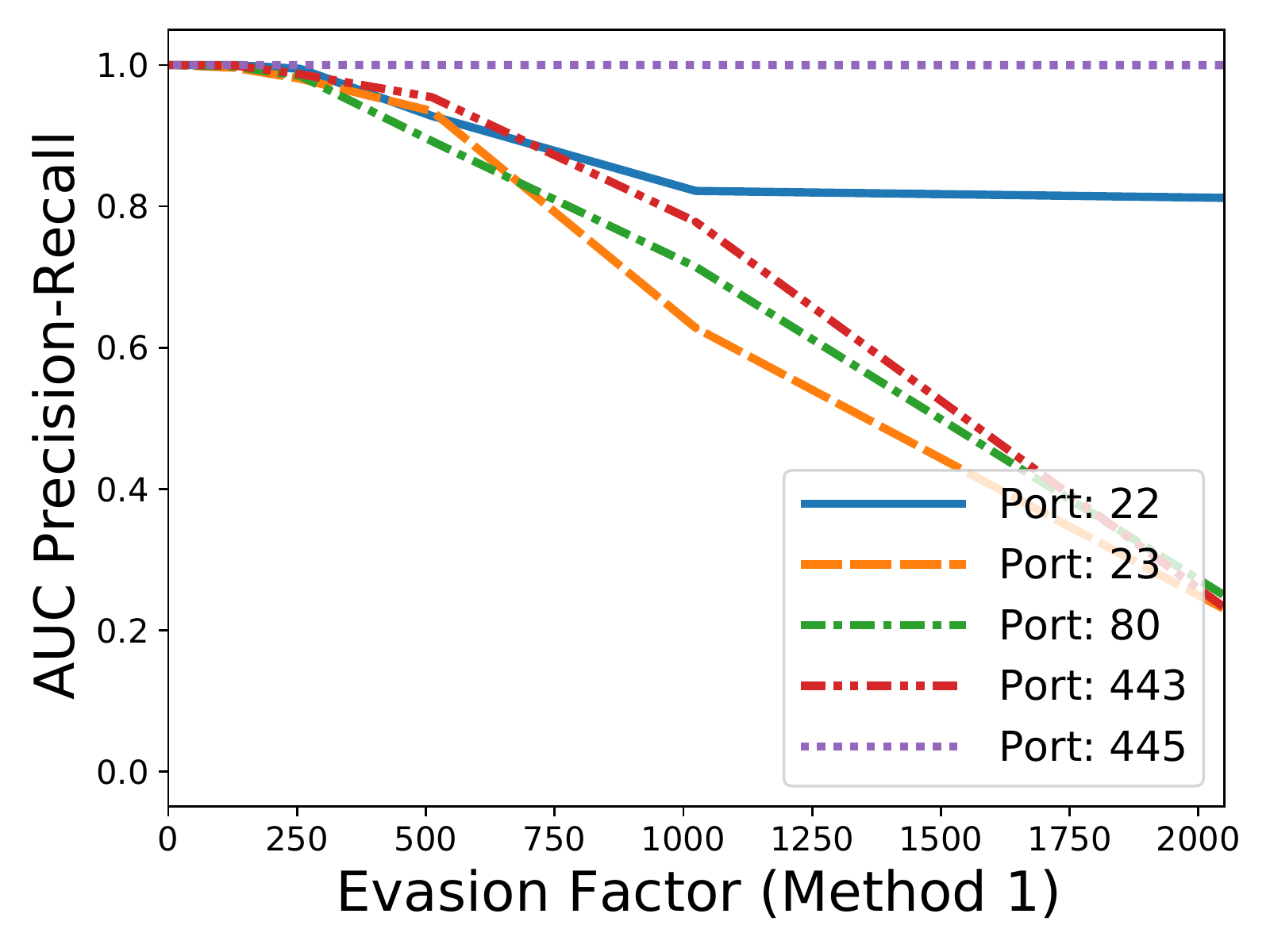}
		\caption{\textit{Multi-feature KDE}}
	\end{subfigure}
	\hspace{0.5cm}
	\begin{subfigure}[b]{0.27\linewidth}
		\centering
		\includegraphics[width=\textwidth]{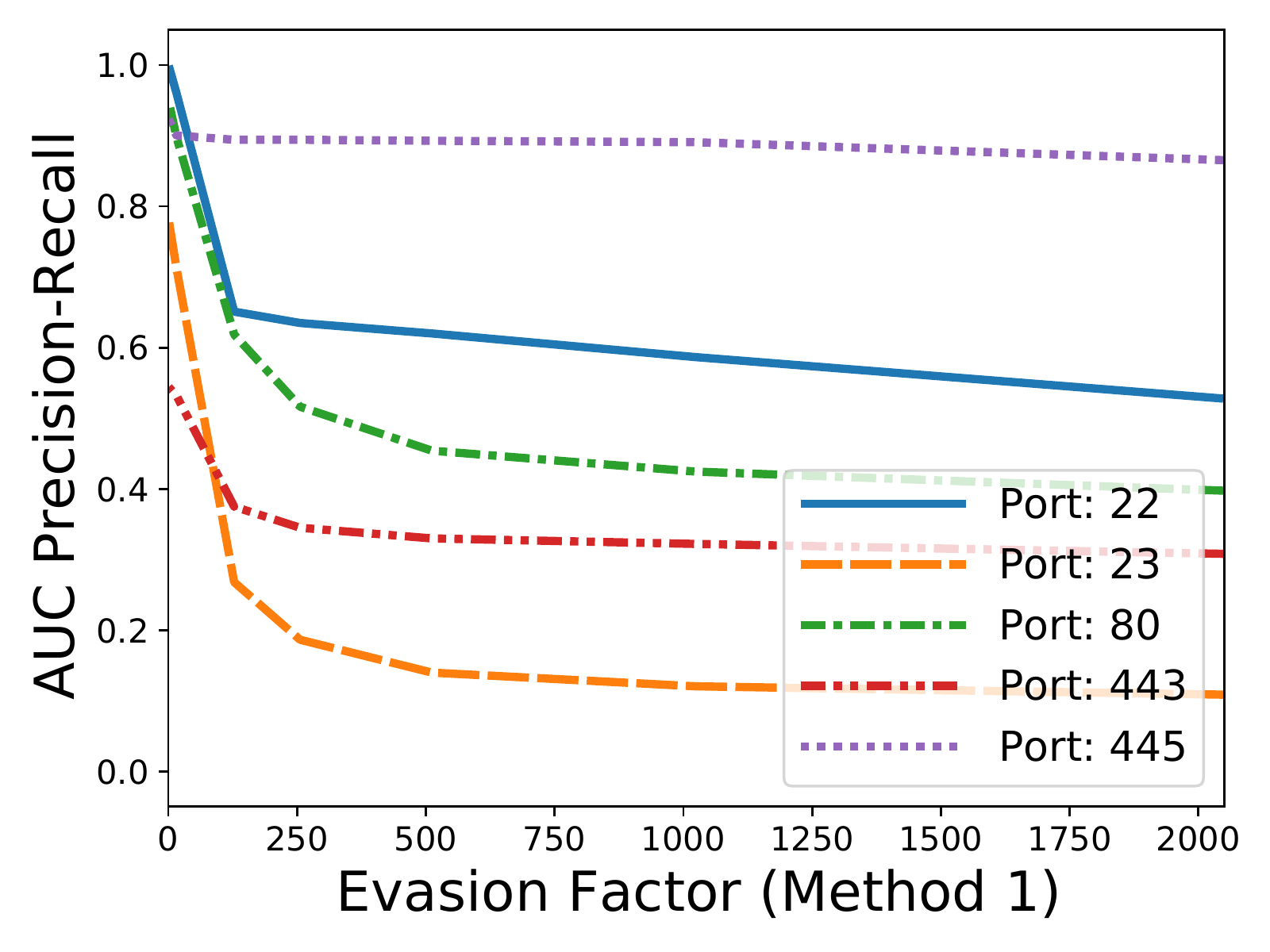}
		\caption{\textit{Ensemble KDE (mean)}}
	\end{subfigure}
	\hspace{0.5cm}
	\begin{subfigure}[b]{0.27\linewidth}
		\centering
		\includegraphics[width=\textwidth]{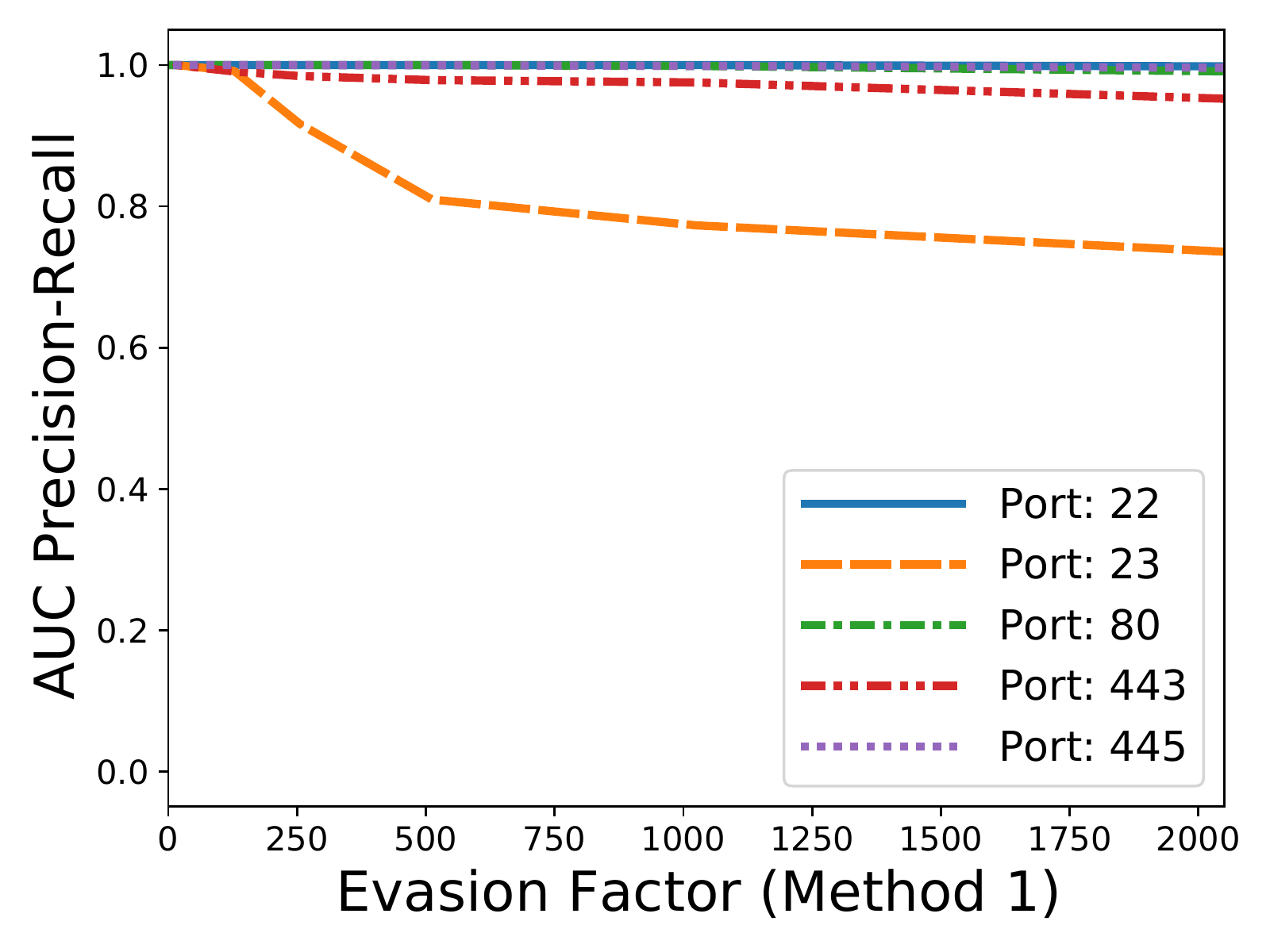}
		\caption{\textit{Ensemble KDE (weighted)}}
	\end{subfigure}
	\caption{Evasion comparison of WannaCry (top) and Mirai (bottom) variants on baseline, mean, and weighted ensemble models.  Ensemble models are more resilient for both malware against Evasion Method 1 — slowing down the probing rate. Evasion factor for Mirai needs to be significantly higher to impact the detection performance due to more aggressive scanning.}
	\label{fig:evasion_mirai_wc}
	%	\vspace{-4mm}
\end{figure*}

Compared to KDE ensembles, the Isolation Forest ensembles have slightly worse performance in terms of their overall resilience to evasion.
The resilience of both ensemble methods is explained by the better generalization achieved by bagging~\cite{Bagging}. The weighted ensembles perform best against evasion, using the additional knowledge of the attack for determining the most relevant features and assigning  weights to different models in the ensemble. In practice, attack knowledge could be obtained from threat intelligence sources, domain experts, or from threat information shared by other networks that detected the attack beforehand.

Figure~\ref{fig:evasion_comparison} presents a direct comparison of weighted ensembles against the baseline KDE model. The \pr\ curves demonstrate the significant improvement of weighted ensembles  on port 443 using the WannaCry variant with the evasion factor 64. The weighted ensemble achieves \prauc\ of  0.93 and 0.86 for the two evasion methods, compared to values of 0.62 and 0.27 for baseline KDE.  

However, of all methods explored, weighted ensembles are the most resilient against evasion.

Robustness evaluation of Mirai detection against evasion produced comparable results. As illustrated on Figure~\ref{fig:evasion_mirai_wc}, the evasion rate has to be significantly greater to impact the results due to the fast scanning rate of original Mirai.

\myparagraph{Ranking Evaluation.} 
%\label{subsec:ranking}
We evaluate our method for ranking the alerts across all ports, as described in Section~\ref{subsec:ranking_method}. In this experiment, the attack sample is merged on one of the ports, and the most suspicious samples on all five ports are ranked to provide a unified alert list. Table~\ref{tab:ranking_100}  shows the statistics of the top-100 ranked alerts, which are ranked using the weighted ensembles of KDE models to detect the original and 8x slower WannaCry variant. The metrics are defined based on 7200 samples over 5 ports, with 116 of them malicious.
We observe that precision in the top-100 alerts is very high (100\% on ports 80, 443, and 445), and the minimum is 94\% on port 23. In addition, the False Positive Rate is lower than $8 \times 10^{-4}$ for the original WannaCry. Results are similarly promising on the slower variant. We also experimented with weighted ensembles of IF models, and the results are slightly worse.
Table~\ref{tab:ranking_comp} shows a comparison of ranking results for the slow WannaCry variants (1/8), and the results demonstrate that KDE models outperform the Isolation Forest models which is in line with the previous experiments where the two models are compared.

\begin{table}[]
	\centering

\begin{tabular}{|c|cc|cc|}
	\multicolumn{1}{c}{}  & \multicolumn{2}{c}{\textit{Original Wannacry}}                         & \multicolumn{2}{c}{\textit{Slow WannaCry}}                            \\ 
	\hline
	\textbf{Port}         & \textbf{Prec}  & \textbf{FPR}   & \textbf{Prec}  & \textbf{FPR}   \\ 
	\hline
	\textbf{80}           & 1.0            & 0      & 0.93           & 0.0009           \\ 
	\hline
	\textbf{443}          & 1.0            & 0       & 1.0            & 0             \\ 
	\hline
	\textbf{22}           & 0.95           & 0.0007      & 0.94           & 0.0008  \\ 
	\hline
	\textbf{23}           & 0.94           & 0.0008    & 0.73           & 0.005     \\ 
	\hline
	\textbf{445}          & 1.0            & 0    & 0.99           & 0.0001         \\
	\hline
\end{tabular}
	
  \caption{Precision (Prec) and False Positive Rate (FPR) in the top-100 ranked alerts across the ports for the original and slow WannaCry (1/8 rate), demonstrating the strength of weighted ensembles of KDEs with low FPR. } 
  \label{tab:ranking_100}
  \vspace{-4mm}
\end{table}

\begin{table*}[]
	\centering
	\begin{tabular}{c|c|c|c|c|c|c|}
			\cline{2-7}
			\multicolumn{1}{l|}{}                        & \multicolumn{3}{c|}{\textbf{KDE (weighted ensemble)}} & \multicolumn{3}{c|}{\textbf{IF (weighted ensemble)}}  \\ \hline
			\multicolumn{1}{|c|}{\textbf{Infected Port}} & \textbf{Precision} & \textbf{FPR} & \textbf{FP Count} & \textbf{Precision} & \textbf{FPR} & \textbf{FP Count} \\ \hline
			\multicolumn{1}{|c|}{\textbf{80}}            & 0.93               & 0.0009       & 7                 & 0.73               & 0.005        & 27                \\ \hline
			\multicolumn{1}{|c|}{\textbf{443}}           & 1.0                & 0            & 0                 & 0.77               & 0.004        & 23                \\ \hline
			\multicolumn{1}{|c|}{\textbf{22}}            & 0.94               & 0.0008       & 6                 & 0.98               & 0.0002       & 2                 \\ \hline
			\multicolumn{1}{|c|}{\textbf{23}}            & 0.73               & 0.005        & 27                & 0.80               & 0.002        & 20                \\ \hline
			\multicolumn{1}{|c|}{\textbf{445}}           & 0.99               & 0.0001       & 1                 & 1.0                & 0            & 0                 \\ \hline
		\end{tabular}
	  \caption{Comparison of ranking metrics between weighted ensembles of KDE and Isolation Forest  for the slow WannaCry variant (1/8 rate). KDE models outperform the Isolation Forest models.} 
	\label{tab:ranking_comp}
	
\end{table*}

%% file: deploy_aggr.tex
\subsection{Deployment on Two University Networks}
\label{sec:deploy_aggr}

\begin{table*}[]
	\begin{center}
		\begin{tabular}{|c|c|c|c|c|c|c|}
			\hline
			VirusTotal & Number of & \multirow{2}{*}{Ports} & \multicolumn{3}{c|}{Connections per external IP} & Total number of \\
			\cline{4-6} Score & external IPs & &max & min & avg &internal IPs \\
			\hline
			8	&1 	&80	&844	&844	&844  &844\\
			6	&3	&22	&1577	&318 	&914  &2726\\
			5	&4	&22	&33726	&561 	&10154.8  &38968\\
			4	&4	&22	&2517	&464	&1440.5  &5506\\
			3	&14	&22, 80	&3234	&235	&963.4  &7864\\
			2	&12	&22, 80, 443	&7509	&5	&1456.1  &14519\\
			1	&9	&22, 80, 443	&11757	&605	&2440.4  &3946\\
			0  &47  &22, 80, 443  &16800 &5 &2290  &38503\\
			\hline
		\end{tabular}
	\end{center}
	\caption{\uvanetwork statistics aggregated by VirusTotal score. We include external IPs with the highest number of connections during the top-ranked anomalous windows. Malicious Incoming scanning activities performed on the internal network are also detected by \thesystem. }
	\label{tab:uva_dec}
\end{table*}

\begin{table*}[]
	\begin{center}
		\begin{tabular}{|c|c|c|c|c|c|c|}
			\hline
			VirusTotal & Number of & \multirow{2}{*}{Ports} & \multicolumn{3}{c|}{Connections per external IP} & Total number of \\
			\cline{4-6} Score & external IPs & &max & min & avg &internal IPs \\
			\hline
			6  &1  &22  &472  &472  &472.0  &470\\
			5  &3  &80, 445  &1545  &843  &1274.3  &3797\\
			4  &5  &80, 445, 443, 22  &927  &405  &682.8  &3374\\
			3  &5  &80, 445  &42535  &136  &10396.0  &48952\\
			2  &6  &80, 445, 443  &9205  &856  &3890.0  &21474\\
			1  &9  &80, 445, 443, 22  &25545  &32  &4238.3  &34701\\
			0  &67  &80, 445, 443, 22  &16265  &48  &2965.6  &96406\\
			\hline
		\end{tabular}
	\end{center}
	\caption{\vtnetwork statistics aggregated by VirusTotal score. We include  external IPs with the highest number of connections during the top-ranked anomalous windows.  Malicious Incoming scanning activities performed on the internal network are also detected by \thesystem. }
	\label{tab:vt_dec}
\end{table*}

\begin{figure}[th]
	\centering
	\begin{subfigure}[b]{0.7\linewidth}
		\includegraphics[width=\linewidth]{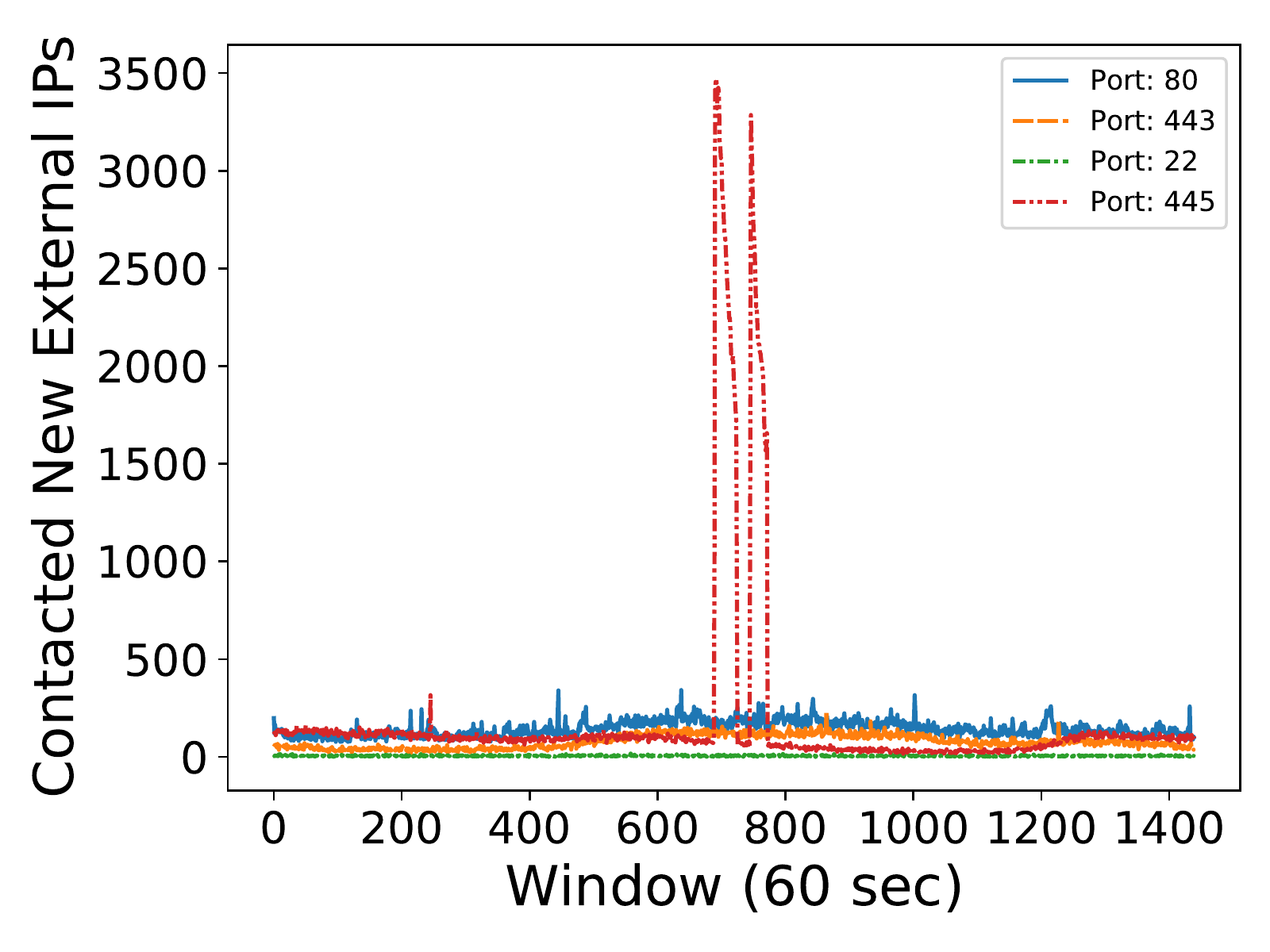}
		
	\end{subfigure}
	%  \begin{subfigure}[b]{0.49\linewidth}
	%  \includegraphics[width=\linewidth]{figs/traffic_plots/uva_new_external_ips_days_sep_10_vt_ts1}
	%  \end{subfigure}
	%\vspace{-1\baselineskip}
	\caption{New external IPs on Sept. 9 on  \vtnetwork. \thesystem\ detected an attack similar to SPM, with a single internal IP attempting to connect to over 15K external destinations. The malicious activity was confirmed by the SOC.}
	
	\label{fig:vt_sep9}
	\vspace{-4mm}
\end{figure}

\myparagraph{Malware Detection ``in the Wild''.} We deployed \thesystem on two university networks, \uvanetwork and \vtnetwork, in order to evaluate our system in a real setting. No merging of malicious traffic was performed in this experiment. We rank the alerts, and consider the top 10 highest risk alerts on both networks. 
At \uvanetwork, we identify the top ranked internal hosts that generate the largest number of connections. Interestingly, 9 out of 10 IP addresses are NAT gateways, VPN servers, or e-mail servers. These servers  aggregate information from multiple clients and have SPM-like behavior. These servers have different behavior than most hosts on the network and are the most anomalous under normal network conditions, when the network is not infected with SPM. 

At \vtnetwork, \thesystem\  detected a series of anomalies on September 9 on port 445, between 11:29am and 12:51pm.  During this time period, the value of the new IP feature significantly increased for some time windows as shown in Figure~\ref{fig:vt_sep9}. After additional investigation, we discovered that all the top 10 detections were part of the same attack. A single internal IP attempted to connect to over 15K external destinations, of which at least 14K failed. This probing behavior appears to be identical to what we observe in the WannaCry logs we collected (in terms of timing patterns, random IP selection, and packet sizes), and we thus suspect this internal IP was infected with WannaCry. The SOC at \vtnetwork\ confirmed that the activity was indeed malicious. 

The \thesystem\ detections on two university networks that were identified by our KDE model include external scanning activities performed on the internal network, as some of the features we use enables our models to detect suspicious behavior in both directions (incoming and outgoing). Tables~\ref{tab:uva_dec} and~\ref{tab:vt_dec} show the statistics on external IPs that were ranked at the top by our model for \uvanetwork and \vtnetwork\ in December, respectively. We confirmed with VirusTotal 47 IPs on \uvanetwork\ and 29 IPs on \vtnetwork\ as having a score at least 1. The December report was shared with the \vtnetwork security team. They confirmed that all external IPs were conducting either malicious or suspicious activities. Several types of attacks were identified: 2 external IPs were carrying out SSH brute force attacks, 15 external IPs were scanning for open SMB, 7 external IPs were scanning for open HTTP / HTTPS ports. These connections were unsuccessful and did not result in bytes transferred.

\myparagraph{Mirai Attack Recreation and Detection.} 
We extended our evaluation by leveraging an actual Mirai attack recreated on the two university networks that lasted 5 days. 
At \uvanetwork, two Openstack networks were set up, consisting of 142 vulnerable VM nodes.
At \vtnetwork, 3 vulnerable VMs with public IPs were configured. In coordination with the security teams at both \uvanetwork and \vtnetwork, the open source Mirai malware was modified to prevent any potential security side effects on the network by allowing infection only in the intended, vulnerable VMs.  After infection, nodes communicated to a command-and-control server under our control on the Chameleon~\cite{keahey2020lessons} service, and attempted to infect other nodes through bi-directional scanning between \uvanetwork and \vtnetwork. The entire public IP space of 500K IPs at both networks was scanned during the attack. 
  Mirai was configured to have a fast scanning behavior on port 2323 (90\% of scanning)  and a slow scanning behavior on port 23 (10\% of scanning). The legitimate and malicious traffic was captured at the network edge via Zeek logs for the entire duration of the experiments. We added capability to Mirai to label the malicious scanning traffic to obtain ground truth for evaluation.

\begin{figure}[]
	\vspace{1mm}
	\centering
	\begin{subfigure}[b]{0.475\linewidth}
		\includegraphics[width=\linewidth]{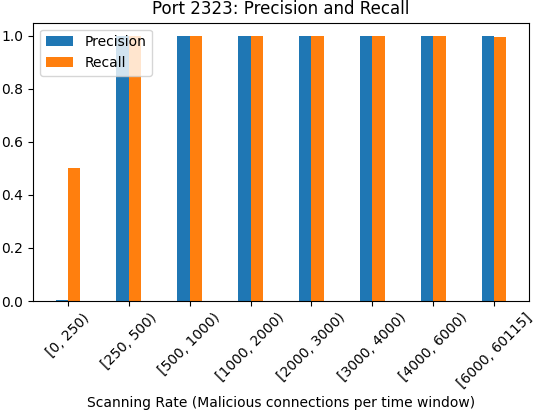}
		
	\end{subfigure}
	\begin{subfigure}[b]{0.49\linewidth}
		\includegraphics[width=\linewidth]{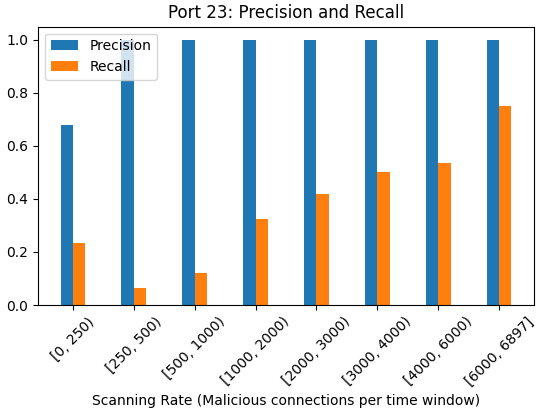}
	\end{subfigure}
	%\vspace{-1\baselineskip}
	\caption{Detection performance by scanning rate for ports 2323 and 23 using Mirai attack recreation. \thesystem detects the attack successfully on port 2323 for scanning rates higher than 250 connections per minute. On port 23, it exhibits increased recall at higher scanning rates.}
	
	\label{fig:mirai_by_speed}
	\vspace{-4mm}
\end{figure}

We deployed \thesystem directly on \uvanetwork using the mean ensemble model, attempting to detect the recreated Mirai attack. The attack starts slowly with a single infected node and the scanning rate (total number of scanning requests per minute) increases as the attack progresses. Figure~\ref{fig:mirai_by_speed} shows the precision and recall metrics at various scanning rates. On port 2323, which has less legitimate traffic, \thesystem is able to detect the attack very well throughout the experiment,  even at low scanning rates. On port 23, performance improves as the scanning rate increases, reaching almost 0.8 recall at a scanning rate of 6000 connections per minute.

\input{eval_autoencoders}

%% file: eval_autoencoders.tex
\begin{figure}[]
	\centering
	%\vspace*{-4mm}
	\begin{subfigure}[b]{0.49\linewidth}
		\includegraphics[width=\linewidth]{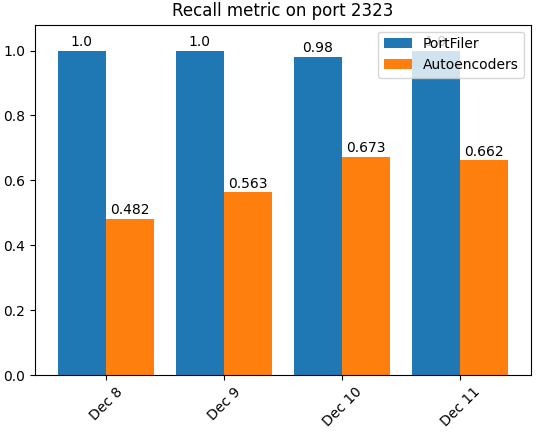}
		
	\end{subfigure}
	\begin{subfigure}[b]{0.49\linewidth}
		\includegraphics[width=\linewidth]{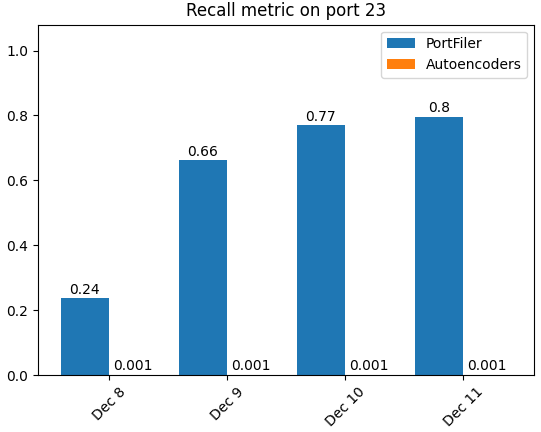}
	\end{subfigure}
	%\vspace{-1\baselineskip}
	\caption{Recall per day showing superior performance of \thesystem\ (mean ensemble method) compared to an unsupervised deep learning approach (autoencoders) on ports 2323 and 23 using Mirai attack recreation.}
	
	\label{fig:autoencoders}
	\vspace{-4mm}
\end{figure}

\myparagraph{Comparison with Unsupervised Deep Learning Models.}
\label{sec:autoencoders}
% \myparagraph{Baseline Unsupervised Deep Learning Detector.}
Using the recreated Mirai attack, we further deepened our analysis of unsupervised malware detection methods. To this end, we conducted several experiments  to provide a baseline comparison between \thesystem\ and an off-the-shelf unsupervised deep learning approach. We designed a feature space equivalent to \thesystem's, by using the same fields from Zeek connection logs. Numerical fields (i.e., duration, bytes, packets) were imported directly and categorical fields (i.e., connection state) were one-hot encoded.  IP fields were omitted given the difficulty of one-hot encoding the extremely large sample space. This resulted in 19 features per connection log. Following a standard autoencoder architecture, we vectorize the input by concatenating a number of feature rows (i.e., 100 connection logs) together, resulting in 1900 features. \ignore{The autoencoder ranks and labels each such input vector using the reconstruction error. }Through an extensive hyperparameter search we established that a one-layer architecture with 128 hidden nodes and no regularization leads to the smallest train/validation loss and reconstruction error. 
We compare this baseline deep learning approach against \thesystem's mean ensemble method, which does not require any information about the attack. We assume a fixed budget of false positive alerts and evaluate the precision-recall metric for the two methods. We experimented with multiple false positive alert budgets, and the findings were consistent across the board: while precision is high for both methods, the recall metric is much lower for autoencoders due to a high number of misclassifications as false negatives. Figure~\ref{fig:autoencoders} shows detection results for a fixed budget of 144 false positive alerts per day. 

There are also practical limitations for deep learning approaches, including higher execution times and memory requirements. The input of autoencoders is not aggregated, in contrast to \thesystem, which uses aggregated features. For ports with a high traffic volume, processing the entire train/test data in memory on a system with 64GB of RAM was not possible, which prompted us to employ batching strategies. These experiments indicate that a basic unsupervised deep learning model is not sufficient to effectively detect SPM attacks. Moreover, deep learning models are much more resource intensive compared to \thesystem.

%% file: discussion.tex
\section{Discussion and Limitations}
\label{sec:discuss}

We have detailed the design and analysis of the \thesystem\ system targeting SPM, and demonstrated its effectiveness to find malicious behavior on large networks and its resilience to two evasion strategies. 
Our system aims to prevent malware infection on a global scale as a result of an unknown vulnerability on an application port. 
However, several challenges in this space remain, as we highlight next. The remainder of this section discusses the scope of our work, as well as the limitations of our approach.

\myparagraph{Deployment in practice.} \thesystem\ uses aggregated port-level features and could be deployed in a feasible manner for any number of selected ports besides the five ports we considered here. The most performance-intense computation is feature extraction, but model training is relatively fast. In practice, \thesystem\ would have to be retrained on a regular basis to adapt to changing network environments and update the learned profiles to changes of legitimate applications. An interesting  future work avenue is to  develop scalable methods for retraining and updating the models gradually.

\myparagraph{Bidirectional Port-Scan Detection} We design our system to collect information on each port independently of the direction of communication. This results in detecting anomalies resulting from both SPM-like activity initiated by internal machines, as well as remote port scanning targeting the monitored network. We show in Section~\ref{sec:deploy_aggr} how we can detect both cases based on the ranking of internal and external communications.

\myparagraph{Coordinated global detection} Our system is placed at the border of a network, monitoring the bipartite communications between internal and external hosts. Self-propagating malware usually attempts to spread on the local network, as well as externally. \thesystem generates alerts once external probing attempts are identified in the network. Reducing the time to detect an SPM is critical for blocking its global spread. We showed through our weighted ensembles that minimal knowledge about an attack can improve its detection. We are interested in how to design coordinated defenses across networks, in which information about detected attacks is shared in a timely manner between different organizations. We envision that sharing attack information would help reduce false positive rates, and decrease the time before an SPM attack is detected.

\myparagraph{Evasion Techniques.} As adversaries are adapting against deployed detectors, further evasion techniques should always be considered for designing a system. We explored two possible techniques that could be employed by malware to evade detection. We showed the resiliency of our system against these techniques at feasible levels. Increasing the evasion factor (lowering the scanning rate to hide the signal completely) inherently cripples the malware's ability to spread successfully. More sophisticated methods could be developed for evasion. Malware that operates on multiple ports, with different spreading mechanisms could be a challenge for our existing models, as the signal will be distributed among different ports. 
Our models and feature set could be extended to work at host level rather than port level, in order to collect more data and thus be resilient against multi-port attacks.

\myparagraph{Poisoning training data.} Poisoning attacks have recently been studied in a variety of machine learning applications~\cite{biggio2012poisoning,Jagielski18}. Advanced adversaries such as advanced persistent threats (APT) could infect the victim network and remain stealthy for long periods of time to poison the models slowly. The traffic profiles on different ports can be changed over time, with a sufficient amount of poisoned data. Attacks of this nature are challenging both for the adversaries to execute, and for the defense mechanisms to detect without expert auditing of changes in the network. We envision the development of complementary techniques that monitor  host-level activities in conjunction with network traffic to detect advanced poisoning attacks.

%% file: related_work.tex
\section{Related Work}
\label{sec:related}

\myparagraph{ML malware detection.} ML for security has been an active area of research, focusing on various attacks and malware types and many vendors have started incorporating machine learning solutions for security~\cite{microsoft2020Feb,symantec2020Jun,fireeye2020Jun}. Perdisci \etal~\cite{perdisci2010behavioral} and Rafique \etal~\cite{rafique2013firma} investigate unsupervised network analysis to build a malware detection system. Mcgrew \etal~\cite{mcgrew2016enhanced} design a supervised network-based malware detection system, with n-gram document analysis. Alahmadi \etal~\cite{alahmadi2018malclassifier} study supervised classification for malware through network flows. 
Botnet detection has been proposed based on network behavior~\cite{gu2008botminer,gu2007bothunter,alahmadi2020botection}  and  C\&C patterns~\cite{bilge2012disclosure,gu2008botsniffer,rossow2013provex}. 
Celik \etal~\cite{celik2015malware} propose building tamper-resistant features for malware detection, for stealthy C2 communication patterns. 
Alahmadi \etal~\cite{alahmadi2020botection} use connection state patterns to detect bots using Markov models. 
Enterprise security analytics has been an active research area~\cite{hu2016baywatch,oprea2018made,yen2013beehive} and ML solutions have been deployed in the industry~\cite{symantec2020Jun}.

\myparagraph{Self-Propagating Malware detection.}
Internet worms have been studied in the past as early examples of SPM~\cite{staniford2002own}. Detection mechanisms based on payload signatures~\cite{singh2003earlybird, kim2004autograph, newsome2005polygraph}, and fine-grained host-level network monitoring~\cite{gu2004worm, xia2006effective, li2014detecting} have been proposed. Our system is designed  independently of specific malware, and is more scalable compared to host-level monitoring.
Recent work proposed detection models for SPM such as WannaCry~\cite{akbanov2019ransomware, chen2017automated} and Mirai~\cite{kumar2019early}, focusing mostly on binary analysis, host-based signals or fingerprinting certain network behavior using DNS requests. Akbanov \etal~\cite{akbanov2019ransomware} propose detection methods against WannaCry using software-defined networking to mitigate the threat. 
Host-based methods that address ransomware attacks in general have also been proposed~\cite{chen2017automated,kharaz2016unveil}. Said \etal~\cite{said2018detection} propose detection of the Mirai using syntactic and behavioral analysis combining binary-level features with dynamic analysis on the host.
 Kumar \etal~\cite{kumar2019early} discuss  detection of Mirai at the probing phase by leveraging Mirai traffic signatures. They sample the packets transmitted by IoT devices both across time and device to find probing activities matching Mirai's behavior.

\myparagraph{Port scanning detection} 
Early studies proposed rule-based detectors by simply counting the number of hosts probed by an external IP~\cite{heberlein1989network,roesch1999snort,lee2003detection,patel2016rule}. These techniques are not resilient against active adversaries that change their scanning behavior, and they result in high false positive rates. 
Staniford \etal~\cite{staniford2002practical} propose complex models to detect slow scans by keeping track of external IPs over long periods of time. Jung \etal~\cite{jung2004fast} build a model to identify scans quickly, based on failed connections using hypothesis testing on each external IP. This method fails in cases of active adversaries alternating between successful and failed connections to evade detection. 
Raftopoulos \etal~\cite{raftopoulos2015dangerous} conduct a study that shows the dangers of the benign-looking port scanning activity on the internet. 
 Ring \etal~\cite{ring2018detection} propose a feature-based approach to build both supervised and unsupervised models to detect slow port scans. Most of these methods require keeping track of individual IP addresses, and either fail against active adversaries, or are computationally expensive. While the well-known SPM attacks performed rapid scanning to have a significant global impact, future attacks can be designed to be more stealthy by performing scanning at slower rates. Our work tries to bridge the gap in the defense side against this new wave of automated, self-propagating attacks.

%% file: conclusion.tex
\section{Conclusion}
\label{sec:conc}

Self-propagating malware is a prevalent threat on the Internet. Recent  campaigns  demonstrate that SPM can propagate fast and cause global disruption. We propose \thesystem, an ML-based anomaly detection system on network traffic for detecting SPM. \thesystem\ extracts port-based features from network logs, profiles legitimate activity on network ports using newly designed unsupervised ensemble methods, and ranks anomalies for SOC investigation. We evaluate \thesystem\ using  Zeek network logs from two university networks and  several SPM families, and show its ability at detecting SPM with high precision and recall, and low false positives. We also show the resilience of our newly-introduced ensemble methods against evasion and its benefits compared to standard ML and deep learning methods.

\thesystem\  detected SPM attacks and malicious scanning activities on the two university networks it was deployed, confirmed by the university SOC.